\documentclass[manuscript]{aastex}

\usepackage{amssymb}
\usepackage{amsmath}
\usepackage{subfigure}
\usepackage{overcite}
\usepackage{lastpage}
\usepackage{fancyhdr}
\usepackage{booktabs}
\usepackage{multirow}

\usepackage{geometry}
\usepackage{pifont}
\usepackage{natbib}
\usepackage{txfonts}
\usepackage{hyperref}
\usepackage{color}

\usepackage{graphicx}
\usepackage{epsfig}
\usepackage{verbatim}
\usepackage{textcomp}
\begin{document}


\title{Convection in Oblate Solar-Type Stars}


\author{Junfeng Wang\altaffilmark{1,2}, Mark S. Miesch\altaffilmark{2} and Chunlei Liang\altaffilmark{1}}

\altaffiltext{1}{High Altitude Observatory, National Center for Atmospheric Research, Boulder, CO 80301, United States}
\altaffiltext{2}{Department of Mechanical and Aerospace Engineering, George Washington University, DC 20052, United States}



\newcommand{\pd}{\partial}
\newcommand{\del}{\nabla}
\newcommand{\bdot}{\mbox{\boldmath $\cdot$}}
\newcommand{\uvn}{\mathbf{n}}
\newcommand{\cross}{\mbox{\boldmath $\nabla \times$}}
\newcommand{\surf}{\mbox{\boldmath ${\cal S}$}}



\begin{abstract}
We present the first global 3D simulations of thermal convection in the oblate envelopes of rapidly-rotating solar-type stars.  This has been achieved by exploiting the capabilities of the new Compressible High-ORder Unstructured Spectral difference (CHORUS) code.  We consider rotation rates up to 85\% of the critical (breakup) rotation rate, which yields an equatorial radius that is up to 17\% larger than the polar radius.  This substantial oblateness enhances the disparity between polar and equatorial modes of convection.  We find that the convection redistributes the heat flux emitted from the outer surface, leading to an enhancement of the heat flux in the polar and equatorial regions.  This finding implies that lower-mass stars with convective envelopes may not have darker equators as predicted by classical gravity darkening arguments.   The vigorous high-latitude convection also establishes elongated axisymmetric circulation cells and zonal jets in the polar regions.  Though the overall amplitude of the surface differential rotation, $\Delta \Omega$, is insensitive to the oblateness, the oblateness does limit the fractional kinetic energy contained in the differential rotation to no more than 61\%.  Furthermore, we argue that this level of differential rotation is not enough to have a significant impact on the oblateness of the star.
\end{abstract}





\section{INTRODUCTION}\label{sec:intro}
Most stars are born spinning rapidly.  Observations of young clusters reveal a wide range of rotation rates, with some stars spinning more than 100 times faster than the Sun \citep[e.g.][]{soder93,meibo09,hartm10}.  This scatter is generally attributed to the vortical nature of the turbulent interstellar clouds from which these stars form.  Though much of the initial angular momentum is lost during the gravitational collapse of a protostellar core, the dramatic decrease in the moment of inertia during the collapse is enough to impart a substantial rotational velocity in a large fraction of young stars.  As the stars age, torques from magnetized stellar winds cause them to spin down and this reduces the scatter of the general population \citep{skuma72,soder01,barne03,meibo15,revil15}.

The rapid rotation of young stars can have a dramatic influence on their internal structure and dynamics.  In the most extreme cases, the centrifugal force can alter the stars' shape, causing it to be significantly oblate.  Models of rotating stars suggest that the effects of rapid rotation and oblateness can alter the stars's convective and radiative heat transport, reduce their luminosities, induce shearing instabilities and global circulations, and modify their rotational and compositional evolution \citep{Zahn1992,Pinsonneault1997,Meynet2000,Maeder2001,MacGregor2007,Lara2013}.  Much of this work has focused on high-mass stars because of their short lifetimes, which makes them intrinsically more likely for fast rotation to play an important evolutionary role from birth to death, when they explode as supernovae and enrich the Galaxy with heavy elements.  High-mass stars are also luminous and large, which makes them relatively easy to observe.  In the last fifteen years, researchers have been able to directly detect the oblate shapes of several rapidly-rotating, massive stars by means of optical and infrared interferometry \citep{van2001,McAlister2005,Monnier2007,Zhao2009,van2012}.  More direct detections are likely to be made in the future as observations continue to improve \citep{nunez15}.

One of the major theoretical predictions that has been addressed by these observations is the phenomenon of gravity darkening, which was introduced nearly a century ago in a pioneering paper by \citet{Zeipel1924a}.   For a massive star with a radiative outer envelope, von Zeipel argued that the effective temperature of the stellar surface should scale as $T_{eff} \propto g_{eff}^\beta$, where $\beta = 1/4$ and $g_{eff}$ is the effective gravity, equal to the Newtonian gravity plus the centrifugal force.  This in turn implies that the radiative energy flux emitted from the stellar surface scales as $F \propto g_{eff}^{4\beta}$.  Since the centrifugal force acts to reduce $g_{eff}$ at the equator, gravity darkening predicts that the equatorial regions should be darker than the polar regions.  Though observations have confirmed the existence of gravity darkening, they suggest that the exponent may be somewhat less than predicted, with $\beta \approx$ 0.13-0.25 \citep{van2001,Domiciano2005,McAlister2005,Monnier2007,Zhao2009,van2012}.   Lower-mass, late-type stars with convective envelopes are also expected to exhibit gravity darkening but again, with a smaller latitudinal variation than that predicted by von Zeipel, with $\beta$ possibly as low as 0.08 \citep{lucy67,elara11}.  However, these expectations are based on simplified models of convection, with little observational guidance.

Substantial progress has been made in recent years in modeling the structure of rapidly rotating stars.  A number of models have been used to investigate the effects of oblateness on stellar structure and evolution, with implications for asteroseismology \citep{Ostriker1968,Jackson2004,Jackson2005,roxbu04,MacGregor2007,reese08,reese09,zahn10,deupr11,ouazz15}.  Though these models are ostensibly $2D$, the thermodynamic quantities only depend on the effective gravitational potential.   Furthermore, like most stellar structure models, these models do not address the internal dynamics of oblate stars; they solve only for the hydrostatic structure using a mixing-length formulation for convection, though some do include idealized cylindrical or shellular differential rotation profiles.  \citet{Kong2013} took a somewhat different approach, calculating the non-spherical shape and internal structure of a rapidly rotating gaseous body using a three-dimensional, finite-element code with an unstructured grid.  But again, the fluid dynamics was not considered. Other analytical and numerical work on fluid dynamics in ellipsoidal geometries has focused on flows induced by precession; for a recent example that employs a finite-element approach see \citet{noir13}.

The first numerical model (to our knowledge) capable of capturing the internal dynamics of oblate stars is the ESTER (Evolution STEllaire en Rotation) code \citep{Lara2007,Rieutor2009,Rieutord2013,Lara2013,Rieutord16}.  ESTER solves the axisymmetric, steady-state fluid equations self-consistently, taking into account nuclear energy generation, differential rotation and meridional circulation.  However, convection is still treated using either a turbulent diffusion approximation \citep{Lara2013} or an isentropic approximation \citep{Rieutord16} in which the effective convective heat flux is inferred from the nuclear energy generation and radiative heat flux under the assumption of a steady state.   Furthermore, published ESTER simulations are currently limited to high-mass stars with convective cores ($M > 1.7M_{\odot}$); they have not yet considered convective envelopes.   There have been many global 3D hydrodynamic (HD) and magnetohydrodynamic (MHD) simulations of convection in the envelopes of lower-mass stars that address the rapid rotation regime \citep[e.g.][]{ballo07,Brown2008,miesc09,kapyl11,guerr13,gasti14,fan14,hotta15,feath15,karak15}.  However, these do not take into account the oblateness caused by the centrifugal force.

In this study, we present the first 3D HD simulations of convection in the oblate envelopes of rapidly-rotating solar-type stars.  These simulations are performed using the Compressible High-ORder Unstructured Spectral difference (CHORUS) code recently introduced by \citet[][hereafter WLM15]{wang2015}.  We focus in particular on the influence of the oblateness on the convective structure, heat transport, mean flows, and the thermodynamic stratification.

This paper is organized as follows. In Section \ref{sec:MP} we discuss the CHORUS code and its application to model oblate spheriodal geometries. We follow this in Section \ref{sec:setup} with a description of the numerical experiments that we have performed, which compare oblate simulations at varying rotation rates with their spherical counterparts.  We consider rotation rates up to $0.85\Omega_{crit}$ where $\Omega_{crit}$ is the critical angular velocity at which the outward centrifugal force at the equator equals the inward gravitational force.  We then present our results, focusing first on the convective structure and energetics (Section \ref{sec:con}), then on the energy transport (Section \ref{sec:transport}) and finally on the mean flows and associated thermal gradients (Section \ref{sec:mean}).  We summarize our results, conclusions and future plans in Section \ref{Sec:summary}.

\section{MODELING THE PROBLEM}\label{sec:MP}
Modeling solar and stellar convection is challenging because the flow is turbulent and a vast range of spatial and temporal scales are involved. The Reynolds number $R_{e} = U d/\nu$ (where $U$ and $d$ are characteristic velocity and length scales and $\nu$ is the kinematic viscosity) is typically very high so direct numerical simulations that effectively resolve all scales in the convective envelope are far beyond today's computational affordability. In this study, we only aim to capture large-scale dynamics which are crucial in order to establish convection, differential rotation, and thermal flux balance. To do so, the three-dimensional CHORUS code is employed to perform a series of global simulations. CHORUS solves the equations of hydrodynamics using the high-order spectral difference method based on unstructured grids (WLM15). It can be naturally extended to deal with complex geometries including oblate spheriodal shells. In what follows we will generally refer to a spherical polar coordinate system $(r,\theta,\phi)$, although the CHORUS code employs a Cartesian coordinate system to represent a spherical or oblate spheriodal geometry using an unstructured mesh (Sec. \ref{sec:der_grid}).

\subsection{Hydrodynamic Equations}\label{sec:hydro}
We consider an ideal gas within an oblate spheroidal shell uniformly rotating at a constant rate $\Omega_{0}$ about the $z-$axis. For simplicity, we assume the bulk of mass is inside the inner radius of the oblate spheroidal shell, giving the gravitational acceleration $\mathbf{g}=-g\mathbf{\hat{r}}=-\frac{GM}{r^{2}}\mathbf{\hat{r}}$, where $G$ is the gravitational constant, $M$ is the interior mass and $\mathbf{\hat{r}}$ is the radial unit vector. In slowly rotating solar-type stars, the centrifugal force has typically been neglected in the momentum equation considering its sufficiently small magnitude relative to the buoyancy force \citep[e.g.][]{Miesch2000,Jones2011}. However, in rapidly rotating stars, the centrifugal force is comparable to gravity and has a significant contribution to stellar structure. Thus, in addition to the Coriolis force, we include the centrifugal force in the hydrodynamic equations.  The resulting equations in the rotating reference frame are
\begin{equation}
\frac{\partial \rho}{\partial t} = -\del \bdot (\rho \mathbf{u}),
\label{eqn:mass_conservation}
\end{equation}
\begin{equation}
\frac{\partial (\rho \mathbf{u})}{\partial{t}} = -\del \bdot \left(\rho \mathbf{u} \mathbf{u}\right) - \del p
+\del \bdot \bold{{\tau}} + \rho \mathbf{g} - 2\rho \mathbf{\Omega}_{0} \times \mathbf{u}
- \rho \mathbf{\Omega_{0}} \times (\mathbf{\Omega_{0}} \times r ~ \mathbf{\hat{r}}),
\label{eqn:monen_convervation}
\end{equation}
\begin{equation}
\frac{\partial E}{\partial{t}} = -\del \bdot ((E + p)\mathbf{u})
+ \del \bdot (\mathbf{u} \bdot \mathbf{{\tau}} - \mathbf{f}) + \rho \mathbf{u} \bdot (\mathbf{g}
- \mathbf{\Omega_{0}} \times (\mathbf{\Omega_{0}} \times r ~ \mathbf{\hat{r}})),
\label{eqn:energy}
\end{equation}
where $t$, $p$, $T$, $\rho$ and $\mathbf{u}$ are time, pressure, temperature, density and the velocity vector respectively. $E$ is the total energy per unit volume and is defined as $E = \frac{p}{\gamma -1} + \frac{1}{2}\rho \mathbf{u}\bdot \mathbf{u}$ where $\gamma$ is the adiabatic index (the ratio of specific heats at fixed pressure and volume). $\mathbf{\tau}$ is the viscous stress tensor for a Newtonian fluid and is given by
\begin{equation}
\tau_{ij} = 2\rho \nu \left [ e_{ij} - \frac{1}{3}\del\bdot \mathbf{u}\delta_{ij} \right ],
\end{equation}
where $e_{ij}$ is the strain rate tensor and $\nu$ is the kinematic viscosity. The term $\mathbf{u}\bdot\mathbf{{\tau}}$ in Equation (\ref{eqn:energy}) represents viscous heating. The diffusive flux $\mathbf{f}$ is treated in the form of $\mathbf{f}= - \kappa_{r}\rho C_{p}\del T - \kappa \rho T \del S$ where the first component is a radiation diffusion flux with the molecular radiation diffusion coefficient $\kappa_{r}$ and $C_{p}$ is the specific heat at constant pressure. The second component is the entropy diffusion due to unresolved, subgrid-scale convective motions. This is a crude but commonly used representation of subgrid-scale (SGS) heat transport \citep{Miesch2000,Jones2011}. $S$ is the specific entropy.

The kinematic viscosity $\nu$ and the thermal diffusivity $\kappa$ are assumed to be constant in time and space (independent of radius), with magnitudes as specified in section \ref{sec:setup}.  The radiative diffusivity $\kappa_r$ peaks at the inner boundary and decreases outward.  It is computed from the temperature and density distribution according to the expression $\kappa_r = \kappa_o T^{6.5} \rho^{-2}$.  Since the initial $\rho$ and $T$ profiles depend only on the Roche potential $\Phi$ (see section \ref{sec:IC}), $\kappa_r$ is also initially a function of $\Phi$ alone.  It does evolve in time along with $\rho$ and $T$ but variations along a potential surface remain relatively small so $\kappa_r \approx \kappa_r(\Phi)$.  The value of $\kappa_o$ is chosen such that the total flux passing through the lower boundary is equal to the stellar luminosity $L$ (see section \ref{sec:bdry}).

The use of the total energy equation (\ref{eqn:energy}) is advantageous from the point of view of the conservation of energy but it can pose problems in stellar interiors where the stratification is adiabatic to within about one part in $10^6$.  To capture the small departures from adiabaticity that drive the convection requires a very accurate numerical method.  However, as discussed in Section \ref{sec:setup}, we have artificially increased the luminosity in our models in order to achieve more tractable values of the Mach number.  This has the additional advantage that it increases the superadiabatic entropy gradient.  Thus, the nondimensional amplitude of the entropy gradient, $(D/C_p) \del S$, is only a factor of 600 less than the nondimensional temperature gradient $(D/T) \del T$.  In other words, the departure from adiabatic stratification is more than one part in $10^3$.  We estimate the accuracy of our spectral difference method (described in Section \ref{sec:method}) to be at least $10^{-7}$ for length scales on the order of the entropy gradient and we see no signs of spurious behavior.

\subsection{Numerical Method}\label{sec:method}

In the CHORUS code, the spectral difference (SD) method is used to discretize Equations (\ref{eqn:mass_conservation})-(\ref{eqn:energy}) in space by using their original strong form (differential form) \citep{liang2009}. In contrast, the traditional Discontinuous Galerkin (DG) method starts from the weak (integrated) form of these equations \citep[e.g.][]{Cockburn1998}. The SD method can be viewed as a particular form of nodal DG method \citep{May2011}.  To discretize these equations in space, the computational domain is divided into a collection of non-overlapping hexahedral elements. These elements in physical coordinates ($x$, $y$, $z$) are then transformed to a unit cube ($\xi$, $\eta$, $\zeta$) using an isoparametric mapping with quadratic relations. Consequently, the governing equations are transformed accordingly through Jacobian matrices. In the unit cube, two sets of points are defined, namely the solution and flux points.  In each coordinate direction of ($\xi$, $\eta$, $\zeta$), $N$ solution points are chosen to be the Chebyshev-Gauss-quadrature points and $(N-1)$ flux points are selected to be roots of the Legendre polynomial plus two end points. Then the solution and flux polynomials can be built through Lagrange interpolation basis functions. In order to deal with flux discontinuities across element interfaces, an approximate Riemann solver (Rusanov or Lax-Friedrichs scheme) and BR1 \citep{Bassi1997} scheme are employed for inviscid and viscous fluxes respectively. Over each elemental interface, $N\times N$ flux points are needed to construct both inviscid Riemman and viscous BR1 fluxes where $N$ is the order of the accuracy of the SD method. The CHORUS code has proven to be particularly attractive for parallel computing (WLM15). The high-order SD method is locally intensive in term of computation associated with $N\times N \times N$ solutions points per element and requires a light amount of communication across elemental interfaces with $N\times N$ flux points. For temporal discretization, a fourth-order accurate, strong-stability preserving five-stage Runge-Kutta scheme is employed \citep{Ruuth2004}.  For further details on the numerical method see WLM15.




\subsection{Initial Conditions}\label{sec:IC}

For anelastic codes, the initial conditions are often based on a 1D, unstable, equilibrium, polytropic solution derived from the governing equations with an imposed superadiabatic entropy gradient, $\partial S/\partial r<0$ \citep{Jones2011}. However, this approach does not apply here for two reasons. One is the exclusion of the centrifugal force.   Taking this into account in the initial hydrostatic balance requires a 2D setup (both radial and latitudinal directions) which poses a problem for anelastic codes that generally expand about a 1D reference state.  The second reason that we cannot use a typical anelastic setup is the compressible nature of CHORUS.  In an anelastic code, the superadiabatic entropy gradient can be incorporated into the thermodynamic perturbations and decoupled from the specification of the reference state.  Since CHORUS does not linearize the equations about a reference state, this does not apply.

For these reasons, we take a different approach.  Instead, we initialize CHORUS with a 2D, hydrostatic, adiabatic stratification ($\partial S/\partial r = 0$).  This has the advantage that it can be defined analytically, even for oblate geometries.  However, in contrast to some anelastic setups, this initial state is not an equilibrium solution to the equations because it does not satisfy the steady-state energy equation.  In particular, there is a non-zero convergence of the radiative flux which heats the plasma and establishes the superadiabatic entropy gradient that eventually triggers convection.  In this section we describe how this initialization is accomplished for both spherical and oblate spheroidal shells.  

The effective gravitational potential of a uniformly rotating body including the
centrifugal term is described by the Roche potential
\begin{equation}
\Phi = -\frac{GM}{r} - \frac{\Omega_{0}^{2}}{2}r^{2}\sin^{2}\theta ,
\label{eqn:potential}
\end{equation}
where $\theta$ is the colatitude. We assume that initial stratification is adiabatic such that
\begin{equation}
P = C\rho^{\gamma},
\label{eqn:adabatci}
\end{equation}
where $C$ is a constant. The hydrostatic equation reduced from the momentum conservation Equation (\ref{eqn:monen_convervation}) in the absence of flow motion ($\partial /\partial t = 0$, $\mathbf{u} = 0$)
is
\begin{equation}
\del P = - \rho \del \Phi.
\label{eqn:reduced_hydro}
\end{equation}
These equations are supplemented with the ideal gas law
\begin{equation}
p = {\mathfrak{R} \rho T},
\label{eqn:gas_relation}
\end{equation}
where $\mathfrak{R}$ is the specific gas constant.  We obtain a solution to Equations (\ref{eqn:potential})-(\ref{eqn:gas_relation}), by introducing a polytropic stratification that can be regarded as an oblate generalization of the spherically symmetric polytropic solutions described by \citet{Jones2011}.  The initial profiles for pressure $P$, density $\rho$, and temperature $T$ are specified as
\begin{eqnarray}
P({\Phi}) &=& P_{i}[1 - \frac{\Phi - \Phi_{i}}{C_{p}T_{i}}]^{\alpha},
\label{eqn:P_initial}\\
\rho({\Phi}) &=& \rho_{i}[1 - \frac{\Phi - \Phi_{i}}{C_{p}T_{i}}]^{\frac{1}{\gamma-1}},
\label{eqn:rho_initial}\\
T({\Phi}) &=& T_{i}[1 - \frac{\Phi - \Phi_{i}}{C_{p}T_{i}}],
\label{eqn:T_initial}
\end{eqnarray}
where $\alpha = \gamma/(\gamma-1)$,  the subscript $i$ denotes the inner boundary, and the inner boundary
temperature is given by
\begin{equation}
T_{i} = \frac{\Phi_{o} - \Phi_{i}}{C_{p}[1-\exp{(-(\gamma-1)N_{\rho)}}]} ~~~.
\label{eqn:initial_temperature}
\end{equation}
Here $\Phi_i$ and $\Phi_{o}$ are the values of $\Phi$ on the inner and outer surfaces respectively.  In terms of the inner and outer radius at the poles, $r_{ip}$ and $r_{op}$, they are given by $\Phi_i = - G M / r_{ip}$ and $\Phi_o = - G M / r_{op}$.  In practice, we specify $r_{ip}$ and $r_{op}$ and then solve Equation (\ref{eqn:potential}) to find the equipotential surfaces $\Phi_i$ and $\Phi_o$ that pass through these polar radii.  We then define these equipotential surfaces as the inner and outer boundary of the domain (see section \ref{sec:der_grid}).

Note that Equations (\ref{eqn:P_initial})-(\ref{eqn:T_initial}) reduce to the 1D polytropic solutions of \citet{Jones2011} if the centrifugal force term is omitted from the effective gravitational potential in Equation (\ref{eqn:potential}).  As shown there, $N_\rho$ corresponds to the number of density scales heights across the shell.  The initial profile is completely specified by choosing the parameters $M$, $r_{ip}$, $r_{op}$, $\Omega_{0}$, $\gamma$, $C_{p}$, $\rho_{i}$, and $N_{\rho}$.  For our oblate simulations, we use the 2D profiles in Equations (\ref{eqn:P_initial})-(\ref{eqn:T_initial}) and for our spherical shell simulations we use the 1D profiles of \citet{Jones2011} which can be obtained from (\ref{eqn:P_initial})-(\ref{eqn:T_initial}) by setting $\Omega_0 = 0$.  Since we wish to model solar-type stars, we use appropriate parameter values of $N_\rho = 3$, $\rho_i = 0.21$ g cm$^{-3}$, $M = M_{\odot} = 1.99 \times 10^{33}$ g, $C_P = 3.5\times 10^8$ erg g$^{-1}$ K$^{-1}$, and $\gamma = 5/3$.  Note also that the initialization procedure described here implies that the oblate geometries will have both a larger volume and a larger mass than their spherical counterparts (see section \ref{sec:setup}).

These polytropic, adiabatic initial conditions do not satisfy the Schwarzschild criterion for convective instability, which requires a negative (superadiabatic) radial entropy gradient ($\partial S/\partial r< 0$).  Furthermore, the Rayleigh number $Ra$ is subcritical. Thus, convection will not occur immediately.  However, as noted at the beginning of this section, there is a fixed radiative heat flux $\propto - \kappa_r \mathbf{\del} T$ that carries the stellar luminosity in through the bottom boundary (see also sections \ref{sec:hydro}, \ref{sec:bdry} and \ref{sec:flux}).  The convergence of this heat flux heats the plasma in the shell, raising its temperature.  Since the temperature at the top is fixed (see section \ref{sec:bdry}), this heating of the plasma increases the temperature gradient.  Soon the temperature gradient becomes both superadiabatic and supercritical, and convection ensues.   The temperature and entropy gradients will eventually equilibrate when the simulation achieves flux balance, such that the combined convective and diffusive energy fluxes carry the full stellar luminosity through the entire shell (see section \ref{sec:flux}).

\begin{figure}[hp!]
\center
\includegraphics[width=0.7 \textwidth, angle=0]{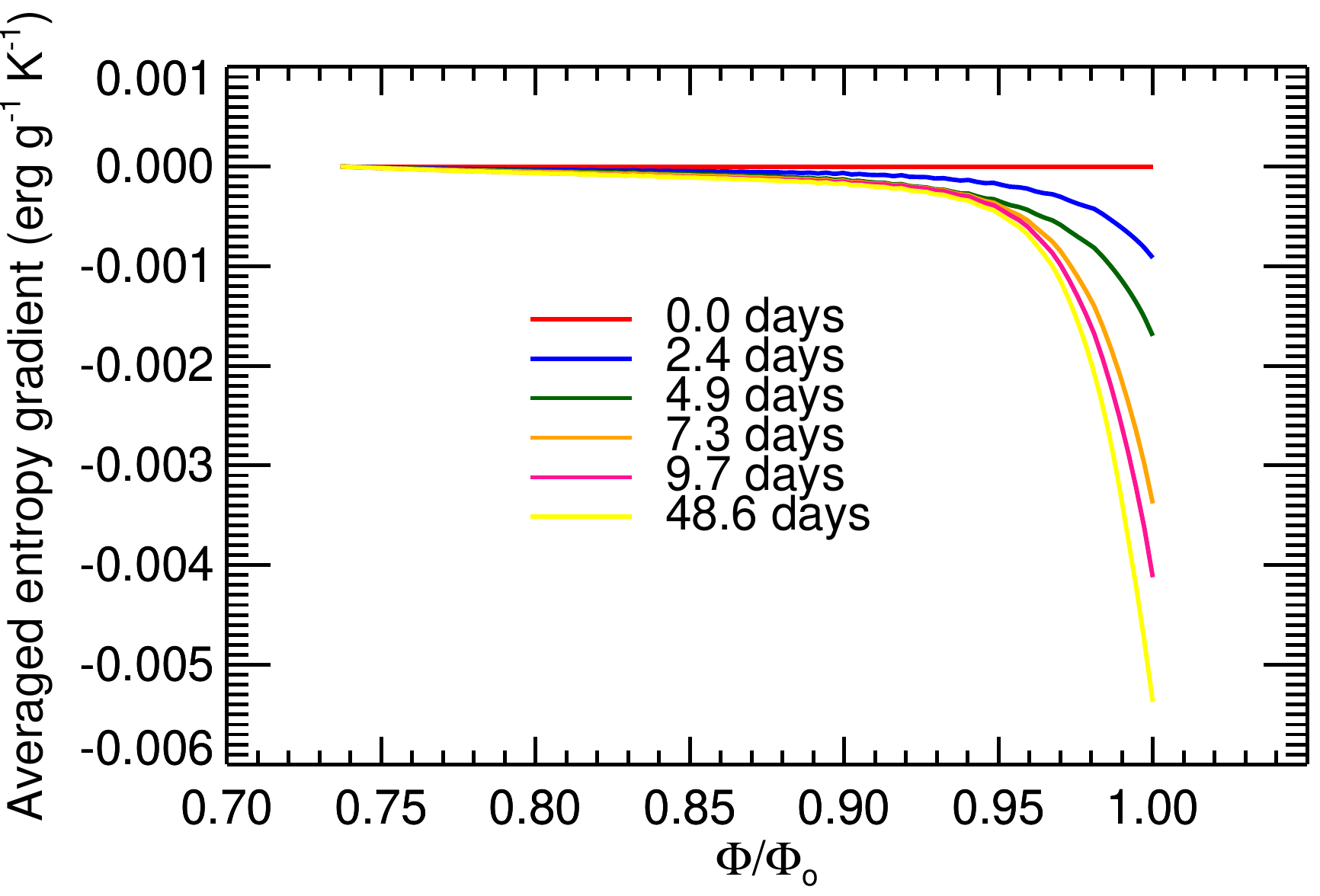}
\caption{Evolution of the vertical entropy gradient in one of the simulations (s60).  Shown is the horizonally-averaged normal entropy gradient $\del S \bdot \uvn$ as a function of $\Phi/\Phi_o$, where $\uvn$ is a unit vector normal to $\Phi$ isosurfaces.  Each curve corresponds to a different time, as indicated.}
\label{fig:entropy_s60}
\end{figure}

This process is illustrated in Fig.\ \ref{fig:entropy_s60}, which shows how the entropy gradient in a typical simulation (s60; see Table \ref{Tab:parameters}) evolves with time.  It starts out adiabatic ($\partial S/\partial r = 0$), but it quickly becomes superadiabatic ($\partial S/\partial r < 0$).  The heating of the shell due to the convergence of the radiative heat flux not only increases the temperature, it also raises the adiabat.  This increases the negative entropy gradient throughout the shell, but particularly near the upper boundary.  By $t \sim $ 6 days, it becomes superadiabatic enough to trigger convection (see Figs.\ \ref{fig:KE} and \ref{fig:KE_SO} below).  By $t \sim $ 50 days, the simulation has achieved flux balance and the entropy gradient has equilibrated.

\subsection{Boundary Conditions}{\label{sec:bdry}}
The inner and outer boundaries, $\Phi = \Phi_i$ and $\Phi = \Phi_o$, are assumed to be impenetrable and free of viscous shear stresses
\begin{equation}
\mathbf{u}\bdot \mathbf{n} = 0,
\end{equation}
and
\begin{equation}
(\mathbf{\tau} \bdot \mathbf{n}) \mathbf{\times} \mathbf{n} = \mathbf{0},
\end{equation}
where $\mathbf{n}$ is the unit vector normal to equipotential surfaces $\Phi$ and pointing radially outward.

At the bottom boundary a fixed heat flux is imposed such that
\begin{equation}\label{eq:Fbot}
\mathbf{f}= - \kappa_{r}\rho C_{p}\del T - \kappa \rho T \del S = \frac{L}{A} ~ \uvn  
\mbox{\hspace{.2in} $\left(\Phi = \Phi_i\right)$,}
\end{equation}
where $L$ is the stellar luminosity and $A$ is the area of the inner surface $\Phi = \Phi_i$.  For the spherical simulations, $A = 4\pi r_{ip}^2$, but for the oblate simulations the area is greater than this and is calculated numerically.  Since $\del S \bdot \uvn \approx 0$ at $\Phi = \Phi_i$ (see Fig.\ \ref{fig:entropy_s60}), this imposed heat flux is almost entirely due to the radiative diffusion term $(\propto - \kappa_r \del T)$.  The magnitude of $\kappa_r$ is specified such that Equation (\ref{eq:Fbot}) is satisfied (see section \ref{sec:hydro}).

At the upper boundary, $\Phi = \Phi_o$, the temperature is fixed at the value specified by the initial conditions (section \ref{sec:IC}).

\subsection{Grids for Oblate Spheroidal Shells}{\label{sec:der_grid}}
In our model, the oblate outer surface of the star is defined by the equipotential surface $\Phi = \Phi_o$. In order to determine the outer surface shape, the polar ($\theta = 0$) radius $r=r_{op}$ is specified and kept unchanged. Then the potential at the pole can be compared to that at an arbitrary polar angle $\theta$. The resulting cubic equation for $r$ is
\begin{equation}
\Phi_o = -\frac{GM}{r_{op}} =-\frac{GM}{r} - \frac{\Omega_{0}^{2}}{2}r^{2}\sin^{2}\theta.
\label{eqn:geo_equipotential2}
\end{equation}
By solving Equation (\ref{eqn:geo_equipotential2}) for $r(\Phi_o,\theta)$, the axisymmetric ($\phi$-independent) surface is obtained. The inner equipotential surface can be computed in a similar way by replacing the outer polar radius $r_{op}$ with the inner polar radius $r_{ip}$ in Equation (\ref{eqn:geo_equipotential2}). The geometry of an oblate spheroidal shell is then defined by both inner and outer equipotential surfaces.

\begin{figure}[hp!]
\subfigure[]{
\includegraphics[width=0.48 \textwidth, angle=0]{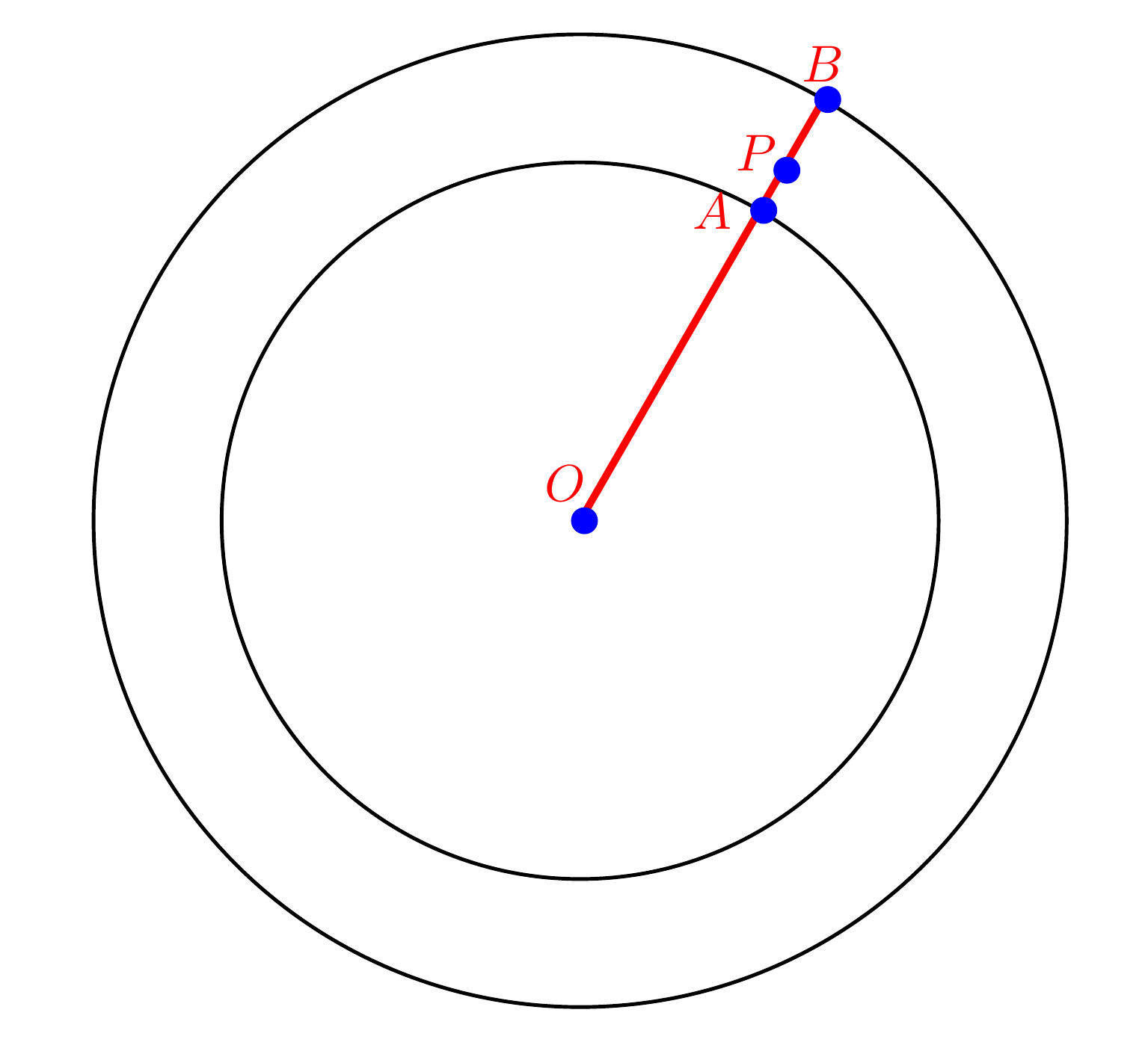}}
\subfigure[]{
\includegraphics[width=0.52 \textwidth, angle=0]{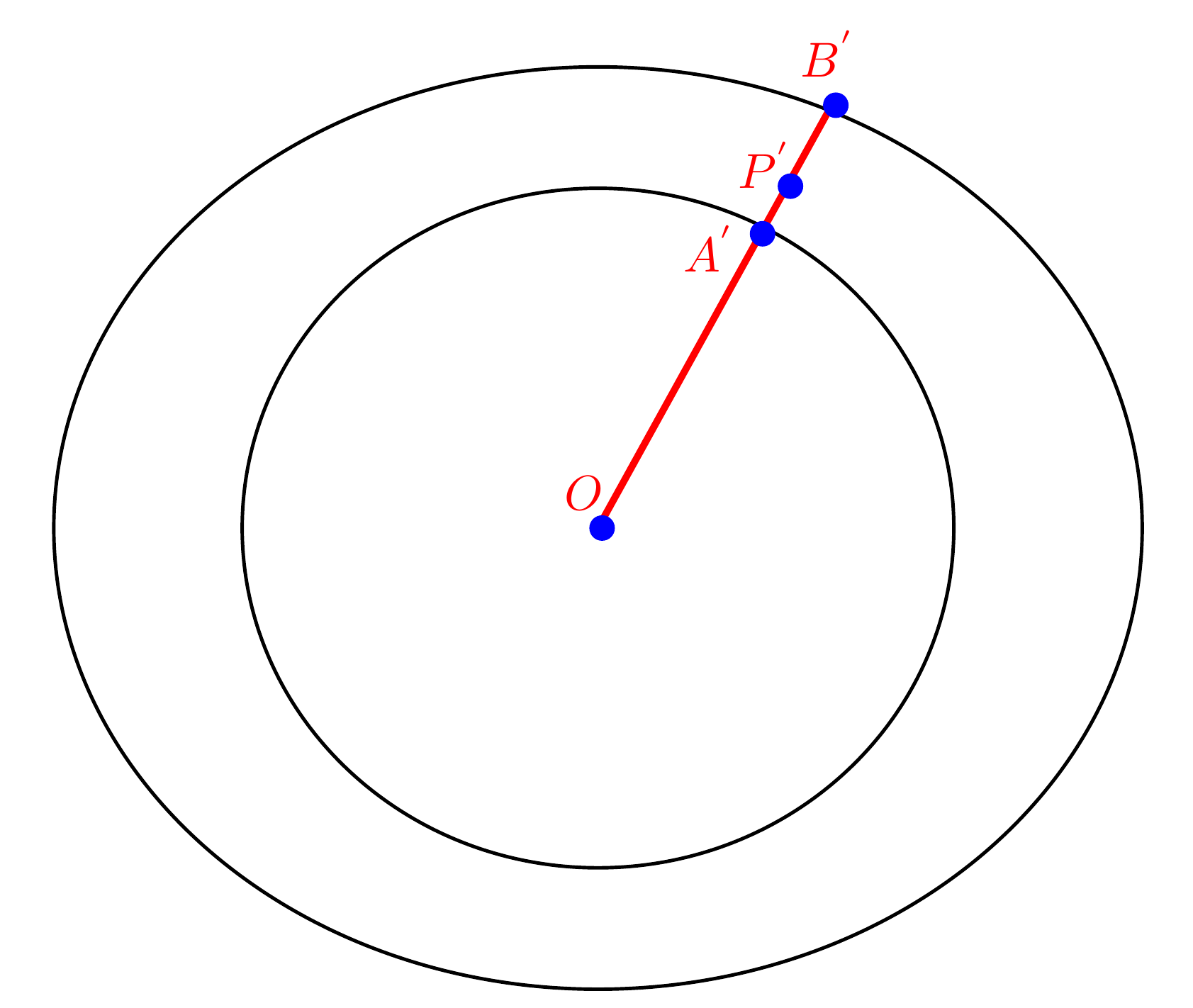}}
\caption{Shown are \textit{(a)} an arbitrary node point $P$ within a spherical shell and \textit{(b)} an updated node point $P^{'}$ in the desired oblate spheroidal shell. Points $A$ and $B$ (and $A^{'}$ and $B^{'}$) lie on the inner and outer boundaries at the same latitude and longitude.}
\label{fig:deform_fig}
\end{figure}

\begin{figure}[!htp]
\centering
\includegraphics[width=0.9\textwidth, angle=-90]{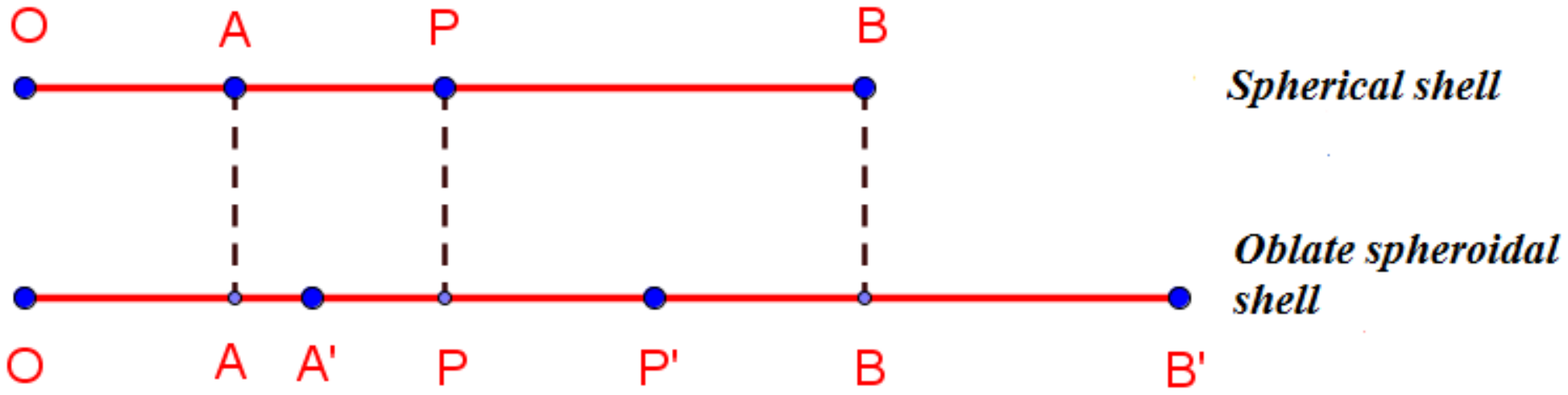}
\caption{Configuration of the intersection points $A$, $B$ and node point $P$ in a spherical shell and
corresponding points $A{'}$, $B{'}$ and $P{'}$ in the desired oblate spheroidal shell.}
\label{fig:move_points}
\end{figure}

It is straightforward to generate grids for spherical shells by using commercial or publicly available grid generation software (e.g. Gmsh). Spherical shells can also be represented by a cubed-sphere grid \citep{McCorquodale2015}. However, it is very challenging to generate grids for oblate spheroidal shells as defined above because we need to represent the curvature on both the bottom and top surfaces very accurately in order to conserve mass, momentum, and energy. Here, an alternative approach has been chosen. We first generate a spherical shell grid with inner radius $r_{i}$ and outer radius $r_{o}$ using Gmsh software \citep{Gmsh2015}. The following procedure is taken to deform this generated spherical shell grid to the desired oblate spheroidal shell grid with the rotating rate $\Omega_{0}$.
\begin{enumerate}
\item In the spherical shell grid, for an arbitrary grid node point $P$, find the points of intersection between line $OP$ and bottom and top surfaces.  Denote these as $A$ and $B$ respectively (see Fig. \ref{fig:deform_fig}a). Note that point $O$ is the center. 
\item Calculate lengths of $\overline{OA}$, $\overline{OP}$, $\overline{OB}$, $\overline{AP}$ and $\overline{PB}$.
\item Define new bottom and top surfaces using Equation (\ref{eqn:geo_equipotential2}).
\item Keep the latitude and longitude of line $OP^{'}$ identical to that of line $OP$ and find the new intersection points $A^{'}$ and $B^{'}$ (see Fig. \ref{fig:deform_fig}b).

\item As shown in Figure \ref{fig:move_points}, the position of point $P^{'}$ is obtained by satisfying
\begin{equation}
\frac{\overline{A^{'}P^{'}}}{P^{'}B^{'}} = \frac{\overline{AP}}{PB}.
\end{equation}
\end{enumerate}

\begin{figure}[hp!]
\subfigure[]{
\includegraphics[width=0.5 \textwidth, angle=-90]{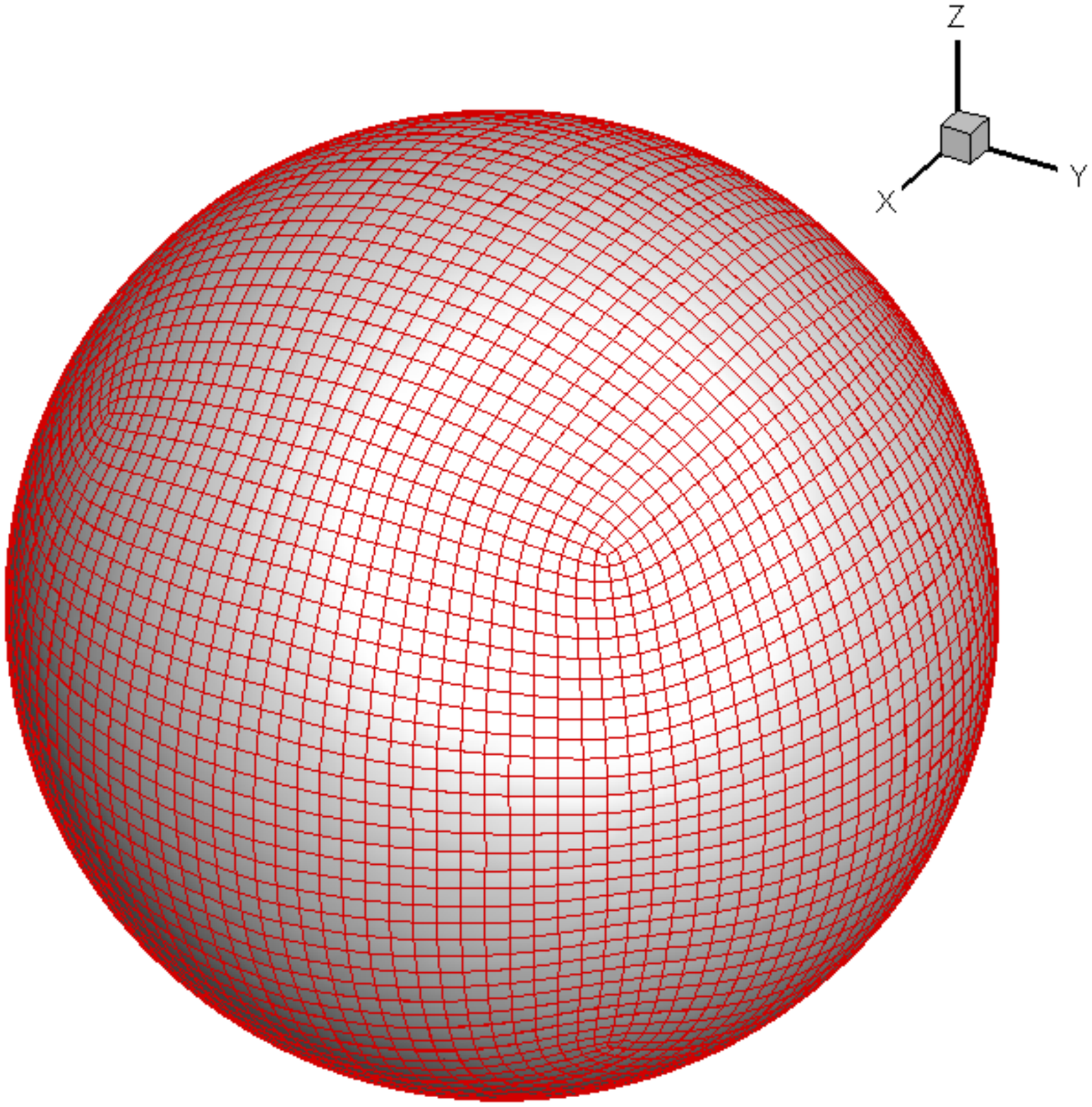}}
\subfigure[]{
\includegraphics[width=0.5 \textwidth, angle=-90]{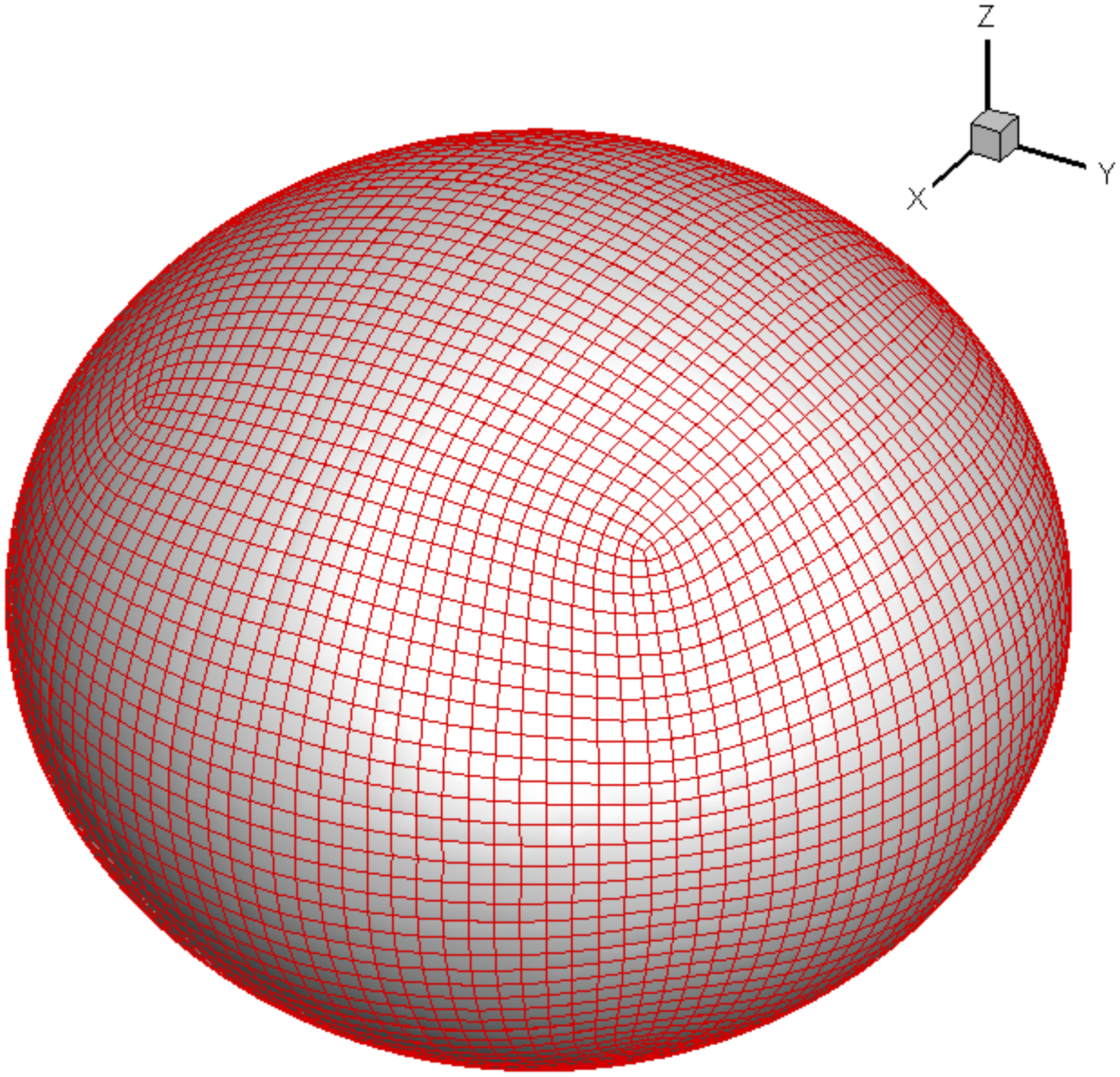}}
\caption{Shown are \textit{(a)} an illustrative spherical shell grid and \textit{(b)} the oblate spheroidal shell grid generated from \textit{(a)} as described in the text.}
\label{fig:sphere_oblate}
\end{figure}

The spherical shell grid and the newly generated oblate spheroidal grid are shown in Figure \ref{fig:sphere_oblate}.

\begin{figure}[htp!]
\subfigure[]{
\includegraphics[width=0.5 \textwidth, angle=-90]{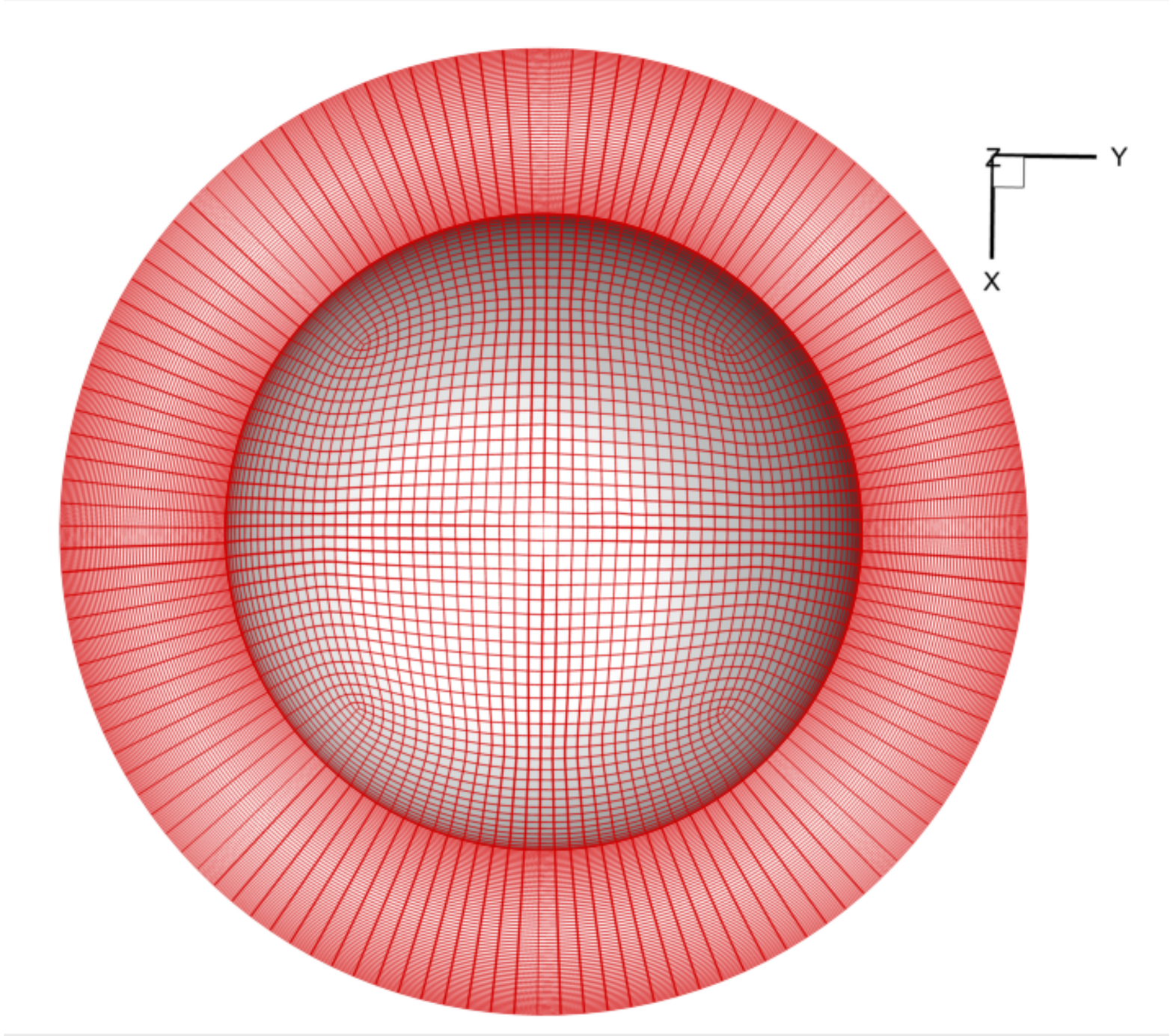}}
\subfigure[]{
\includegraphics[width=0.5 \textwidth, angle=-90]{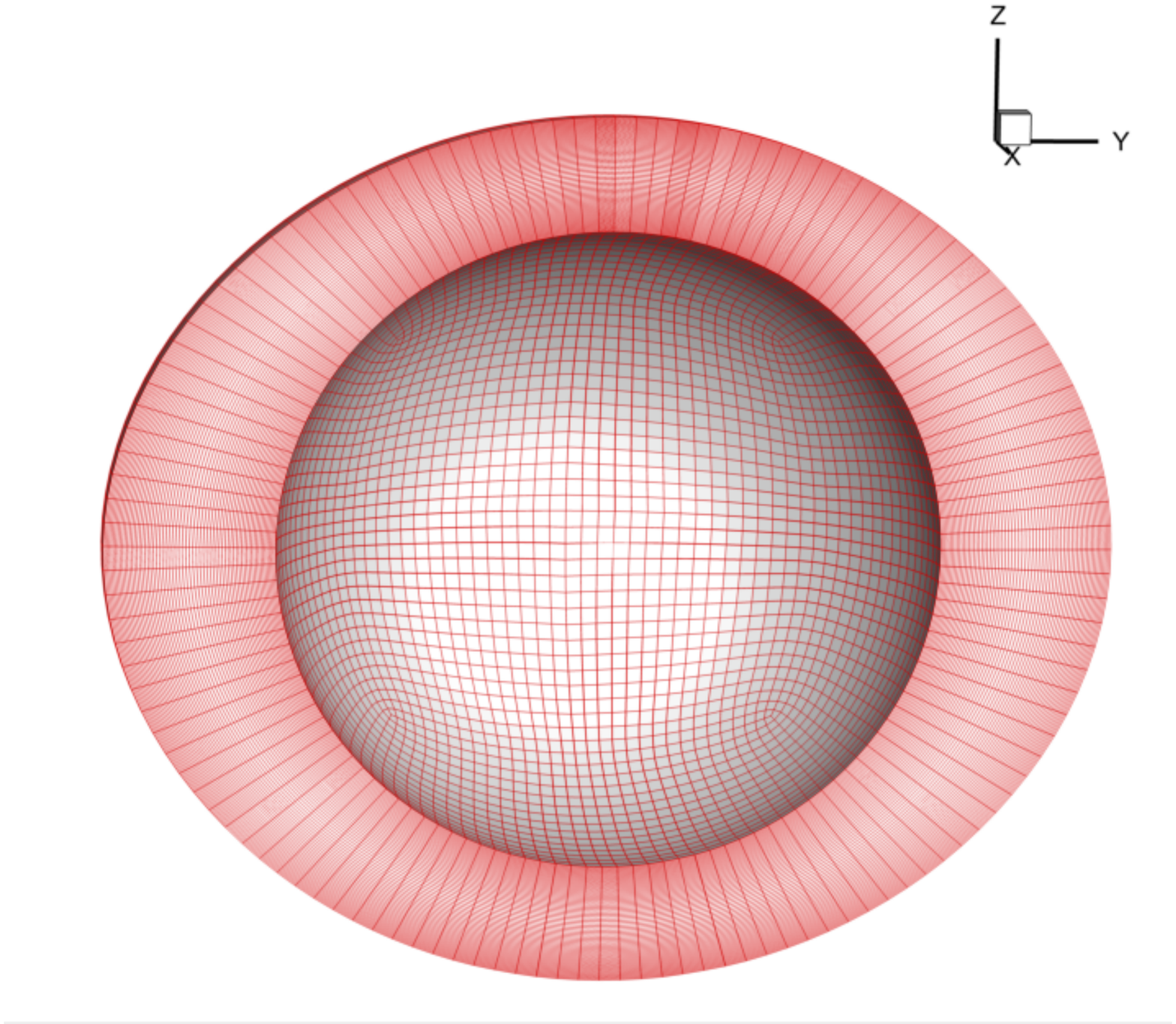}}
\caption{Shown are internal grid node points in \textit{(a)} the equatorial plane and \textit{(b)} a meridional plane for a typical oblate spheroidal shell grid.}
\label{fig:mesh_plane}
\end{figure}

This procedure is performed at the preprocessing stage and all desired oblate spheroidal shell
grids with varying $\Omega_{0}$ (discussed in Section \ref{sec:setup}) are generated from a single spherical shell grid. The generated oblate spheroidal shell grids are smooth as shown in Figure \ref{fig:mesh_plane}; the node positions on both top and bottom surfaces are exactly specified (to machine precision) by solving Equation (\ref{eqn:geo_equipotential2}).

\section{SIMULATION SUMMARY}\label{sec:setup}
The objective of this study is to investigate the effects of oblateness on the properties of the convection, heat transfer, and mean flows in rapidly-rotating solar-type stars. To this end, we have performed global-scale numerical simulations within spherical shells (neglecting the centrifugal force) as well as oblate spheroidal shells that include the centrifugal force in order to investigate the difference. Simulations within spherical shells are labeled with leading letter $s$ while those within oblate spheroidal shells are marked with leading letter $o$ as shown in Tables \ref{Tab:parameters} and \ref{Tab:diagnostics}. The computational domain for the spherical shell extends from $r_{ip} = 0.70R_{\odot}$ to $r_{op} = 0.95R_{\odot}$ where $R_{\odot}=6.9599\times 10^{10}$ cm is the solar radius. The computational domains for the oblate spheroidal shells are derived from the spherical shell using the approach described in Section \ref{sec:der_grid}.  As described there, the inner and outer radii at the poles are the same as in the spherical cases, but the depth of the convection zone becomes wider at low latitudes due to the centrifugal distortion.

The low Mach numbers of deep stellar interiors pose a significant challenge for fully compressible codes.  In particular, if explicit timestepping is used as is the case here, they set strict limits on the size of the time step needed to satisfy the Courant-Freidrichs-Lewy (CFL) condition.  They also imply long thermal equilibration times.  We address this challenge as in previous simulations by artificially increasing the luminosity \citep[e.g.][WLM15]{kapyl11}.  This has the effect of increasing the velocity amplitudes (which scale roughly as $L^{1/3}$) and decreasing the thermal relaxation time.  For all simulations reported here we use $L = 6.4\times 10^{4} L_{\odot}$, where $L_{\odot}=3.846 \times 10^{33}$ erg s$^{-1}$ is the solar luminosity.  Thus, we estimate that the velocity amplitudes in our simulations are artificially enhanced by about a factor of 40.

Our studies here explore a range of rotation rates from 60$\Omega_{\odot}$ to $120 \Omega_{\odot}$ (cases $s60$-$s120$ and $o60$-$o120$) where $\Omega_{\odot} = 2.6\times 10^{-6}$ rad $s^{-1}$ is the fiducial solar rotation rate.   However, the effective Rossby number Ro $= U / (2 \Omega_0 d)$ in our simulations is artificially increased by roughly a factor of forty because of the larger convective velocity amplitudes (relative to a star rotating at the same rotation rate).  This problem is sometimes addressed by artificially increasing the rotation rate along with the luminosity to achieve a desired Rossby number \citep[e.g.][]{kapyl11}.  However, we do not have that luxury here.  Since we include the centrifugal force, there is a critical rotation rate $\Omega_{crit} = \sqrt{\frac{8}{27}\frac{GM}{r^{3}_{op}}}$ where the outward centrifugal acceleration of rotation equals the inward gravitational acceleration \citep{van2012}.  If the star were to rotate faster than $\Omega_{crit}$ it would break up.  Our range of rotation rates extends up to 85\% of $\Omega_{crit}$ (see Table \ref{Tab:parameters}).  In other words, the oblateness of these stars is comparable to a real star rotating at these rotation rates but the Rossby number is about a factor of forty higher.  We could in principle remedy this situation by adjusting $M$ and $G$ or we could simply accept that, though the convection in our simulations is strongly rotationally constrained (Ro $<< 1$; see Table \ref{Tab:diagnostics}), it is less so than in a real star with the same rotation rate.  For this initial study we follow the latter approach and take it into account in our interpretation.

All of the simulations discussed here employ the third-order accuracy spectral difference method on 393,216 hexahedral elements, resulting in a total of 10,616,832 degrees of freedom ($DOFs$).  This corresponds to 96 solution points in radius and 192 solution points in latitude.  The number of solution points in longitude is twice this at the equator, 384, but diminished toward higher latitudes to achieve a nearly uniform horizontal resolution.  The values of $\nu$ and $\kappa$ are selected to be small, but large enough to ensure that the flows are well resolved.  Since we wish to focus on the effects of the oblateness, $\nu$ and $\kappa$ are held fixed at a value of $\nu = \kappa = 2.4 \times 10^{14}$ cm$^2$ s$^{-1}$, implying a Prandtl number $Pr=\nu/\kappa=1$.  However, in order to make contact with the previous global simulations of \citet{Brown2008}, we also run a series of spherical simulations in which $\nu$ and $\kappa$ are reduced as the rotation rate is increased in order to maintain the supercriticality of the convection.  These are listed in Tables \ref{Tab:parameters} and \ref{Tab:diagnostics} as cases $s100a$, $s80a$, and $s60a$.  In these simulations, $\nu = \kappa \propto \Omega_{0}^{-2/3}$, where the normalization is chosen such that $\nu = \kappa = 2.4 \times 10^{14}$ cm$^2$ s$^{-1}$ when $\Omega_0 = 120 \Omega_{\odot}$.  Thus, this $a$ series of simulations also has $P_r = 1$ and follows a parameter path that includes $s120$.  

All of the simulations considered here are relatively laminar.  This is appropriate for the innovative nature of this study; these set the baseline for future simulations that can explore the parameter sensitivities more thoroughly.  Reynolds numbers are on the order of 10-20, as shown in Table \ref{Tab:parameters}.  Recall that the convective velocity is artificially enhanced so the Reynolds numbers are higher than might otherwise be expected for this value of $\nu$.   The supercriticality of these simulations is not easy to establish due to the lack of a linear theory for convection in oblate, compressible spheroids and due to the nature of the initial conditions, which do not express an unstable equilibrium (Sec.\ \ref{sec:IC}).  However, we have determined the critical Rayleigh number Ra$_{crit}$ for the least supercritical simulation, namely case o120.  Here convection is inhibited by both the rapid rotation and the oblateness.  We determine Ra$_{crit}$ by progressively increasing $\nu$ and $\kappa$ until convection is suppressed.  The results indicate a relatively small value of Ra/Ra$_{crit} \sim 2.5$.  All other simulations are expected to have a higher supercriticality.  For example, in Boussinesq systems, the critical Rayleigh number for the onset of convection in rotating spherical shells scales as $\Omega_0^{4/3}$ \citep{dormy04}.  This would imply a supercriticality on the order of 6 for case o60.

Other quantities listed in Table \ref{Tab:diagnostics} include the volume-averaged kinetic energy density relative to the rotating reference frame, KE, as well as the contributions to KE from the convection (CKE), differential rotation (DRKE), and meridional circulation (MCKE).  These quantities are defined as follows:
\begin{equation}
KE = \frac{1}{V}\int_{V} \frac{1}{2}\rho (\mathbf{u}\bdot \mathbf{u}) dV,
\end{equation}
\begin{equation}
DRKE = \frac{1}{V}\int_{V}\frac{1}{2}\rho \langle V_{\phi} \rangle^{2}r^{2}\sin \theta dr d\theta d\phi,
\end{equation}
\begin{equation}
MCKE = \frac{1}{V}\int_{V}\frac{1}{2}\rho(\langle V_{r} \rangle^{2} + \langle V_{\theta} \rangle^{2})r^{2}\sin \theta dr d\theta d\phi ~~,
\end{equation}
and
\begin{equation}
CKE = KE - DRKE - MCKE,
\end{equation}
where $V$ denotes the volume of the computational domain and angular brackets denote averages over longitude.

The thermal diffusion time $\tau_d \sim d^2 / \kappa$ is roughly 15 days, the convective turnover time scale is roughly 1.2 days, and the rotation period ranges from 0.23-0.47 days.  All simulations have been run for at least 48 days so this spans at least three diffusion times, at least 40 turnover times, and at least 100 rotation periods.  All appear to be equilibrated (see Figs.\ \ref{fig:KE} and \ref{fig:KE_SO} below).

One more comment is in order before proceeding to the main results.  When interpreting the KE data in Table \ref{Tab:diagnostics} and below, it is important to realize that the oblate shells have more mass and volume than their spherical counterparts.    This is a consequence of our initialization approach discussed in section \ref{sec:IC}.  In particular, the density on the inner equipotential surface, $\rho_i$ is the same for all simulations.  However, the oblate geometries have a greater volume, which gives rise to a larger mass.  The mass of the convective envelope $M_c$ for each simulation is listed in Table \ref{Tab:parameters} relative to the central mass $M$.  The most extreme case, o120, has about 37\% more mass in the convection zone than s120.  This is somewhat less than the volume difference, which is about 51\% greater in case o120 than in s120 (Table \ref{Tab:parameters}).  Thus, although the volume-averaged KE density listed in Table \ref{Tab:diagnostics} is slightly smaller in case o120 compared to s120, the total kinetic energy integrated over the entire volume (in ergs) is larger by about 36\%, comparable to the difference in mass.  In other words, the mean kinetic energy per unit mass ($KE$ multiplied by $V$ and divided by $M_c$) is almost the same in cases o120 and s120 (within 1\%).  This is a measure of the density-weighted velocity amplitude.  The other oblate cases have slightly higher kinetic energy per unit mass than their spherial counterparts, ranging from 3\% higher in o60 to 6\% higher in o70, o100, and o110 to 11\% and 15\% higher in cases o90 and o80.  The downward trend in this ratio for $\Omega_0 > 80 \Omega_\odot$ may be due to a saturation of DRKE/KE in the fastest-rotating oblate stars (see Section \ref{sec:energetics}).  Note also that $M_{c}/M << 1$ in all cases (Table \ref{Tab:parameters}), justifying our assumption in section \ref{sec:IC} that the bulk of the mass lies inside the inner radius.

\begin{deluxetable}{lccccc}
\tablecolumns{12}
\tablewidth{0pc}
\tablecaption{Simulation Parameters}
\tablehead{
\colhead{Case} & \colhead{$\textrm{E}\textrm{k}$ $(\times 10^{-3})$} & 
\colhead{$\frac{\Omega_{0}}{\Omega_{crit}}$} & \colhead{$\frac{V}{V_s}$} & \colhead{$\frac{M_{c}}{M}$}    & \colhead{$\textit{O}$}
}
\startdata
$o120$ & $2.54$  & $0.846$ & 1.51 & $3.84\%$ & $16.95\%$ \\
$o110$ & $2.77$ & $0.775$ & 1.38 & $3.60\%$ & $12.77\%$\\
$o100$ & $3.049$ & $0.706$ & 1.29 & $3.42\%$ & $9.72\%$\\
$o90$  & $3.39$ &  $0.634$  & 1.21 & $3.27\%$ & $7.38\%$ \\
$o80$  & $3.81$ &  $0.564$  & 1.16 & $3.15\%$ & $5.54\%$  \\
$o70$  & $4.36$ &  $0.493$ & 1.12 & $3.06\%$ & $4.06\%$ \\
$o60$  & $5.08$ & $0.423$ & 1.08 & $2.98\%$ & $2.89\%$ \\
\cline{1-6}
$s120$ & $2.54$ & $0.846$ & 1 & $2.80\%$ & $0.0$ \\
$s110$ & $2.77$ & $0.775$ & 1 & $2.80\%$ & $0.0$\\
$s100$ & $3.05$ &  $0.705$ & 1 & $2.80\%$ & $0.0$\\
$s90$  & $3.39$ &  $0.634$ & 1 & $2.80\%$ & $0.0$ \\
$s80$  & $3.81$ &  $0.564$ & 1 & $2.80\%$ & $0.0$  \\
$s70$  & $4.36$ &  $0.493$ & 1 & $2.80\%$ & $0.0$ \\
$s60$  & $5.08$ &  $0.423$ & 1 & $2.80\%$ & $0.0$ \\
\cline{1-6}
$s100a$  & $3.44$ & $0.705$ & 1 & $2.80\%$ & $0.0$ \\
$s80a$  & $4.98$ &  $0.564$ & 1 & $2.80\%$ & $0.0$ \\
$s60a$   & $8.06$ &  $0.423$ & 1 & $2.80\%$ & $0.0$ \\
\enddata
\tablecomments{The Ekman number is defined as Ek $=\nu/(\Omega_0 d^2)$ where $d=r_{op} - r_{ip}$ is the polar depth of the shell. The Prandtl number $Pr = \nu/\kappa=1$ and the number of density scales heights across the shell $N_{\rho}=3$ for all cases ($\rho_i/\rho_o = 20$). $V$ and $M_{c}$ denote the volume and mass of the shell and $V_s$ denotes the volume of a spherical shell with the same values of $r_{ip}$ and $r_{op}$.  The oblateness $\textit{O}$ is defined as $\textit{O} = (r_{oe}-r_{op})/r_{op}$.} 
\label{Tab:parameters}
\end{deluxetable}

\begin{deluxetable}{lcccccc}
\tablecolumns{12}
\tablewidth{0pc}
\tablecaption{Simulation Diagnostics}
\tablehead{
\colhead{Case} & \colhead{$KE$}   & \colhead{$\frac{DRKE}{KE}$} & \colhead{$\frac{MCKE}{KE}$} &
\colhead{$\textrm{M}\textrm{a}$}  & \colhead{$\textrm{R}\textrm{e}$} & \colhead{$\textrm{R}\textrm{o}$} \\
\colhead{} & \colhead{$(\times 10^9)$}   & \colhead{} & \colhead{$(\times 10^{-3})$} &
\colhead{$(\times 10^{-2})$} & \colhead{}  & \colhead{$(\times 10^{-2})$} 
}
\startdata
$o120$ & $1.22$ & $0.607$ & $0.434$ & $0.832$ & $10.08$ & $1.29$ \\ 
$o110$ & $1.60$ & $0.613$ & $0.713$ & $1.03$ & $12.59$ & $1.64$ \\ 
$o100$ & $1.94$ & $0.613$ & $0.876$ & $1.11$ & $13.62$ & $2.08$ \\ 
$o90$  & $2.39$ & $0.615$ & $1.17$ & $1.32$ & $16.09$ & $2.72$ \\ 
$o80$  & $2.82$ & $0.610$ & $1.21$ & $1.45$ & $17.70$ & $3.37$ \\ 
$o70$  & $3.05$ & $0.593$ & $1.88$ & $1.51$ & $18.43$ & $4.01$ \\ 
$o60$  & $3.36$ & $0.536$ & $2.45$ & $1.61$ & $19.64$ & $4.99$ \\ 
\cline{1-7}
$s120$ & $1.35$ & $0.681$ & $0.533$ & $1.03$ & $12.80$ & $1.52$ \\ 
$s110$ & $1.61$ & $0.658$ & $0.720$ & $1.11$ & $13.50$ & $1.82$ \\ 
$s100$ & $1.92$ & $0.646$ & $1.18$ & $1.20$ & $14.60$ & $2.22$ \\ 
$s90$  & $2.24$ & $0.625$ & $1.17$ & $1.35$ & $16.49$ & $2.79$ \\ 
$s80$  & $2.52$ & $0.603$ & $1.94$ & $1.37$ & $16.70$ & $3.18$ \\ 
$s70$  & $2.93$ & $0.570$ & $2.50$ & $1.44$ & $17.55$ & $3.82$ \\ 
$s60$  & $3.33$ & $0.526$ & $2.61$ & $1.67$ & $20.29$ & $5.15$ \\ 
\cline{1-7}
$s100a$  & $0.945$ & $0.605$ & $0.600$ & $0.814$ & $12.24$ & $1.55$ \\ 
$s80a$  & $0.958$ & $0.575$ & $0.990$ & $0.819$ & $10.56$ & $2.21$ \\ 
$s60a$   & $0.958$ & $0.503$ & $2.58$ & $0.820$ & $8.71$ & $2.97$ \\ 
\enddata
\tablecomments{$KE$ is in units of erg cm$^{-3}$. The Mach, Reynolds, and Rossby numbers are defined as Ma = $u_{rms}^\prime/c_s$, Re $= u_{rms}^\prime d/\nu$, and  Ro $ = u_{rms}^\prime/(2\Omega_0 d)$ where $d$ is the polar depth of the shell and $u_{rms}^{\prime}=(\mathbf{u} -\langle V_{\phi} \rangle)_{rms}$ is the convective velocity evaluated at mid-depth, with the differential rotation removed. All values are averaged over a period of five days, computed after the simulations have equilibrated.}
\label{Tab:diagnostics}
\end{deluxetable}

\section{OVERVIEW OF CONVECTIVE STRUCTURE}\label{sec:con}

\subsection{Energetics}\label{sec:energetics}

\begin{figure}[hp!]
\subfigure[]{
\includegraphics[width=0.5 \textwidth, angle=0]{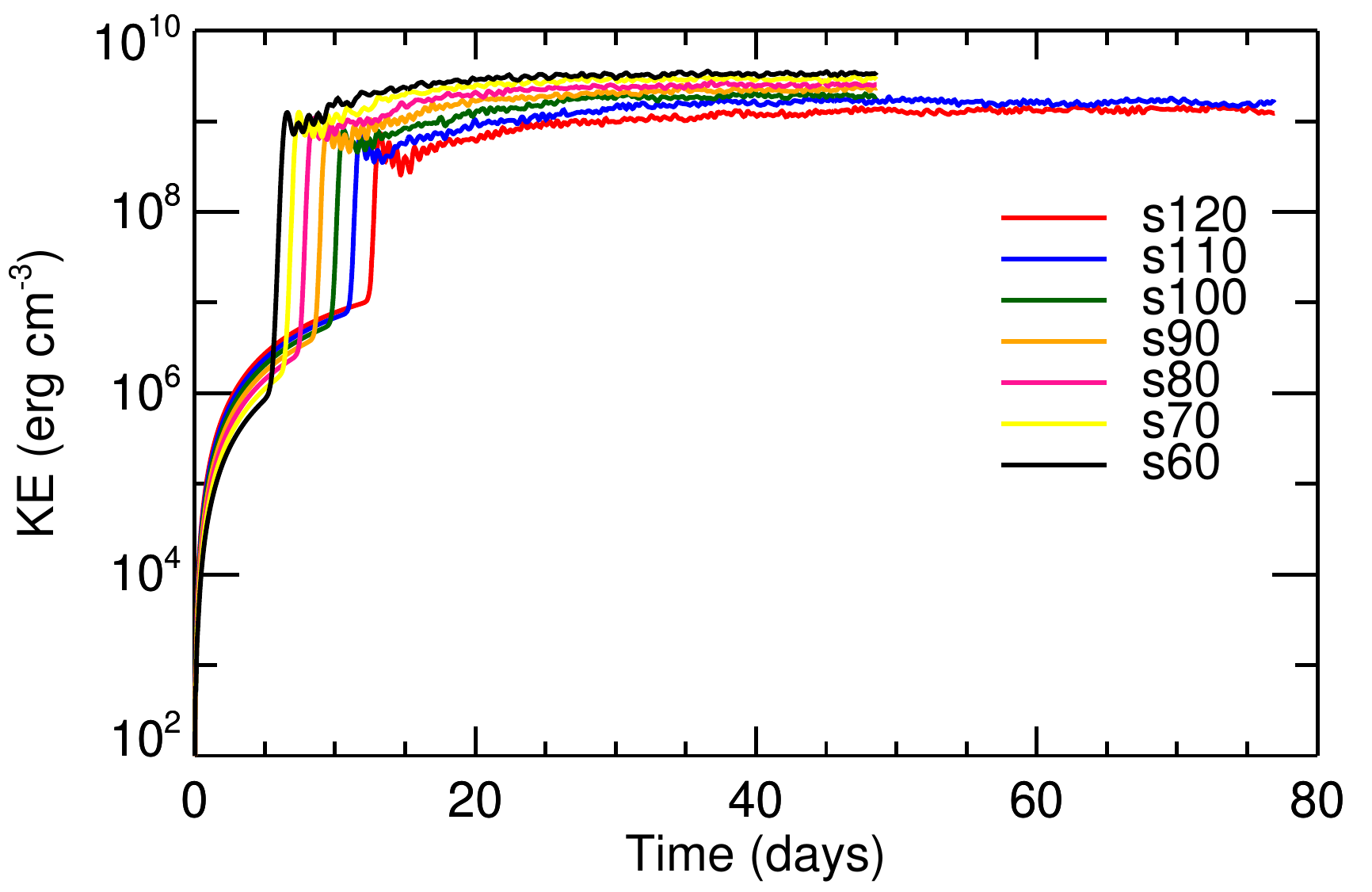}}
\subfigure[]{
\includegraphics[width=0.5 \textwidth, angle=0]{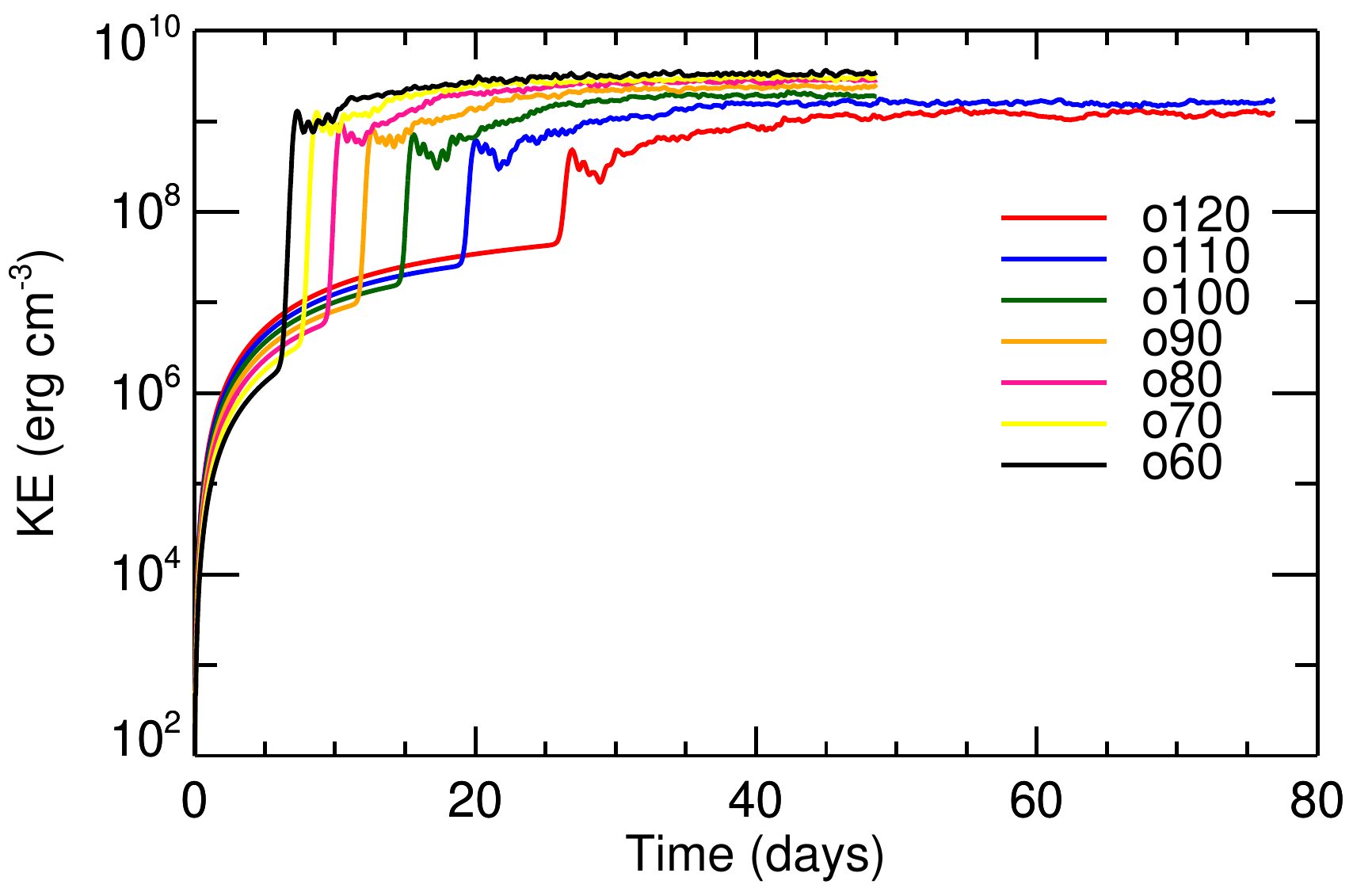}}
\caption{Volume averaged kinetic energy densities for \textit{(a)} cases in spherical shells and \textit{(b)} cases in oblate spheroidal shells.}
\label{fig:KE}
\end{figure}

The evolution of the kinetic energy for the $o$ and $s$ series of simulations is illustrated in Figure \ref{fig:KE}. All of these simulations start from the polytropic, adiabatic stratification described in Section \ref{sec:IC}. Therefore, the initial entropy gradient is zero for all cases. With the heating due to the radiative heat flux at the bottom and the constant temperature at the top, the entropy gradient becomes superadiabatic.  Once the negative entropy gradient is built up and the Rayleigh number $Ra$ exceeds a critical value, the flow locks on to the fastest-growing eigenmode which then grows exponentially.

However, even before this exponential growth phase, there is a slow initial algebraic growth in the KE in response to the radiative heating.  We have verified that this growth does not occur when the luminosity is set to zero so it is not due to numerical errors, such as those associated with the conservation of angular momentum or hydrostatic balance, which can pose problems on unstructured grids. We find our spectral difference method and grid generation procedure described in Section \ref{sec:MP} to be very effective at minimizing these numerical errors.   To demonstrate this, consider the three Cartesian components of the angular momentum vector, $L_x$, $L_y$, and $L_z$, integrated over the volume of the shell.  If the numerics were perfect, these three values would remain zero relative to the rotating frame throughout the simulation.   Our test cases with zero luminosity indicate that the errors associated with angular momentum conservation and hydrostatic balance are on the order of one part in $10^7$.  In order to prevent the gradual accumulation of these errors over tens of thousands of iterations, we apply an angular momentum correction procedure every 5000 iterations as described in WLM15.  For our most oblate simulation, case o120, near the end of the simulation interval ($t = $ 77 days), we find that $L_x/L_0 = 5.60\times10^{-9}$, $L_y/L_0 = 5.28\times10^{-8}$ and $L_z/L_0 = 2.02\times10^{-6}$, where $L_0$ is the initial angular momentum of the shell.  The structure of the flow in this initial establishement period suggests that it is indeed weak convective motions that are induced by the heating.

\begin{figure}[hp!]
\includegraphics[width=0.9 \textwidth, angle=0]{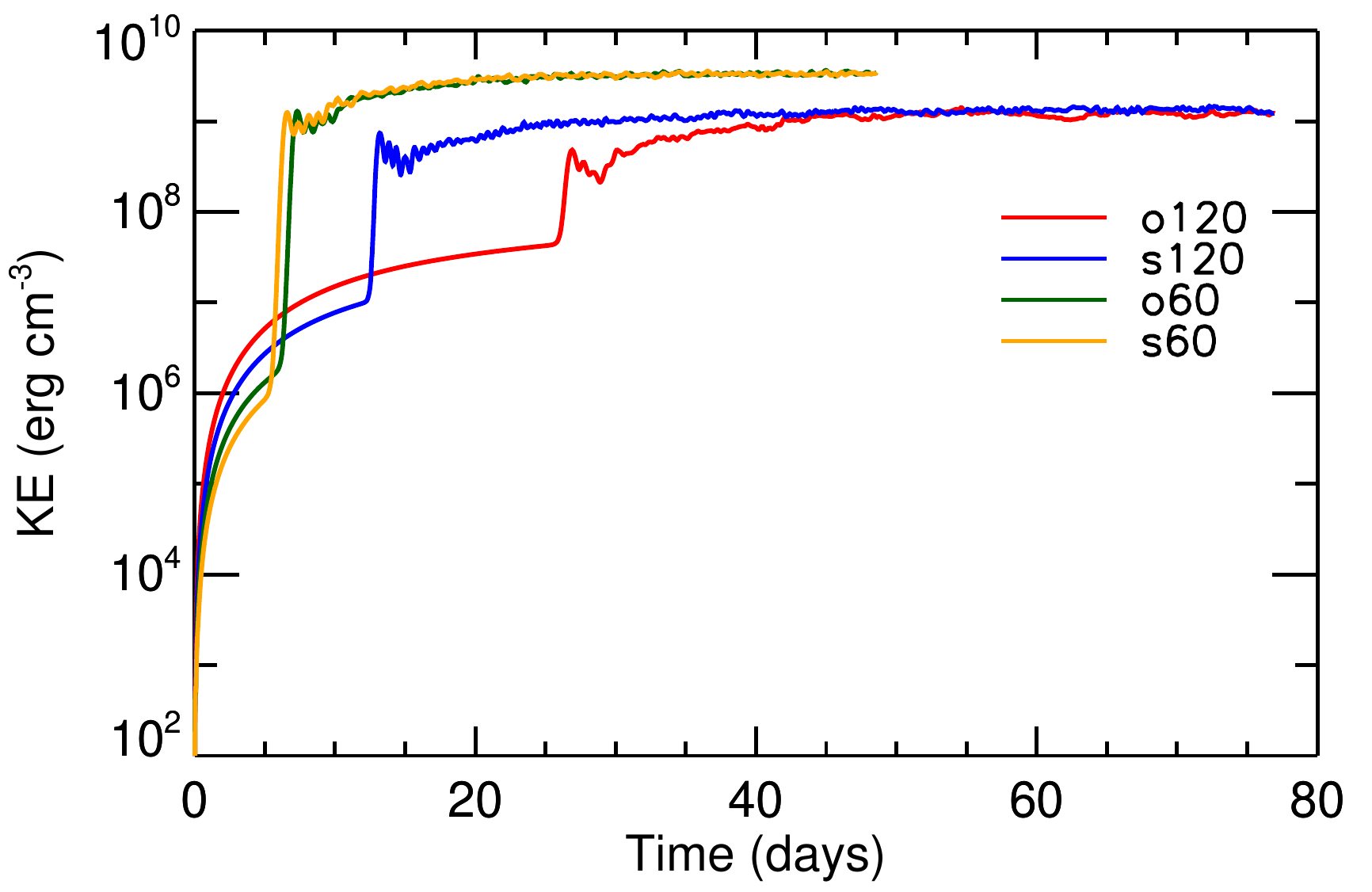}
\caption{Volume averaged kinetic energy density time histories for cases $o120$, $s120$, $o60$, and $s60$.}
\label{fig:KE_SO}
\end{figure}

The duration of this initial establishment period before the exponential growth phase varies with the rotation rate and oblateness. As the rotation rate is increased, this duration increases from 6 days for case $s60$ to 12 days for case $s120$ (Figure \ref{fig:KE}a).  This can be attributed to the stabilizing influence of rotation, which requires a greater entropy gradient to trigger convection.  A similar effect is also observed in the oblate spheroidal shell simulations shown in Figure \ref{fig:KE}b.   In order to assess the effect of oblateness on the growth and saturation of the convection, four cases are compared in Figure \ref{fig:KE_SO}. The initial establishment period in case $o60$ is only a little longer than that in case $s60$, which is expected because the oblateness in case $o60$ is minimal as shown in Table \ref{Tab:parameters}.  In contrast, the difference between cases $s120$ and $o120$ is pronounced (about $13$ days), implying that the significant oblateness in case $o120$ has a substantial influence on the establishment of a supercritical entropy gradient.  This can be attributed to the greater depth of the convection zone at the equator for the oblate cases, which dilutes the flux convergence and weakens thermal gradients.

It is clear from Figure \ref{fig:KE} and Table \ref{Tab:diagnostics} that the more rapidly-rotating cases saturate at a lower level of KE than the more slowly-rotating cases.  This is a consequence of the stabilizing effects of rotation and rotational shear (differential rotation) and it occurs in both the spherical and oblate cases.  However, the stabilizing influence on KE is somewhat larger in the oblate cases.  For example, the KE in o120 is about 36\% of that in o60 whereas the ratio between s120 and s60 is 41\%.  However, recall that KE is the mean kinetic energy density.  If the difference in volume is taken into account, then the total kinetic energy in case o120 is about 51\% of the total kinetic energy in o60.  We also note that the decrease in total KE with increasing $\Omega_0$ is a result that is sensitive to our chosen path through parameter space, which is to keep $\nu$ and $\kappa$ constant.  Recall from Section \ref{sec:setup} that we decrease $\nu$ and $\kappa$ as we increase $\Omega_0$ in the {\em a} series of simulations (s60a, s80a, s100a, and s120) in order to maintain a relatively constant supercriticality.  In these simulations, the total KE decreases only slightly from s60a to s100a and is in fact largest for s120, due largely to a strong differential rotation (Table \ref{Tab:diagnostics}).

For both the spherical and the oblate cases, more of the KE is contained in the differential rotation as $\Omega_0$ is increased (Table \ref{Tab:diagnostics}).  This is a common feature of global convection simulations and is attributed to the enhancement of the convective Reynolds stress \citep[e.g.][]{Brown2008}.  In the spherical cases, the ratio of DRKE/KE increases steadily over the entire range in rotation rates, from 60 to 120 $\Omega_{\odot}$.  However, interestingly, the DRKE/KE ratio seems to saturate in the oblate cases at about 0.61 for $\Omega_0 \gtrsim 80 \Omega_{\odot}$.  In this range of rotation rates, the DRKE/KE ratio is less than in the corresponding spherical cases, indicating that pronounced oblateness tends to suppress differential rotation.  This result may or may not be related to the observed saturation of the differential rotation $\Delta \Omega$ in rapidly-rotating stars, which will be discussed in section \ref{sec:DR1}.  For more moderate rotation rates, $\Omega_0 < 80 \Omega_\odot$, the oblate cases tend to have a slightly higher ratio of DRKE/KE compared to their spherical counterparts.

In contrast to DRKE/KE, the MCKE/KE ratio decreases as $\Omega_0$ is increased.  This again is commonly seen in spherical convection simulations and is attributed to the nature of the Reynolds stress and the tendency for flows to align with the rotation axis as expressed by the Taylor-Proudman theorem \citep{Brown2008}.  In short, the meridional circulation is established in response to the convective angular momentum flux in a process known as gyroscopic pumping \citep{miesc11,feath15}.  As the rotational influence is increased, this flux becomes perpendicular to the $z$ axis and is less effective at inducing a meridional circulation.  Though both the spherical and oblate cases exhibit a decrease in the MCKE/KE ratio as $\Omega_0$ is increased, the effect is somewhat more pronounced in the oblate cases.  Furthermore, the MCKE/KE ratio is generally less in the oblate cases than in the spherical cases with the same rotation rate.  In other words, the oblateness tends to suppress the MC, at least in terms of its kinetic energy.  This is somewhat of a surprise; one might have expected the oblateness to induce strong baroclinic circulations but we see no evidence of this.  We will return to these issues in section \ref{sec:DR1}.

\subsection{Convective Patterns}\label{sec:patterns}
\begin{figure}[hp!]
\subfigure[case $s120$]{
\includegraphics[width=0.5 \textwidth, angle=0]{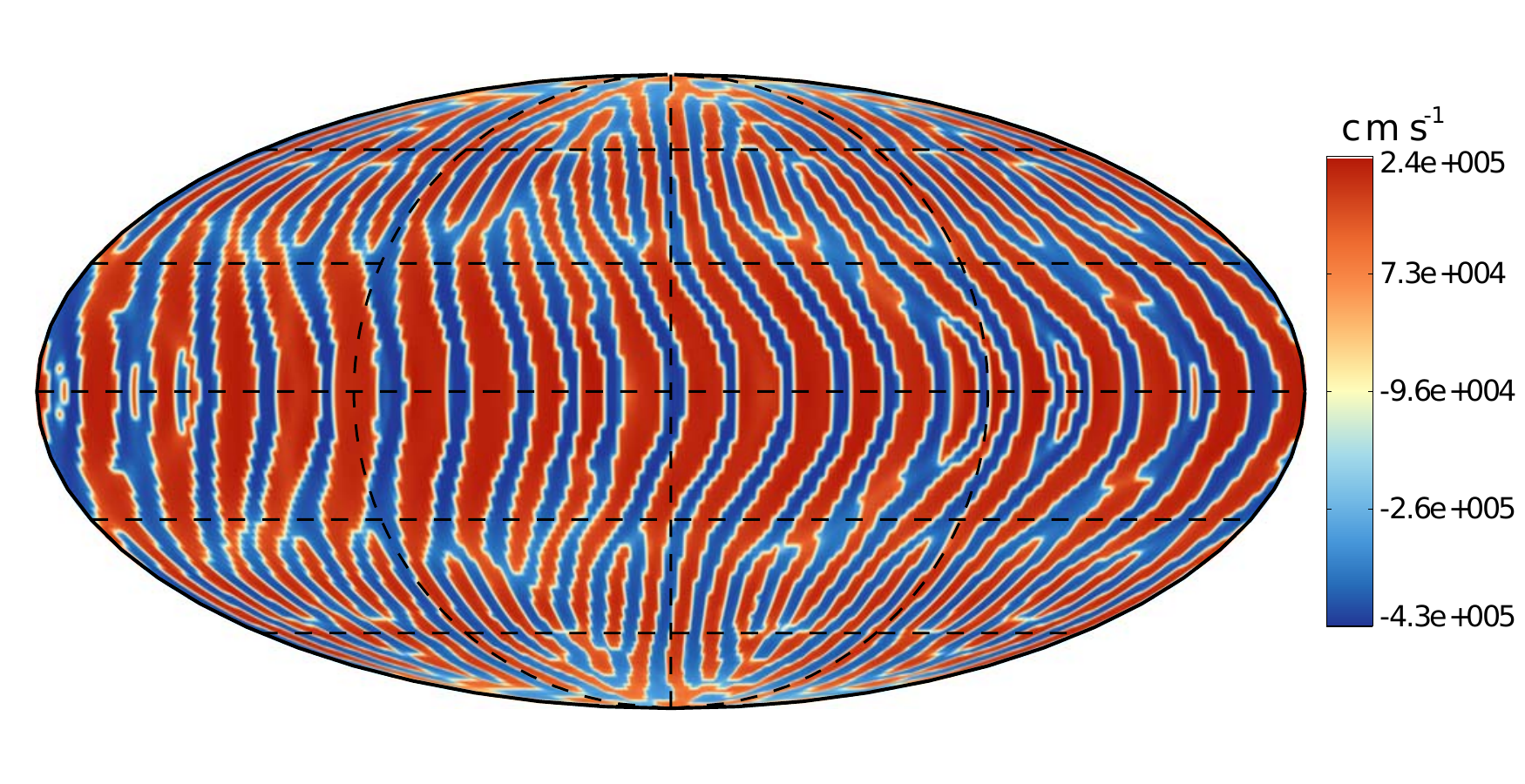}}
\subfigure[case $o120$]{
\includegraphics[width=0.5 \textwidth, angle=0]{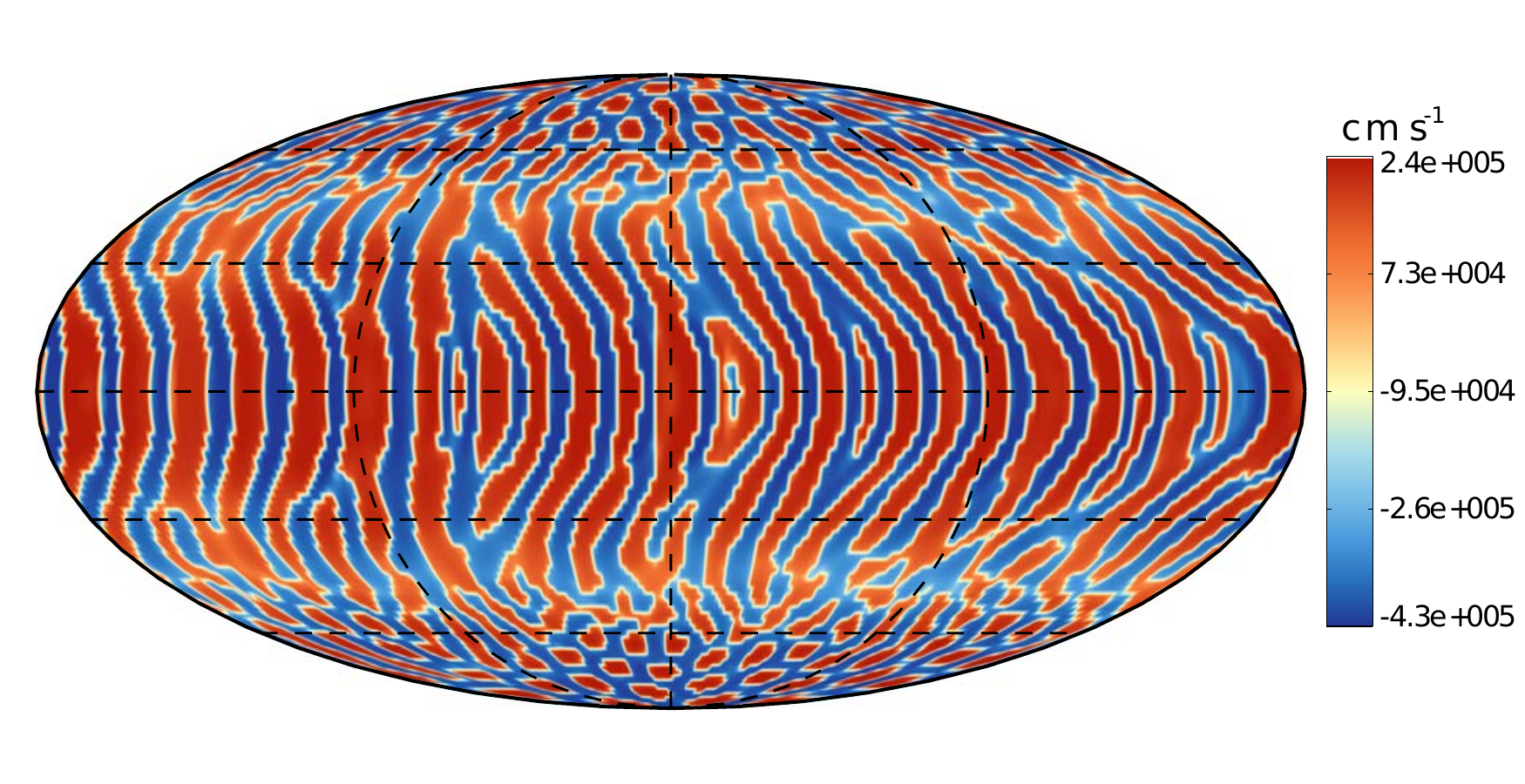}}
\subfigure[case $s90$]{
\includegraphics[width=0.5 \textwidth, angle=0]{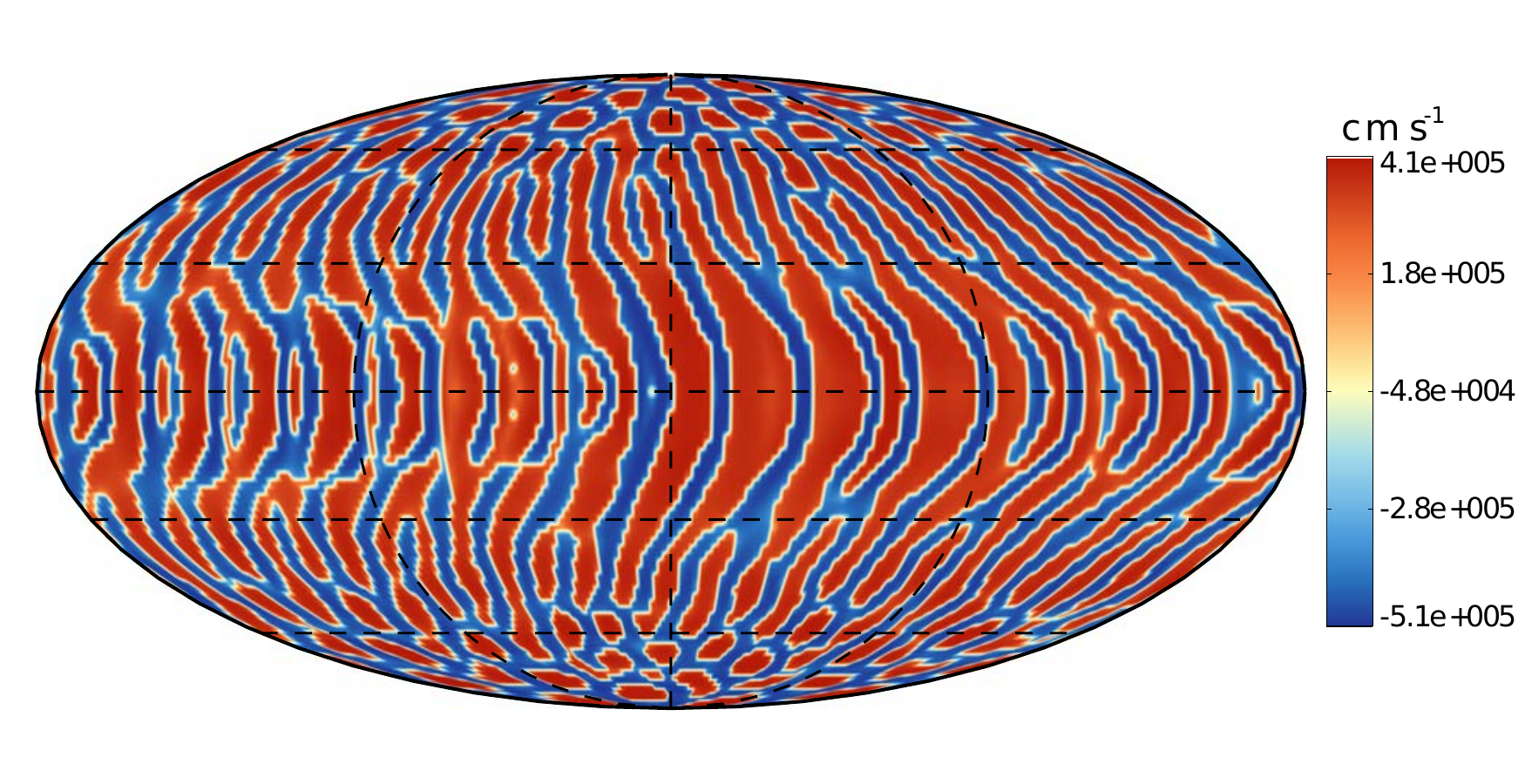}}
\subfigure[case $o90$]{
\includegraphics[width=0.5 \textwidth, angle=0]{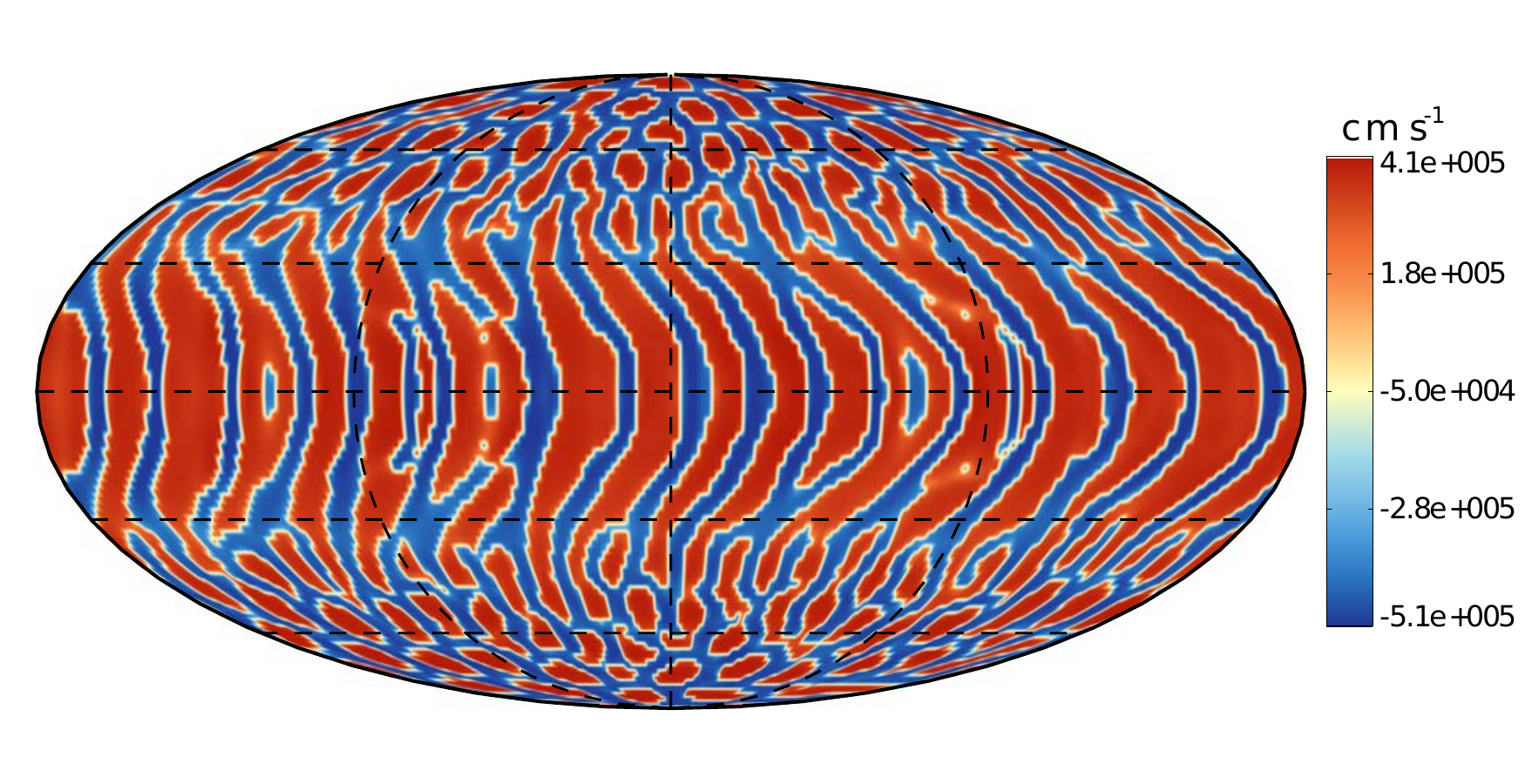}}
\subfigure[case $s60$]{
\includegraphics[width=0.5 \textwidth, angle=0]{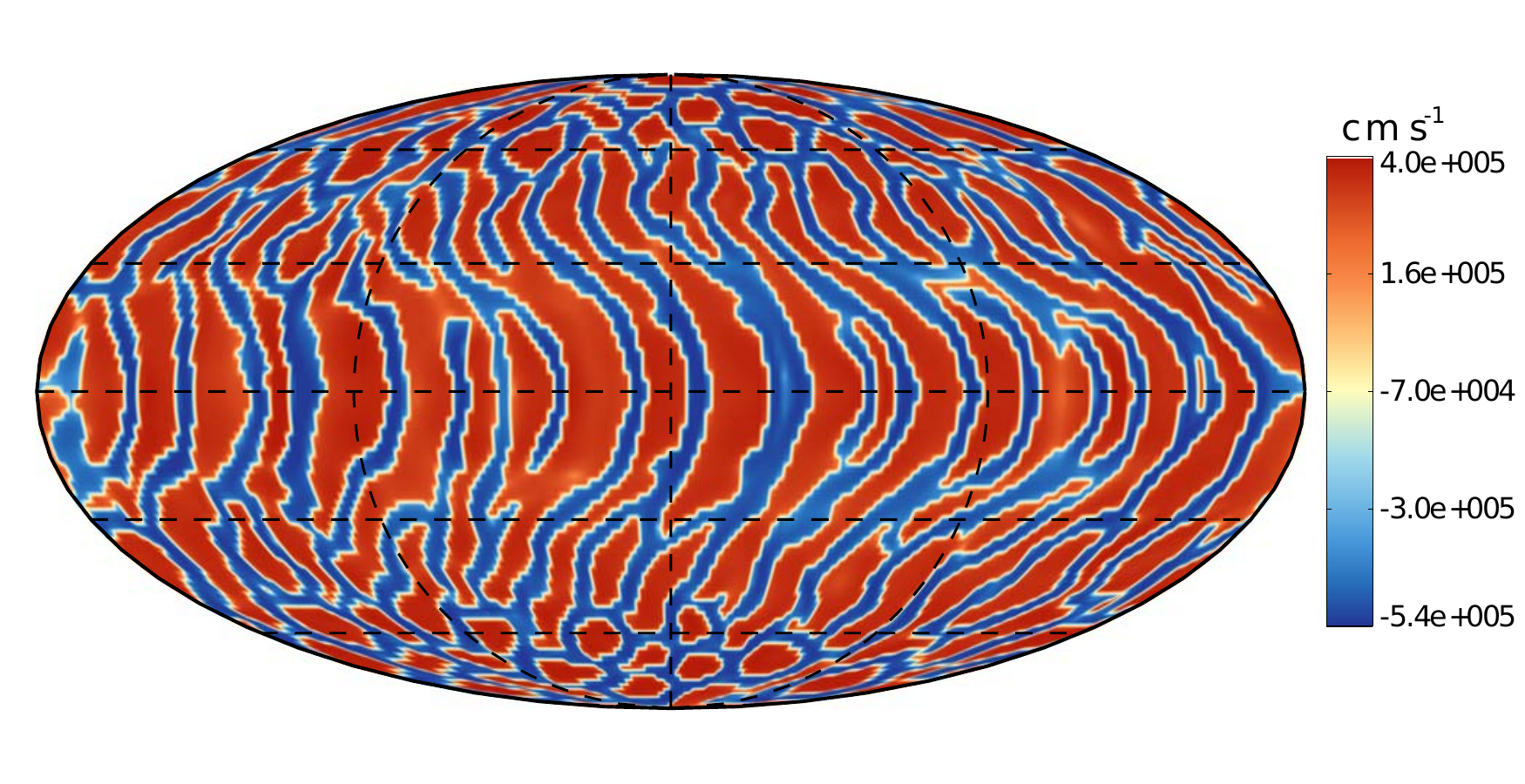}}
\subfigure[case $o60$]{
\includegraphics[width=0.5 \textwidth, angle=0]{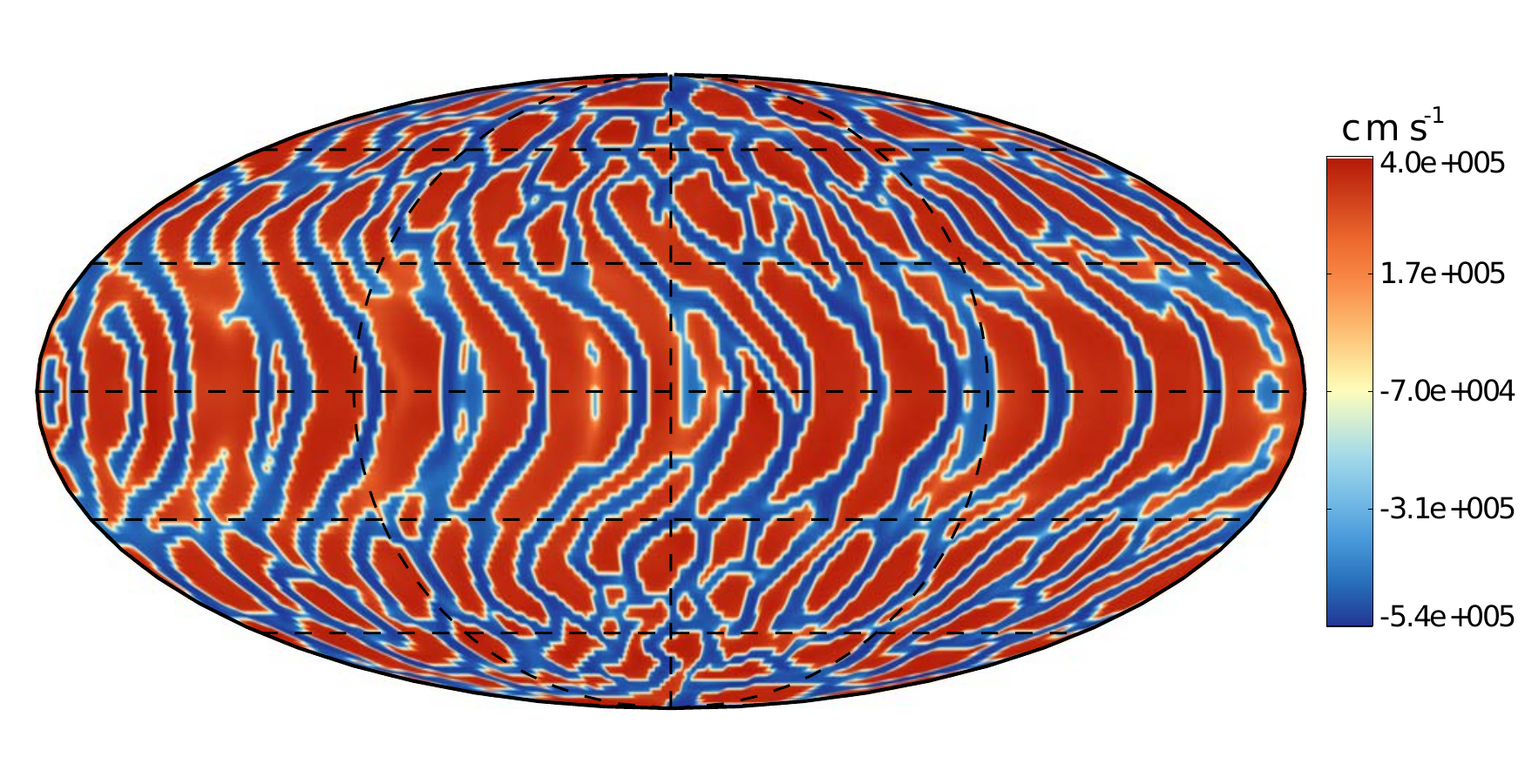}}
\caption{Mollweide projection snapshots of the normal velocity $u_{n}$ at the equipotential surface $\Phi$ corresponding to a polar radius of $0.98r_{op}$.  Cases within spherical shells and oblate spheroidal shells are shown in the left and right columns respectively. Red and blue tones denote the upflow and downflow as indicated by the color bar.}
\label{fig:con_pattern}
\end{figure}

In Figure \ref{fig:con_pattern}, the structure of the convection is illustrated. Shown is the velocity normal to equipotential surfaces $u_n = {\bf u} \bdot \uvn$, in Mollweide projection for a horizontal surface in the upper convection zone.  All simulations are dominated by a series of columnar convective rolls approximately aligned with the rotation axis but sheared slightly in the prograde direction at low latitudes by the differential rotation. These are the well-known 'banana cells' characteristics of convection in rotating shells for laminar parameter regimes \citep{Miesch2005}.   There is little apparent difference between simulations in spherical shells and those in oblate spheroidal shells.  In particular, the azimuthal wavenumber of the banana cell pattern is similar, on the order of $m \sim 22$ for both s120 and o120.   

\begin{figure}[!htp]
\centering
\subfigure[]{\includegraphics[width=0.325 \textwidth, angle=0]{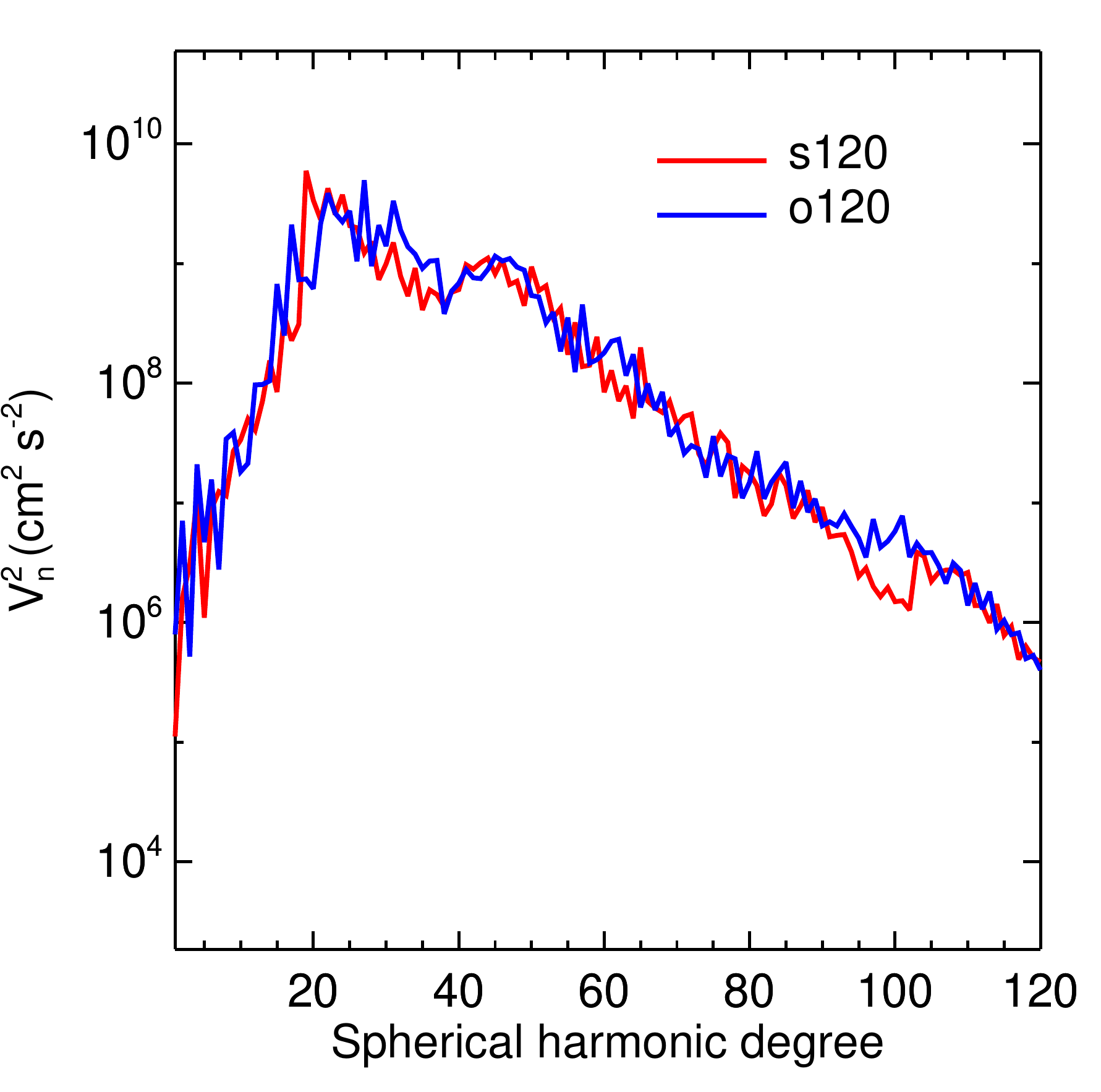}}
\subfigure[]{\includegraphics[width=0.325 \textwidth, angle=0]{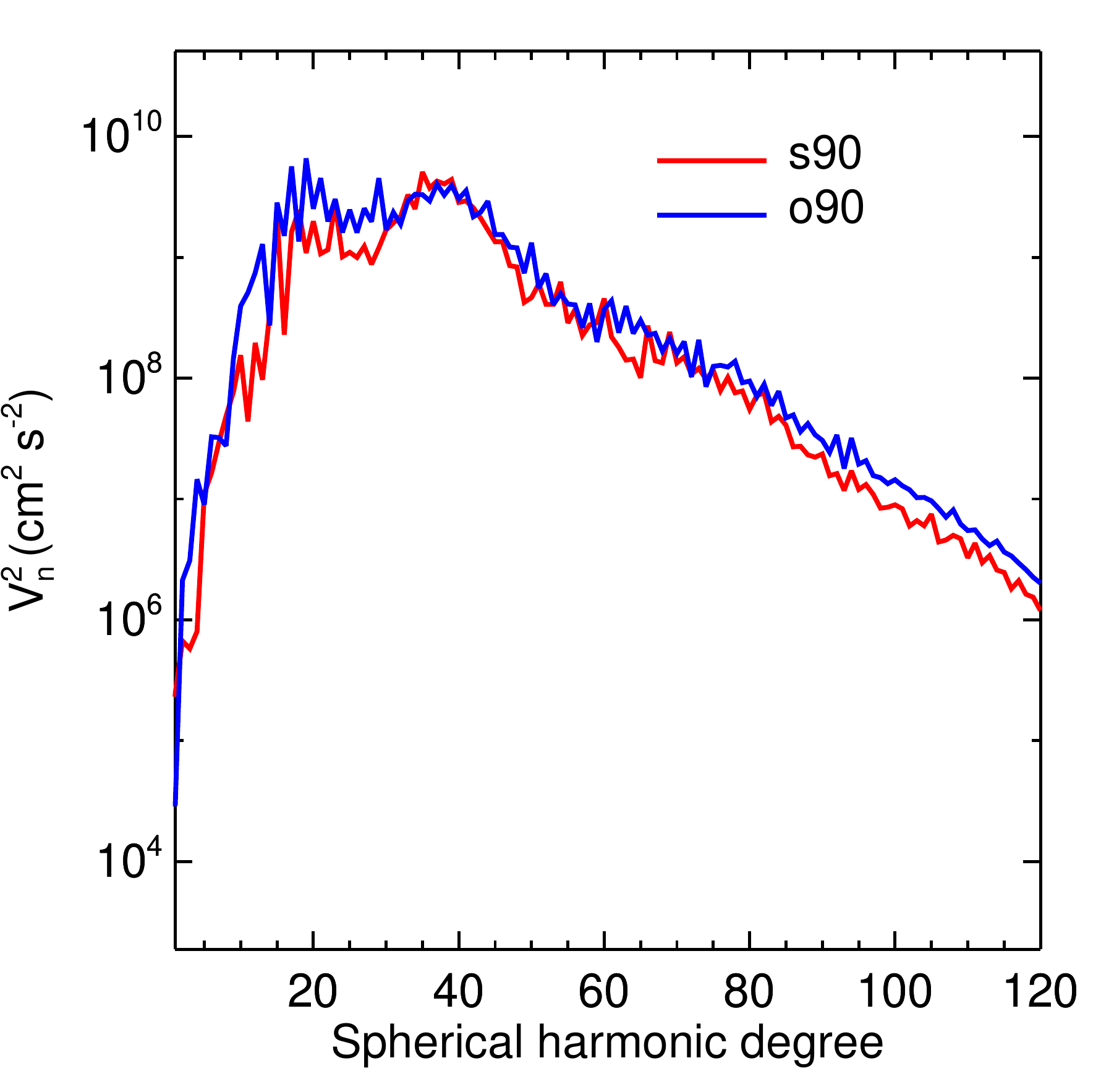}}
\subfigure[]{\includegraphics[width=0.325 \textwidth, angle=0]{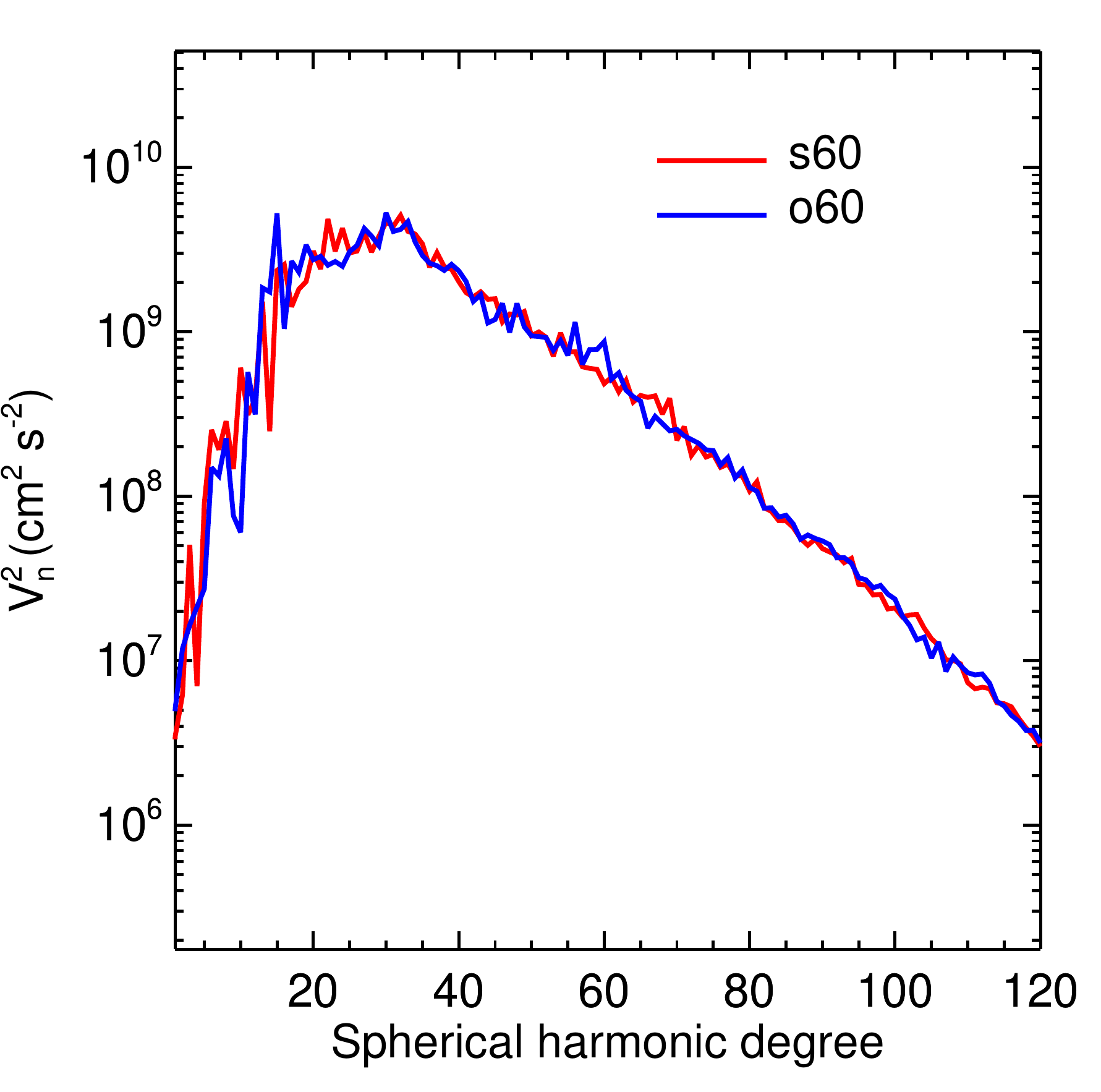}}
\caption{Power spectra of the normal velocity $u_{n}$ as a function of spherical harmonic degree at the equipotential surface $\Phi$ corresponding to a polar radius of $0.98r_{op}$. Shown are (\textit{a}) cases $s120$ and $o120$, (\textit{b}) cases $s90$ and $o90$, and (\textit{c}) cases $s60$ and $o60$.}

\label{fig:spectrum}.
\end{figure}

This qualitative similarity in appearance is confirmed by the quantitative power spectra, which are compared in Figure \ref{fig:spectrum}.   Each curve represents the power spectrum obtained by applying a spherical harmonic transformation to the normal velocity fields shown in Figure  \ref{fig:con_pattern} and then plotting the resulting power as a function of the spherical harmonic degree, which is, in effect, the horizontal wavenumber.  The differences between the spherical and oblate cases are insignificant.

However, closer scrutiny of Figure \ref{fig:con_pattern} does reveal some notable differences, particularly in the polar regions.  It is well known that convection in rotating spherical shells can be divided into two distinct dynamical regions, delineated by the so-called tangent cylinder \citep{Miesch2005}.  The tangent cylinder is a 2D cylindrical surface with a central axis aligned with the rotation axis and a cylindrical radius that is equal to the radius of the base of the convection zone at the equator.  Outside the tangent cylinder, the convection is generally dominated by banana cells or related vortex sheets.  These are the equatorial modes.  Meanwhile, the polar convection modes inside the tangent cylinder tend to be more horizontally isotropic, particularly at moderate Rossby numbers and high Rayleigh numbers.

Though all simulations exhibit a clear difference between polar and equatorial modes, the difference is much more pronounced in the fastest-rotating oblate cases.  This can be seen most clearly by comparing cases s120 and o120 in Figure \ref{fig:con_pattern}$a$ and $b$.   In case s120, the polar modes are highly anisotropic and linked with the equatorial banana modes.  By contrast, in case o120, the polar modes clearly decouple from the equatorial modes and exhibit a more isotropic cellular structure
(see also Figs.\ \ref{fig:Vn_SO} and \ref{fig:Vn_fat} below).  The same effect can also be seen in cases s90 and o90 (Figure \ref{fig:con_pattern}$c$, $d$).  

\begin{figure}[t!]
\subfigure[]{
\includegraphics[width=0.5 \textwidth, angle=0]{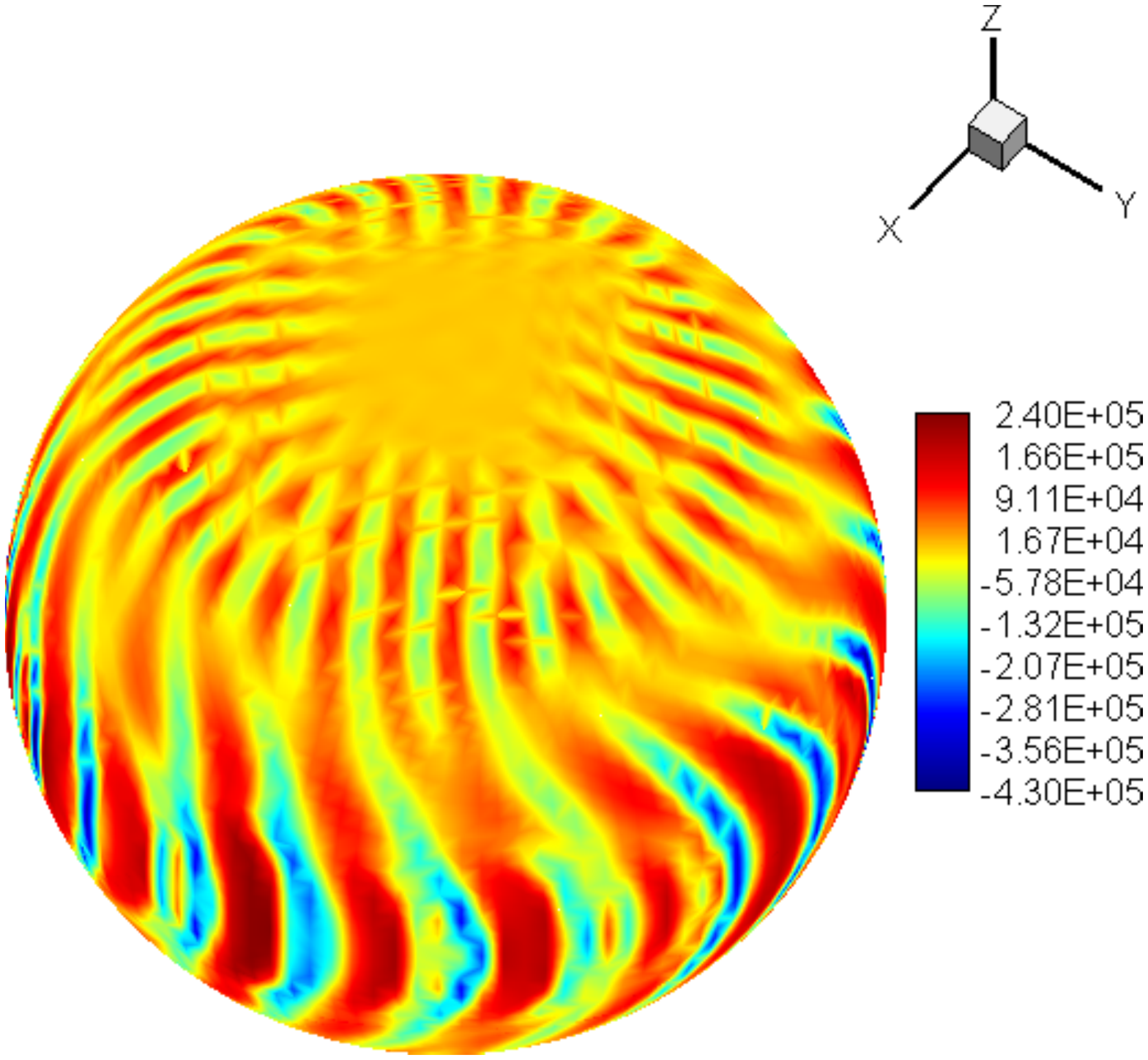}}
\subfigure[]{
\includegraphics[width=0.5 \textwidth, angle=0]{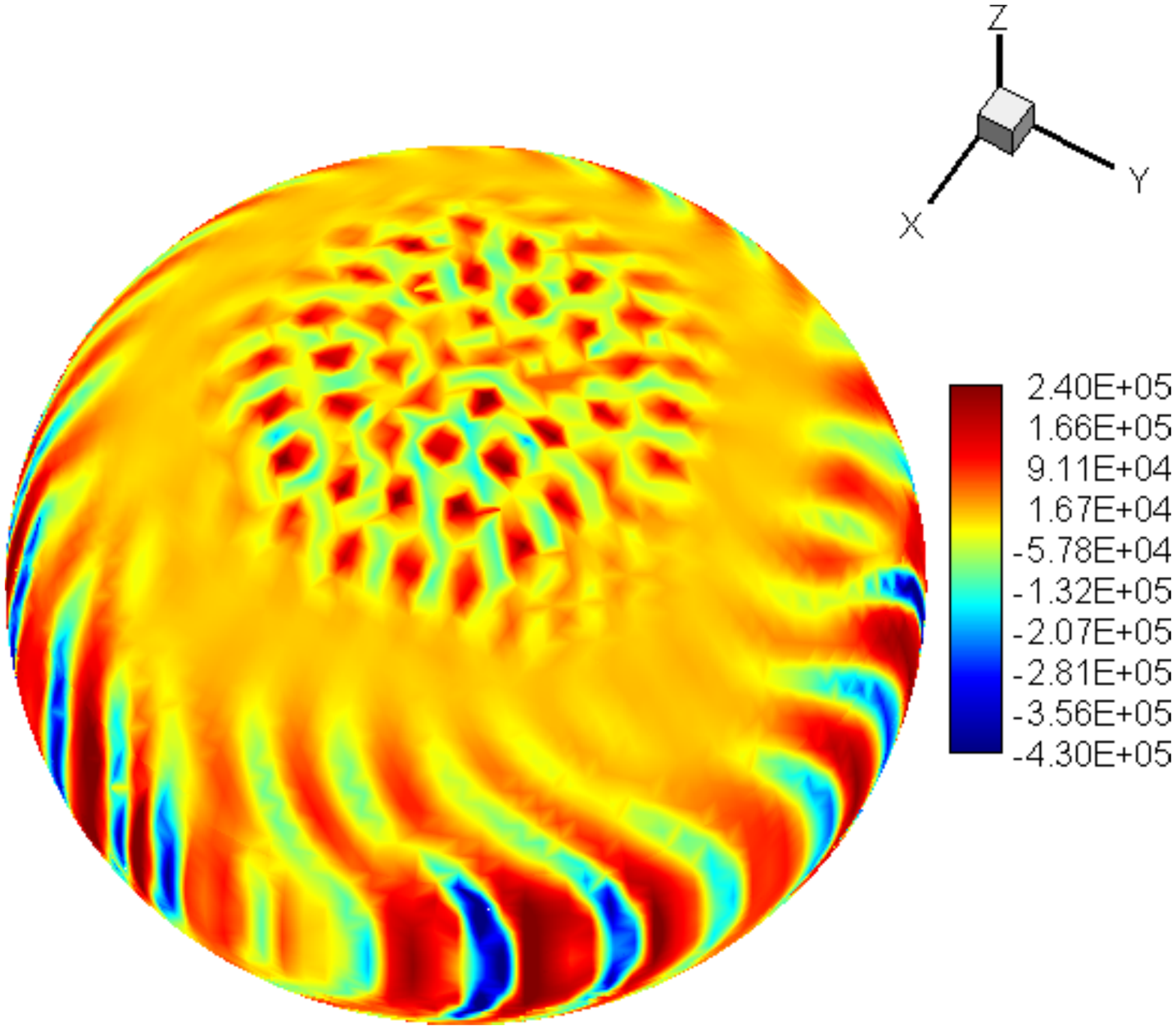}}
\caption{Snapshots of the normal velocity $u_{n}$ at the equipotential surface  $\Phi$ corresponding to a polar radius of $0.98r_{op}$ for (\textit{a}) case $s120$ and (\textit{b}) case $o120$.}
\label{fig:Vn_SO}
\end{figure}

The difference in the structure of the polar convection modes in cases s120 and o120 is further illustrated in Figure \ref{fig:Vn_SO}.  Here it can be seen that the polar convection in case o120 is not only more isotropic and more decoupled from lower latitudes, but it is also more vigorous.  We will return to this issue in Section \ref{sec:flux}.  We close this section with an orthographic view of case o120 from a point closer to the equatorial plane, highlighting its oblateness.

\begin{figure}[h!]
\begin{center}
\includegraphics[width=0.5 \textwidth, angle=0]{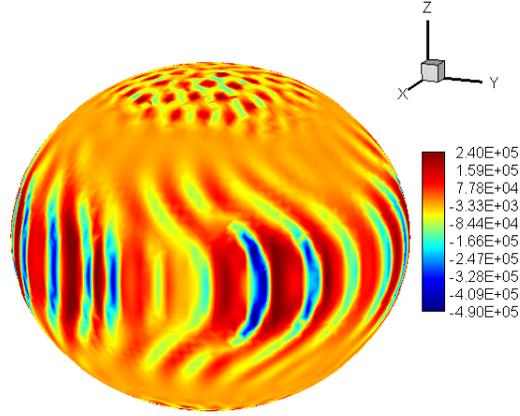}
\end{center}
\caption{As in Figure \ref{fig:Vn_SO}$b$ but viewed closer to the equatorial plane.
\label{fig:Vn_fat}}
\end{figure}

\subsection{Differential Rotation}\label{sec:DR1}
Differential rotation established by convection is a common feature in stars. The latitudinal angular velocity contrast on the surface, $\Delta \Omega$, is an important quantity both for interpreting stellar observations and for dynamo theories. Various observational techniques have been employed to measure $\Delta \Omega$, and to investigate the relationship between $\Delta \Omega$ and the stellar angular velocity $\Omega_{0}$. These observational techniques range from photometric variability studies \citep{DONAHUE1996} to Doppler imaging techniques \citep{Donati2003} and Fourier transform methods \citep{Reiners2003}. Global numerical simulations and mean-field models of convection in spherical shells have also been used to study that relationship \citep{Brown2008,Hotta2011}. Specifically, $\Delta \Omega$ is often defined as the difference in angular velocity between the equator and $60^{\circ}$ latitude, namely,
\begin{equation}
\Delta \Omega = \Omega_{eq} - \Omega_{60} .
\end{equation}

\begin{figure}[t!]
\center
{\includegraphics[width=0.9 \textwidth, angle=0]{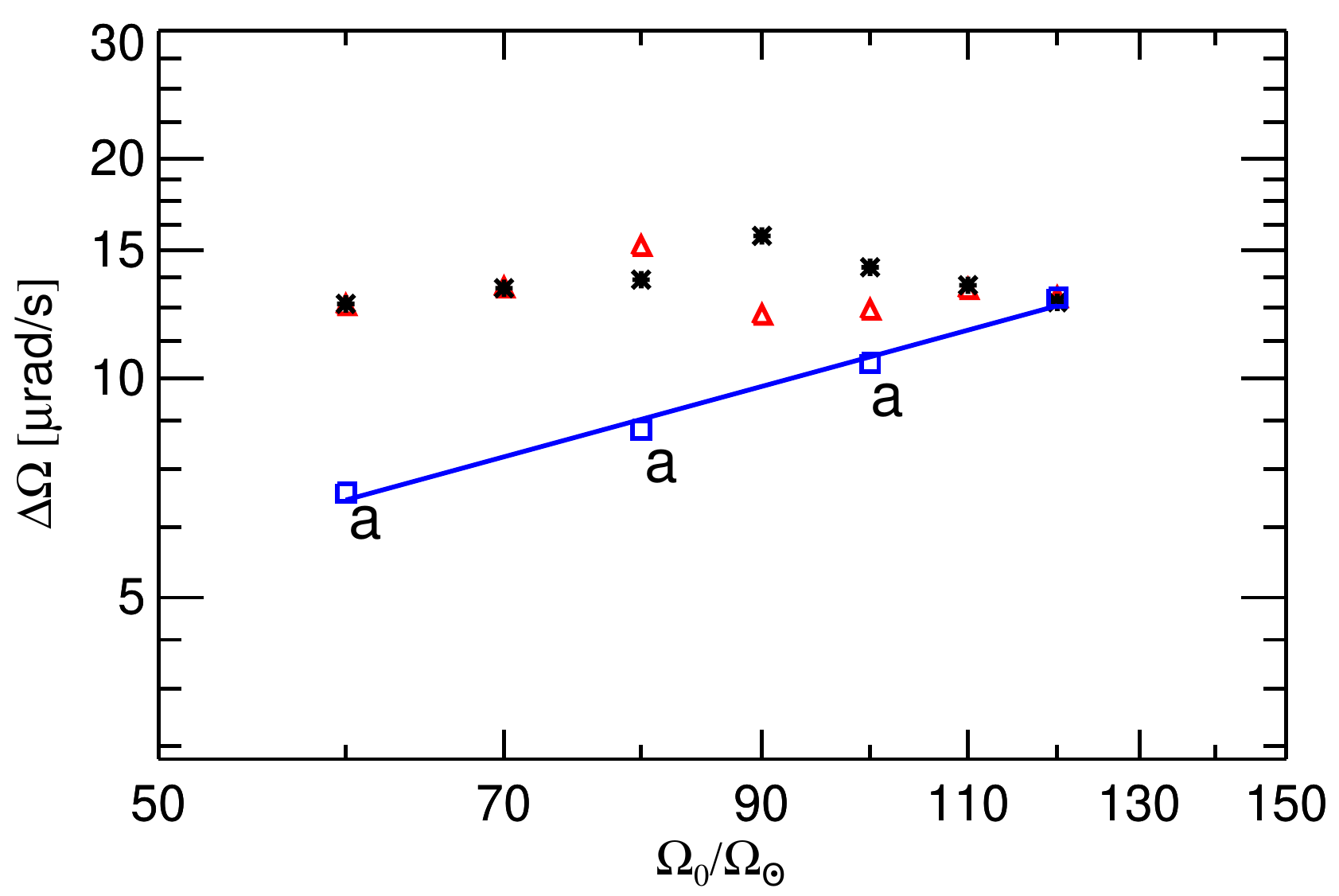}}
\caption{Latitudinal angular velocity contrast $\Delta \Omega$ for the spherical simulations $s60$-$s120$ (red triangles) and the oblate simulations $o60$-$o120$ (black asterisks).  Also shown are the {\em a} series of simulations $s60a$-$s100a$, marked with blue squares.  These latter simulations (shown with blue squares) follow a power law, with a best-fit exponent of $n = 0.89$.}
\label{fig:delta_omega}
\end{figure}

The differential rotation profiles for all of the simulations considered in this paper are solar-like in the sense that the equator rotates more rapidly than the polar regions ($\Delta \Omega > 0$).   Detailed rotation profiles will be discussed in Section \ref{sec:mean}.  Here we focus on the dependence of $\Delta \Omega$ (computed using eq.\ \ref{eq:omega} below) on $\Omega_0$, which is shown in Figure \ref{fig:delta_omega}.

For the spherical simulations $s60$-$s120$, $\Delta \Omega$ is insensitive to $\Omega_0$, showing no clear trend.  This is somewhat of a surprise, since previous studies of convection in rotating spherical shells exhibit an increase of $\Delta \Omega$ with $\Omega_0$ \citep[e.g.][]{Brown2008}.  However, recall from Section \ref{sec:setup} that in this series of simulations $\nu$ and $\kappa$ were kept constant as $\Omega_{0}$ was increased.  By contrast, \citet{Brown2008} decreased $\nu$ and $\kappa$ ($\propto$ $\Omega_{0}^{-2/3}$) in order to maintain a relatively constant supercriticality.  This is the motivation behind our {\em a} series of simulations, represented in Figure \ref{fig:delta_omega} by the blue boxes.  These use the same scaling for $\nu$ and $\kappa$ as in \citet{Brown2008}, and we find similar results.  In particular, $\Delta \Omega$ monotonically increases with $\Omega_0$ approximately as a power law; $\Delta \Omega \propto \Omega_0^n$ with $n \sim 0.89$.  This is qualitatively consistent with the results of Brown et al.\ \citep[see also][]{Hotta2011}, though our exponent is somewhat steeper than their value of $n \sim 0.3$, valid for Rossby numbers below about 0.4. 

These trends are generally consistent with observations, which suggest $0.15 \lesssim n < 1$ for solar-type stars \citep{DONAHUE1996,Donati2003,Reiners2003}.  However, this observed scaling relation only holds out to about $20 \Omega_{\odot}$, after which $\Delta \Omega$ may remain constant or even decrease with $\Omega_0$, though observational evidence for this is still only preliminary \citep{saar09}.  This saturation and perhaps reversal of the $\Delta \Omega$ vs $\Omega_0$ scaling relation appears to be associated with a similar saturation of the magnetic field strength, $B$, which increases with $\Omega_0$ until about the same point, $\Omega_0 \sim 20 \Omega_\odot$, beyond which $B$ becomes approximately independent of $\Omega_0$ \citep[e.g.][]{pizzo03}.  This suggests that the saturation of the $\Delta \Omega$ vs $\Omega_0$ relationship for rapidly-rotating solar like stars is likely due to strong Lorentz-force feedbacks associated with vigorous dynamo action.

However, our simulations suggest that oblateness may also contribute to the saturation of $\Delta \Omega$ for very fast rotation rates.  We already noted in Section \ref{sec:energetics} that the DRKE/KE ratio saturates at about 0.61 for $\Omega_0 \gtrsim 80 \Omega_\odot$ ($0.56 \Omega_{crit}$).  This result is also reflected in the $\Delta \Omega$ plot of Figure \ref{fig:delta_omega}, which shows a peak value at $\Omega_0 \sim 90 \Omega_{\odot}$ ($0.63 \Omega_{crit}$), only for the oblate cases.  At more rapid rotation rates, the oblateness seems to supress the differential rotation.  This is likely due to the decoupling between polar and equatorial modes noted in Section \ref{sec:patterns}. If banana cells reach beyond the tangent cylinder and into the polar regions as in case s120, the associated Reynolds stress can efficiently spin down the poles, enhancing the differential rotation.  However, if the polar and equatorial modes decouple due to oblateness as in case o120, this can inhibit the angular momentum transport and diminish $\Delta \Omega$.  

We will explore the mean flows in more detail in Section \ref{sec:mean}.  However, before we proceed, one more comment is warranted in order to relate our simulations to previous work.  Our numerical approach described in Section \ref{sec:MP} neglects the effect of differential rotation on oblateness.  Is this justified?  \citet{MacGregor2007} showed that differential rotation could significantly distort the shape of rapidly rotating stars compared to corresponding cases with uniform rotation. However, in their model the reference rotation rate $\Omega_{0}$ was allowed to be larger than the critical rotation rate $\Omega_{crit}$ and an anti-solar differential rotation profile was introduced to achieve stable solutions.  This extreme situation is likely to be rare in real stars.  Furthermore, \citet{MacGregor2007} considered a full, 2D stellar structure model that included differential rotation down to the region of energy generation in the stellar core.  Given the sensitivity of energy generation to the thermal structure, even a slight oblateness of the core could have a substantial influence on the stellar luminosity.  If the differential rotation is solar-like in the sense that the equator spins faster than the pole and if it is confined to the convective envelope, then we expect the differential rotation to have a negligible influence on the oblateness.  This is demonstrated in Appendix \ref{Appen:A}.

\section{CONVECTIVE ENERGY TRANSPORT}\label{sec:transport}

\subsection{Flux Balance and Entropy Stratification}\label{sec:flux}

From a star's perspective, perhaps the most important function of convection is energy transport; it carries the energy liberated by fusion in the core to the surface, where that energy is then radiated into space.  Since this paper is the first to consider 3D simulations of convection in oblate stars, we believe that it is useful to consider the issue of outward energy transport with some care.

A careful derivation is given in Appendix \ref{Appen:B} but the result is analogous to previous models of convection in spherical shells. Four principle components of the energy flux are involved in transporting energy in the outward direction, namely the enthalpy flux $F_{e}$, the kinetic energy flux $F_{k}$, the radiative flux $F_{r}$ and the diffusive entropy flux $F_{u}$. In a statistically steady state, these four fluxes together must account for the full luminosity imposed at the bottom boundary such that
\begin{equation}\label{eq:Fbalance}
F_{e} + F_{k} + F_{r} + F_{u} = F_{*} = \frac{L}{A_{\cal S}}  ~~~.
\end{equation}
Explicit expresstions for $F_e$, $F_k$, $F_r$, and $F_u$ are given in Appendix \ref{Appen:B}.  For a spherical shell $A_{\cal S} = 4 \pi r^2$ but this is no longer true for an oblate shell where we compute $A_{\cal S}$ numerically.

\begin{figure}[hpt!]
\subfigure[]{
\includegraphics[width=0.5 \textwidth, angle=0]{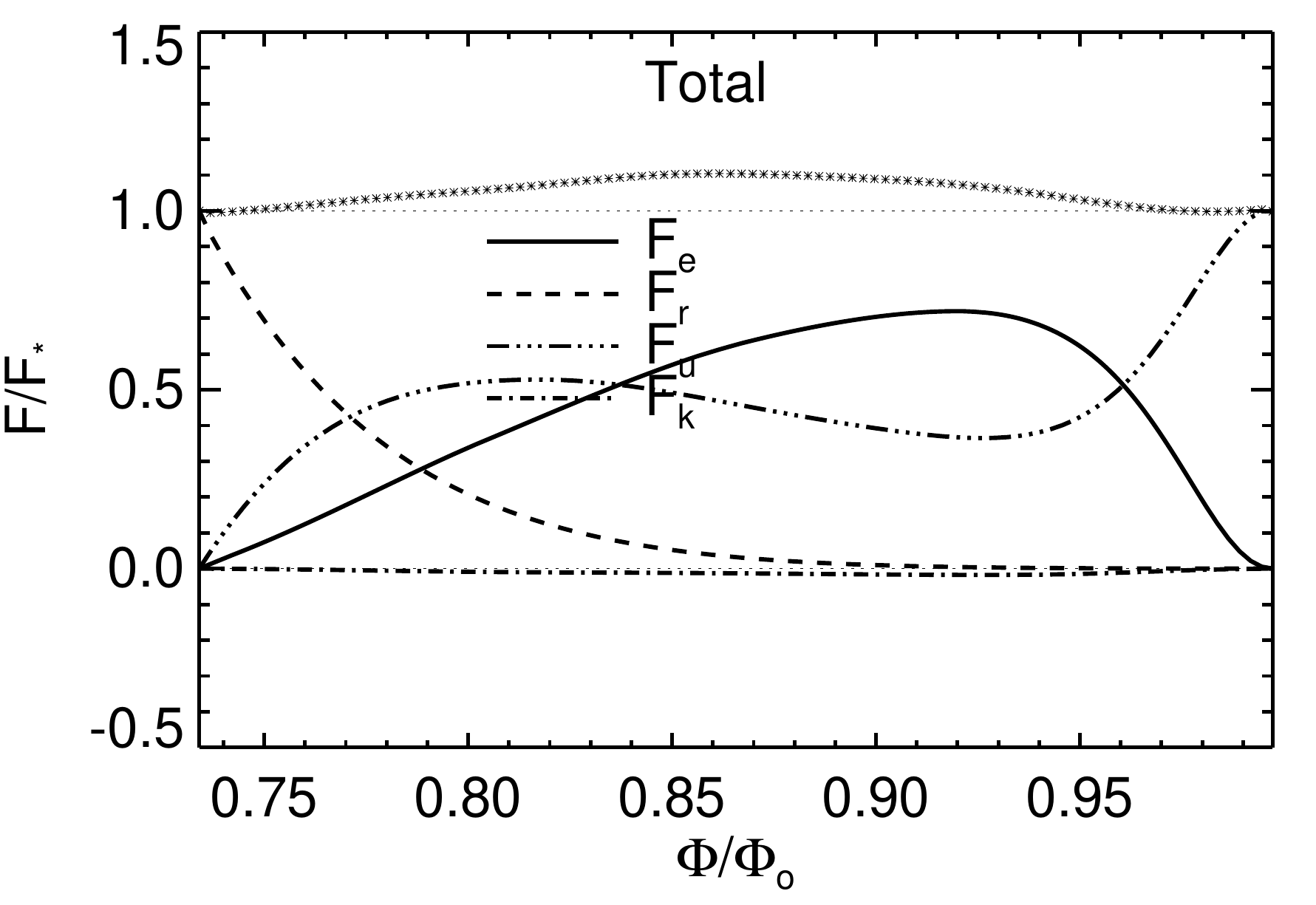}}
\subfigure[]{
\includegraphics[width=0.5 \textwidth, angle=0]{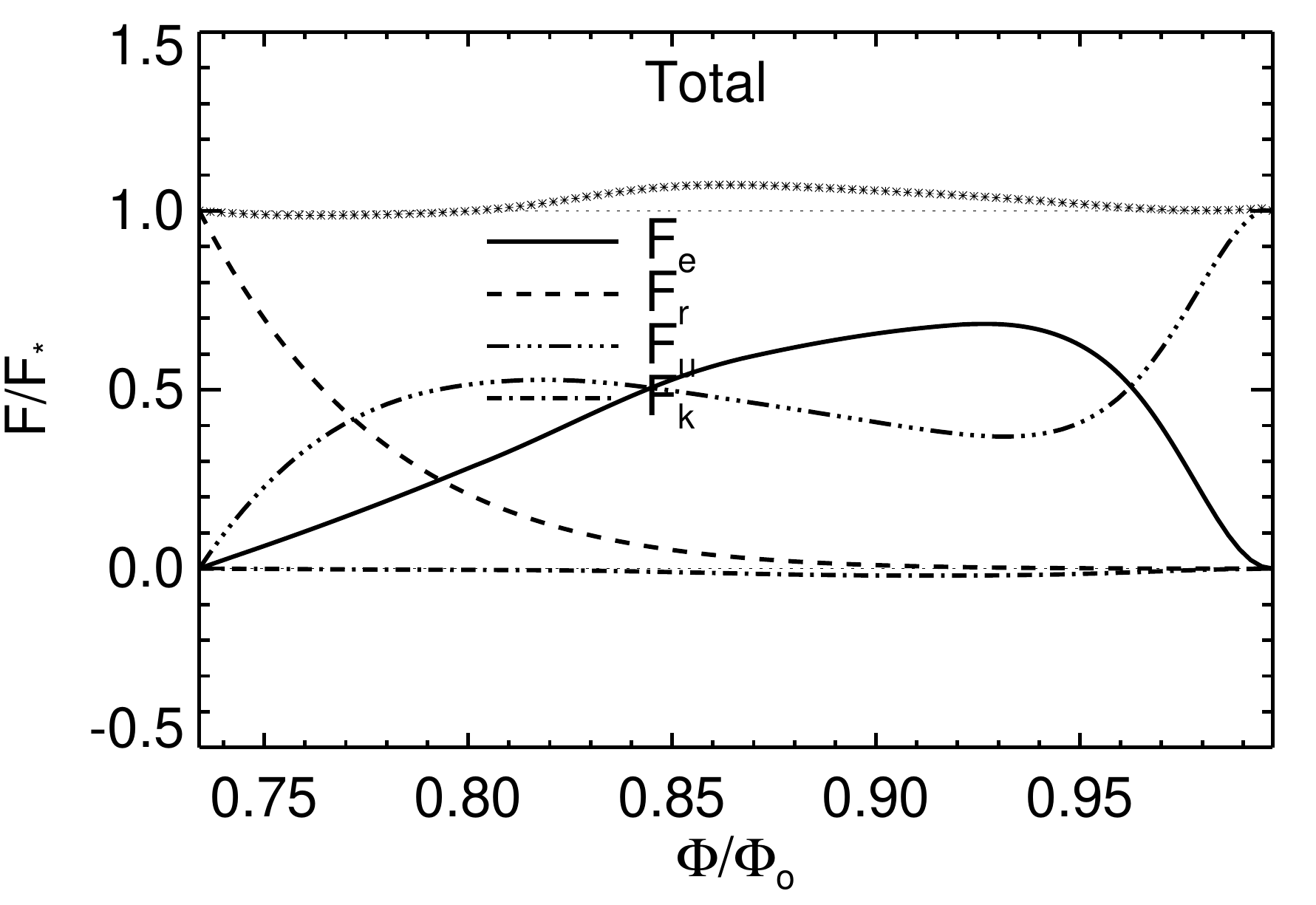}}
\caption{Energy fluxes through equipotential surfaces for cases \textit{(a)} $s60$ and \textit{(b)} $o60$. Shown are
the enthalpy flux $F_{e}$, radiative flux $F_{r}$, entropy flux $F_{u}$, and the kinetic energy flux $F_{k}$, all normalized by $F_{*}$.  Asterisks denote the sum of the other four curves and a dotted line is included at $F/F_* = 1$ for reference.}
\label{fig:flux_balance-60}
\end{figure}

Figure \ref{fig:flux_balance-60} shows the flux balance through equipotential surfaces for cases $s60$ and $o60$. It is apparent that the energy flux distribution in case $s60$ is very similar to that in case $o60$, which is expected because of the small oblateness. By design, the radiative flux $F_{r}$ carries all of the imposed flux at the bottom where the normal velocity vanishes, as do $F_k$ and $F_e$.  The radiative flux $F_{r}$ also dominates the heat transport in the lower convective envelope while the entropy flux $F_{u}$ carries energy through the top boundary and the upper convective envelope. The strong correlations between normal velocities and temperature fluctuations yield the enthalpy flux $F_{e}$, which gradually increases towards the top from the bottom until it peaks near $\Phi/\Phi_o=0.93$. It then drops down to zero rapidly as it approaches the impenetrable top boundary. 

\begin{figure}[hpt!]
\subfigure[]{
\includegraphics[width=0.5 \textwidth, angle=0]{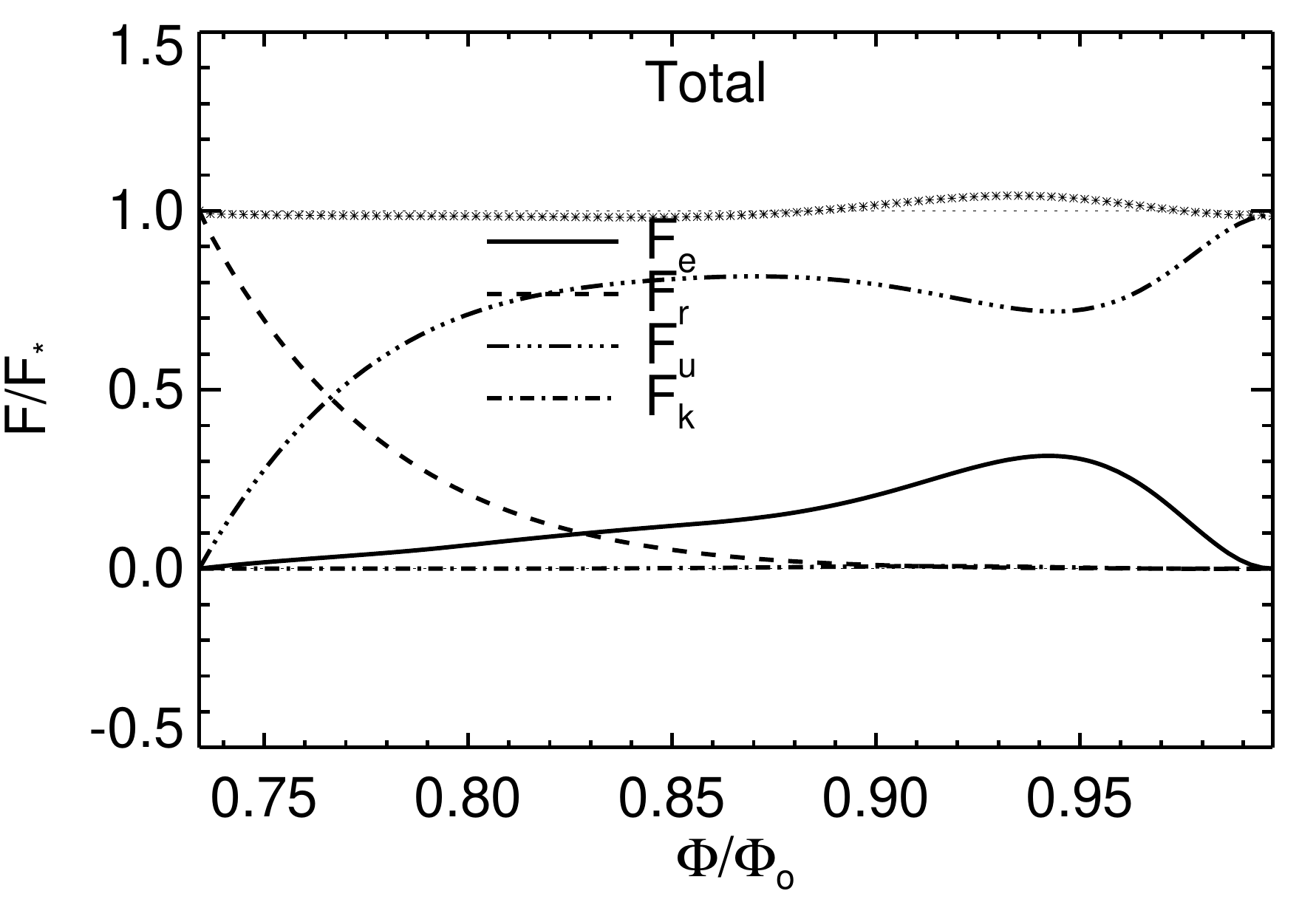}}
\subfigure[]{
\includegraphics[width=0.5 \textwidth, angle=0]{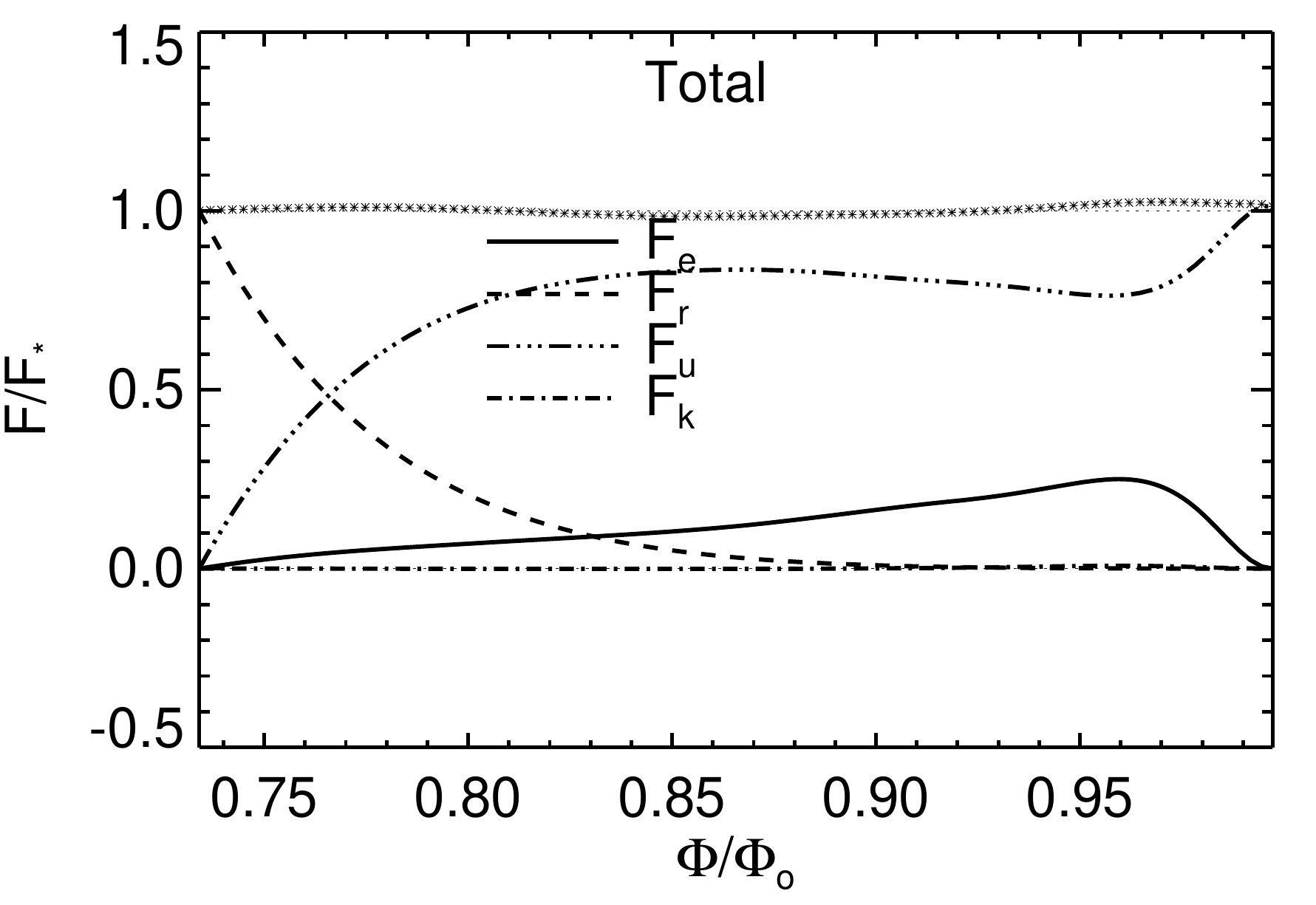}}
\caption{Energy fluxes as in Figure \ref{fig:flux_balance-60}, but for cases \textit{(a)} $s120$ and \textit{(b)} $o120$.}
\label{fig:flux_balance-120}
\end{figure}

As seen in Figure \ref{fig:flux_balance-120}, the difference in the energy flux distribution between cases $s120$ and $o120$ is more significant.  This is especially true for the enthalpy flux $F_{e}$ and the entropy flux $F_{u}$. In case $s120$, $F_{e}$ reaches a maximum value of 0.32 around $\Phi/\Phi_{r_{op}}=0.95$ while the peak $F_e$ value in case $o120$ is only 0.25, achieved around $\Phi/\Phi_o=0.97$.  This suggests that the oblateness is slightly inhibiting the net convective energy transport through the shell.  The normalized sum of the four principle flux components in all of our simulations is close to one throughout the shell, signifying that they have achieved an equilibrated state.

\begin{figure}[hpt!]
\subfigure[]{
\includegraphics[width=0.5 \textwidth, angle=0]{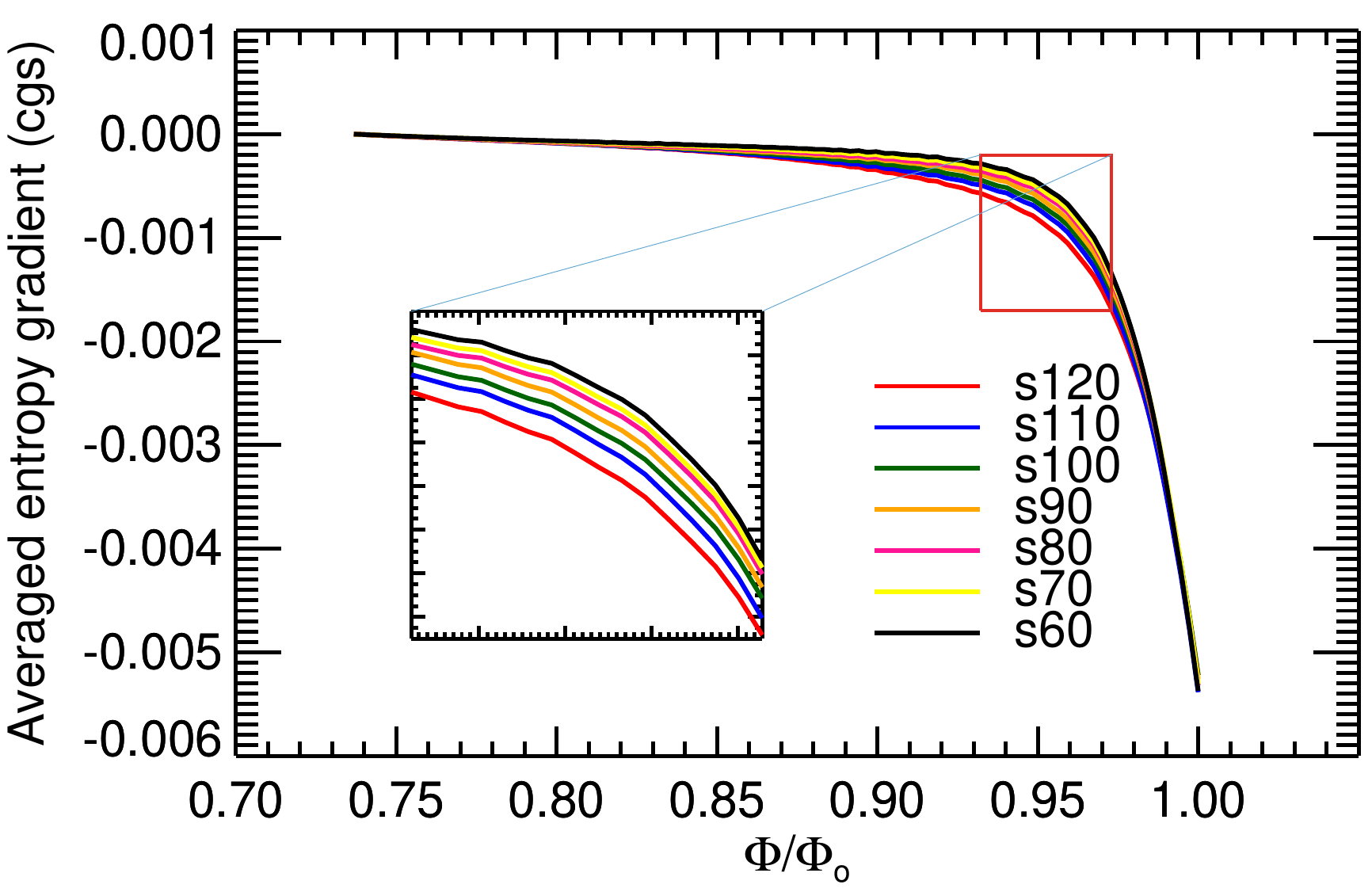}}
\subfigure[]{
\includegraphics[width=0.5 \textwidth, angle=0]{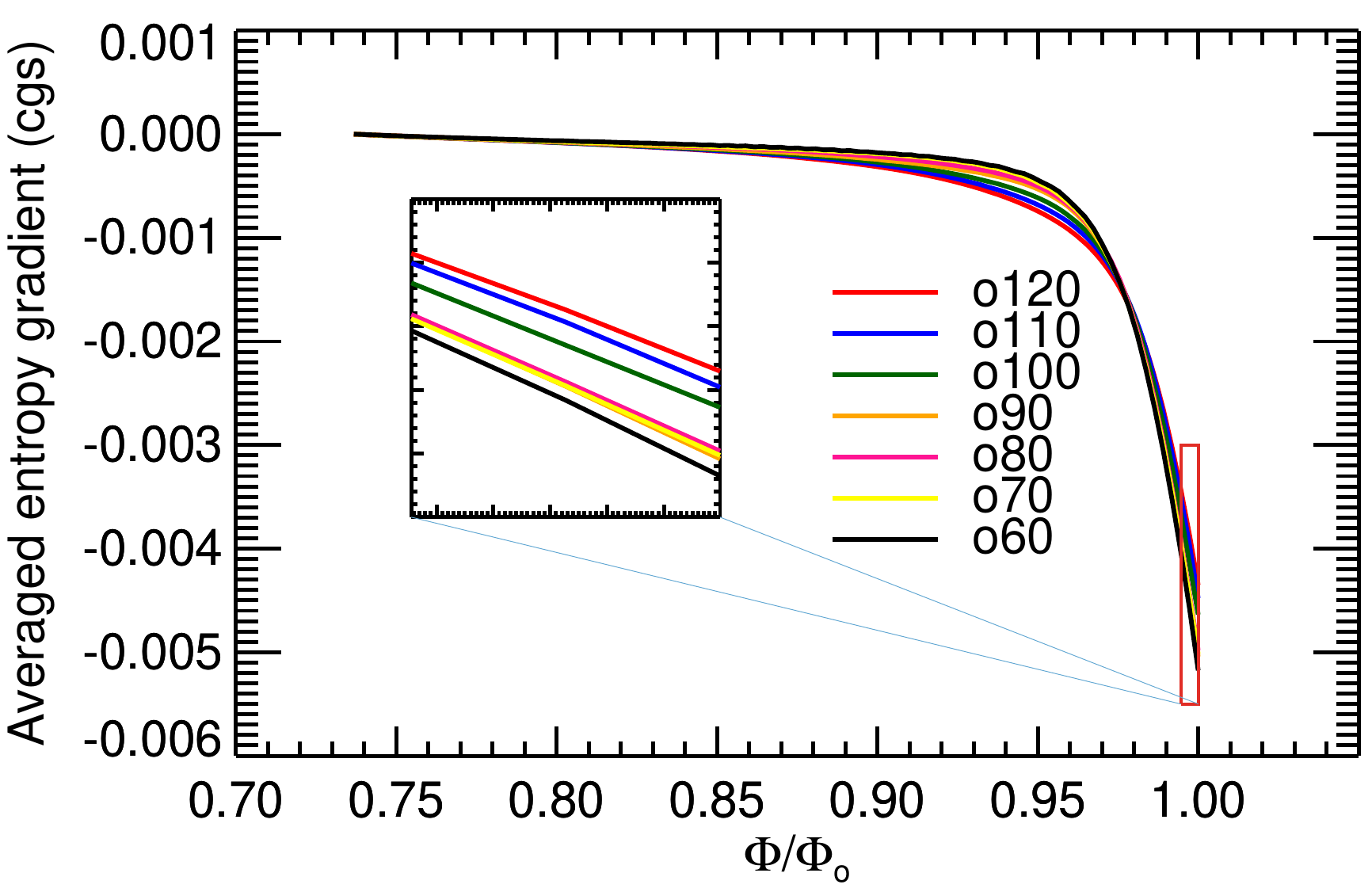}}
\caption{Normal entropy gradient $\del S \bdot \uvn$ averaged over equipotential surfaces and over a time period of $10$ days. The units are $erg$ $g^{-1}$ $K^{-1}$. Shown are \textit{(a)} spherical shell simulations and \textit{(b)} oblate spheroidal shell simulations.  The inset boxes highlight portions of each plot as indicated.}
\label{fig:dsdr}
\end{figure}

It is well known that rapid rotation also tends to inhibit convection \citep{Miesch2005}.  This can be seen in our simulations through its effect on the mean entropy stratification.  We find that the convection is less able to establish an adiabatic profile throughout the convection zone in the rapidly rotating simulations. Instead, much of the convection zone remains slightly superadiabatic ($\del S \bdot \mathbf{n}< 0$).   The radial profile of the normal entropy gradient $\del S \bdot \mathbf{n}$ for our simulations is shown in Figure \ref{fig:dsdr}, highlighting the dependence on the rotation rate. For the spherical simulations (Fig. \ref{fig:dsdr}a), the normal entropy gradient exhibits a greater departure from the adiabatic profile as the rotation rate is increased from $60\Omega_{0}$ to $120\Omega_{0}$.  This signifies a decrease in the efficiency of the convective energy transport for two reasons.  First, the convection is less able to mix entropy to establish an adiabatic stratification.  Second, the reduced enthalpy flux ($F_e$) in the upper convection zone requires a steeper entropy gradient (larger $F_u$) in order to satisfy the flux balance equation (\ref{eq:Fbalance}).

A similar trend is observed in the upper convection zone for the oblate simulations (Fig. \ref{fig:dsdr}b).  Here again, the magnitude of the superadiabatic entropy gradient increases with increasing $\Omega_0$.  However, this trend reverses near the outer boundary as highlighted in the inset.  This can be attributed to the larger surface area of the oblate spheroids.  In order for each simulation to achieve equilibrium, the total flux through the outer boundary must equal the total flux imposed through the inner boundary, which is in turn equal to the luminosity, $L$. In our simulations, the flux through the outer boundary is carried by the entropy flux $F_u$, which is proportional to the normal entropy gradient, as expressed in equation (\ref{eqn:Fu}).  Thus in order to achieve equilibrium, $F_u$ must be equal to $L/A_o$ at $\Phi = \Phi_o$, where $A_o$ is the area of the outer surface.  Thus, for a given $L$, the $F_u$ is reduced as $A_o$ is increased.  In other words, the total luminosity is spread over a larger area in oblate stars so the flux through any fixed area on the surface is reduced. 

\begin{figure}[hp!]
\center
{\includegraphics[width=0.9 \textwidth, angle=0]{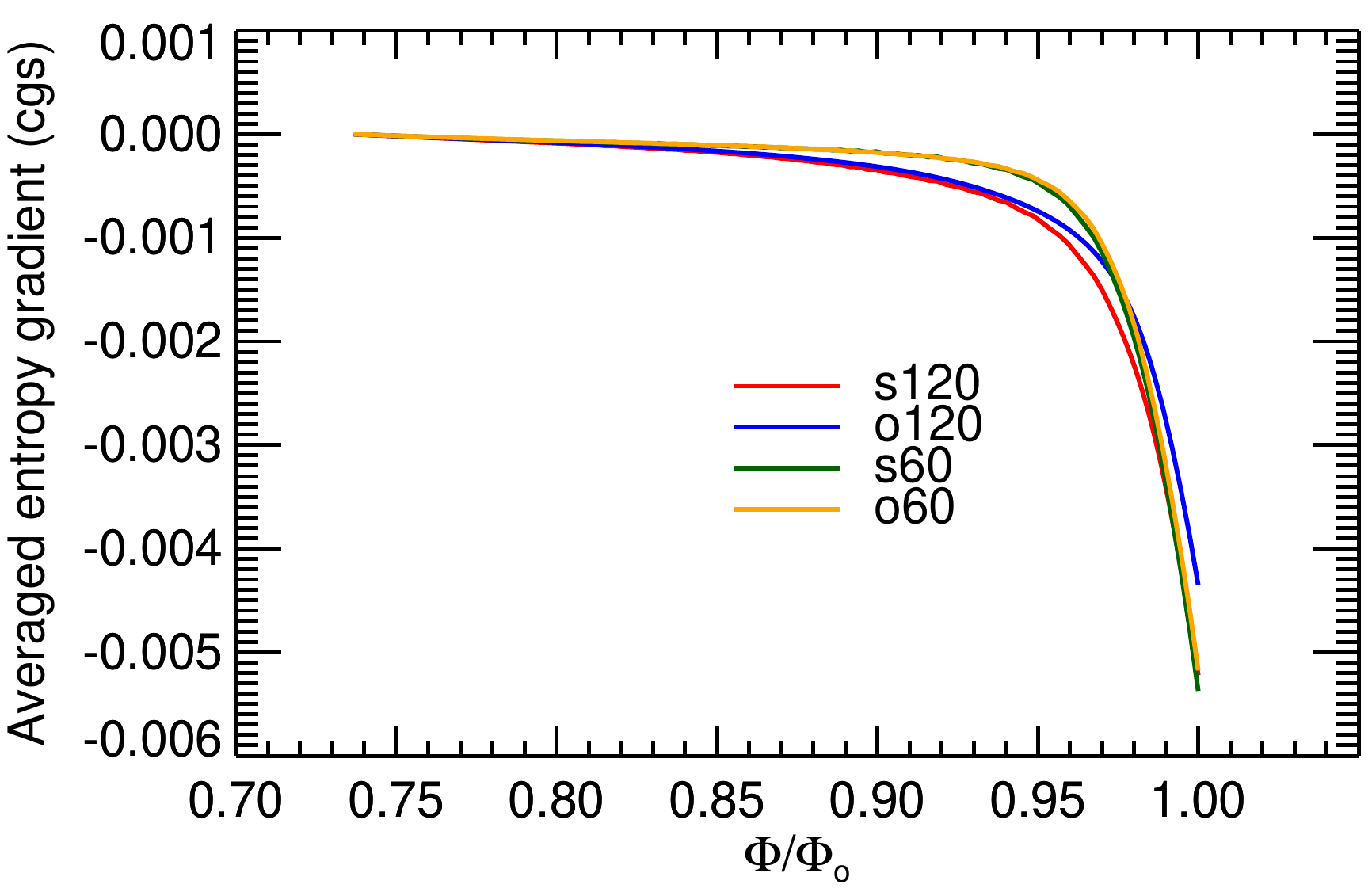}}
\caption{As in Figure \ref{fig:dsdr} but comparing spherical and oblate simulations.  Cases $s120$, $o120$, $s60$ and $o60$ are included, as indicated by the legend.}
\label{fig:dsdr_SO}
\end{figure}

In order to further investigate the effect of oblateness on the entropy stratification, we compare four cases in Figure \ref{fig:dsdr_SO}.  As expected, there is little difference between cases $s60$ and $o60$ since the oblateness is small (see Table \ref{Tab:parameters}).   However, the normal entropy gradient in case $o120$ is significantly smaller in magnitude than in case $s120$.  Several factors contribute to this result.  One is the larger surface area, which dilutes $F_u$ in the upper convection zone as noted in the previous paragraph.  Another is the somewhat less efficient convective transport, as noted above in the discussion of Figure \ref{fig:flux_balance-120}.  Yet another is the larger depth of the convection zone, which tends to weaken thermal gradients in general.  For example, consider the normal entropy flux at the equator where $\uvn$ is parallel to the radial direction.  There $\del S \bdot \uvn = \pd S/\pd r = C_P \left[H_\rho^{-1} - (\gamma H_P)^{-1}\right]$, where $H_P$ and $H_\rho$ are the pressure and density scale heights.  The weaker effective gravity due to the centrifugal force tends to increase both $H_P$ and $H_\rho$, which decreases $\pd S/\pd r$.  Viewed another way, a given entropy difference $\Delta S$ is stretched out over a wider radial distance at low latitudes in an oblate star because of the relatively large spacing between the potential surfaces.

\subsection{Emergent Flux and Gravity Darkening}\label{sec:darkening}

As mentioned in Section \ref{sec:intro}, one of the major theoretical results associated with rapidly-rotating, oblate stars is the concept of gravity darkening.  \textbf{This was originally predicted by \citet{Zeipel1924a} for the particular case of massive stars with radiative envelopes.   In this case the surface flux is dominated by the radiative flux $F_r \propto \del T$.  If $T$ is constant on equipotential surfaces, $T = T(\Phi)$ (see \S\ref{sec:IC} for the case of adiabatic stratification), then $F_r \propto \del T = (dT/d\Phi) \del \Phi \propto g_{eff}$, where $g_{eff}$ is the effective gravity (see Section \ref{sec:intro}).  Since $g_{eff}$ is reduced at the equator, the Von Zeipel model predicts that the equatorial regions should possess less energy flux than the polar regions and should therefore appear darker.  One problem with this result is that the radiative envelopes of oblate stars are unlikely to be barotropic ($T \neq T(\Phi)$), as emphasized by \citet{Rieutord16}.  Still, the qualitative prediction of gravity darkening has indeed been verified observationally for massive stars, though the quantitative contrast between equator and pole is less than predicted \citep{van2001,Domiciano2005,McAlister2005,Monnier2007,Zhao2009,van2012}.}   

However, the von Zeipel model does not apply to late-type stars with convective envelopes.  In these stars, the surface flux is not dominated by $F_r$ so the line of reasoning that gives rise to gravity darkening should be regarded as questionable.  Several attempts have been made to extend the Von Zeipel model to late-type stars but they all involve assumptions about the nature of the convective heat transport.  \citet{lucy67} used mixing-length models of convection to predict a result that resembles gravity darkening but with a weaker latitudinal dependence, such that the emergent flux is $\propto g_{eff}^{0.32}$.  However these mixing-length models are local in nature and do not take into account coherent structures such as banana cells that are known to dominate convective heat transport in global simulations of convection in rotating spherical (or spheriodal) geometries.  \citet{elara11} and \textbf{\citet{Rieutord16}} also presented a model that in principle should apply to convective envelopes as well as radiative envelopes.  However, this model assumes that the convective heat flux is directed normal to $\Phi$ surfaces, which may or may not be true.  Global convection simulations often exhibit a significant latitudinal heat flux that can have a substantial influence on the flux that eventually emerges through the outer boundary \citep[e.g.][]{Miesch2005}.

In this paper, we are the first to address the issue of gravity darkening from the perspective of global, 3D simulations of convection in oblate stars.

\begin{figure}[!htp]
\centering
\subfigure[]{\includegraphics[width=0.45 \textwidth, angle=0]{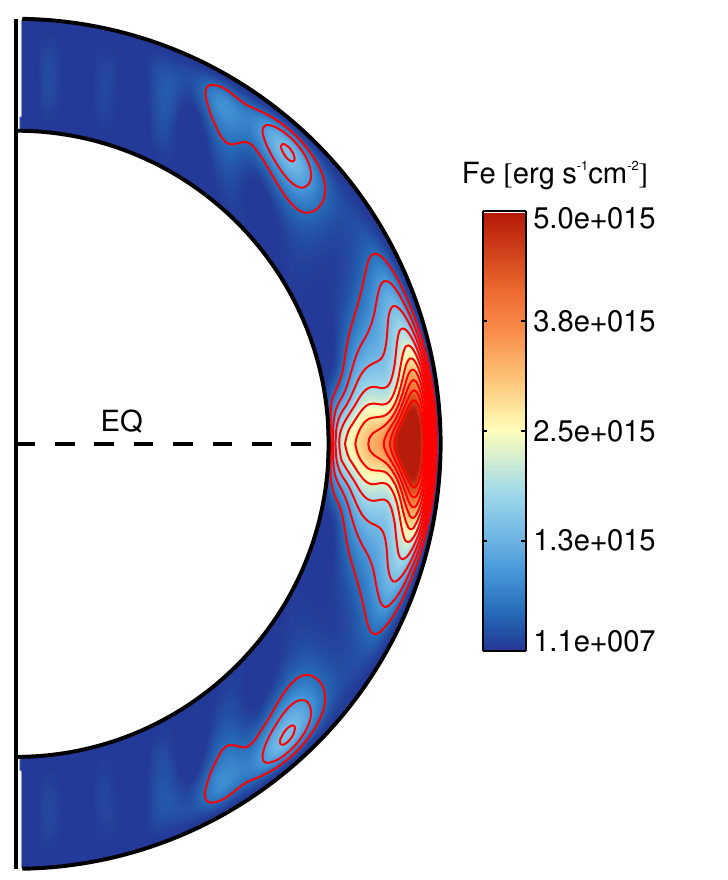}}
\subfigure[]{\includegraphics[width=0.51 \textwidth, angle=0]{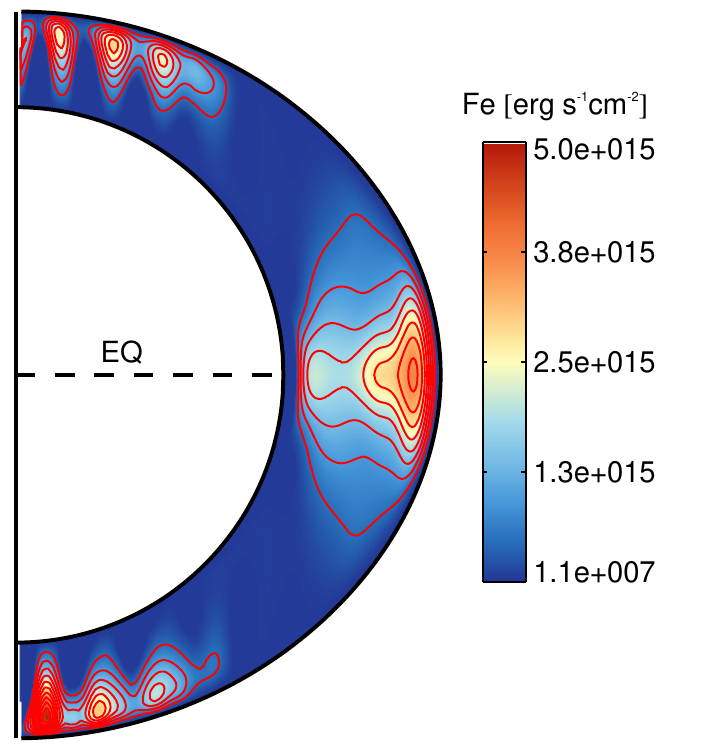}}
\caption{Shown are snapshots of the enthalpy flux $F_{e}$ in cases \textit{(a)} $s120$ and \textit{(b)} $o120$, averaged over longitude.}
\label{fig:Fe_SO}
\end{figure}

As noted in Section \ref{sec:flux}, the convective enthalpy flux $F_{e}$ in our simulations peaks in the upper convection zone (see Fig.\ \ref{fig:flux_balance-120}). Both warm upflows and cool downflows serve to transport flux out of the star.   However, this enthalpy flux is not uniform in latitude, as demonstrated in Figure \ref{fig:Fe_SO} which shows the zonally-averaged $F_e$ in cases $s120$ and $o120$. This shows efficient heat transport through the convective envelope near the equatorial region in both cases.  This is attributed to the banana cells that arise because of the influence of rotation on convection.  The alignment of convective banana cells with the rotation axis allows for more efficient heat transport. 

We also find that more enthalpy flux flows through the polar regions in case $o120$ than in case $s120$. This difference arises because of the thinner depth of the convection zone at high latitudes, which tends to increase the radial temperature and entropy gradient as discussed in Section \ref{sec:flux}.  This in turn increases the buoyant driving, making the convection more vigorous.  The greater vigor of polar convection in case $o120$ is likely also associated with the decoupling of the polar and equatorial convection modes as discussed in Section \ref{sec:patterns}.  The prominent vertical striping of the polar enthalpy flux apparent in Figure \ref{fig:Fe_SO}$b$ is associated with strong mean flows as discussed in Section \ref{sec:mean}.

\begin{figure}[!htp]
\includegraphics[width=0.9 \textwidth, angle=0]{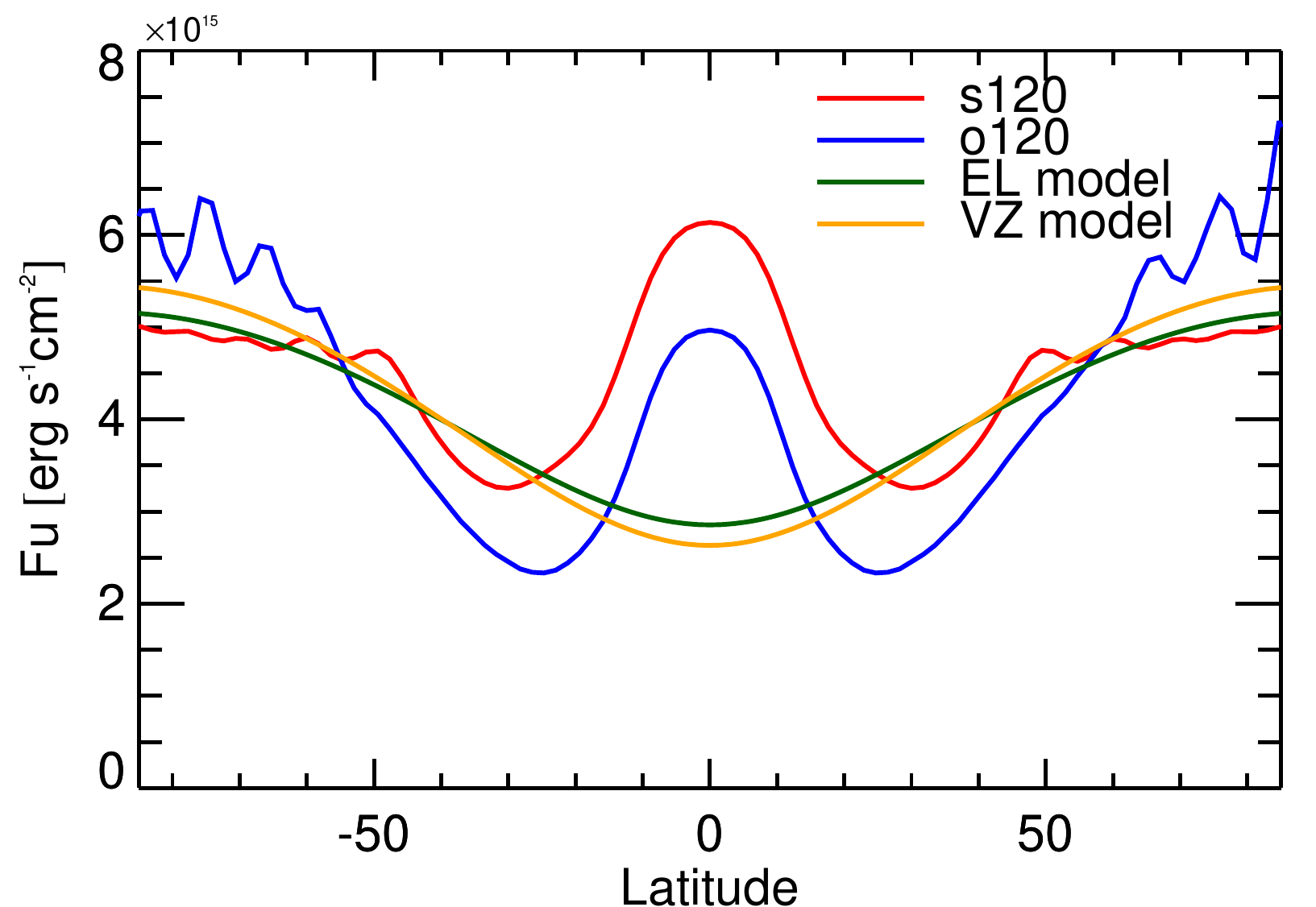}
\caption{The emergent (normal) flux $F_u$ through the outer equipotential surface ($\Phi = \Phi_o$) is shown as a function of latitude for cases $s120$ (red line) and $o120$ (blue line).  Each curve is averaged over longitude and over a time period of 10 days.  Also shown are the emergent flux profiles predicted by \citet[][orange line]{Zeipel1924a} and \citet[][green line]{elara11}.}
\label{fig:Fu_SO}
\end{figure}

The enthalpy flux is within the star but only the heat flux coming out at the surface can be observed.  In our model, the heat flux passing through the outer surface of the star is carried by the diffusive entropy flux, $F_u$.  Thus, this is the quantity to consider in order to assess the latitudinal dependence of the emergent flux, within the context of gravity darkening.  Figure \ref{fig:Fu_SO} shows the comparison of latitudinal entropy flux $F_{u}$ at the outer surface simulated in cases $s120$ and $o120$. For comparison, we have also plotted the profiles predicted by \citet{Zeipel1924a} and \citet{elara11} in Figure \ref{fig:Fu_SO}. 

There are two noteworthy results to be gleaned from Figure \ref{fig:Fu_SO}.  First, a comparison between cases $s120$ and $o120$ clearly shows an enhanced polar heat flux in the latter case.  The oblateness will tend to make the poles brighter, as in the standard gravity darkening paradigm \citep{Zeipel1924a,elara11}.  However, in stark contrast to the standard paradigm, case $o120$ exhibits a secondary peak in the emergent heat flux at the equator.  This is the second noteworthy result and, unlike the first, is somewhat unexpected.

Though the emergent flux is carried by $F_u$, both of these results can be ultimately attributed to the nature of the deeper enthalpy flux $F_e$ illustrated in Figure \ref{fig:Fe_SO}.  Since $F_e$ must vanish at the surface due to the impenetrable boundary condition, there must be a convergence of the outward enthalpy flux just below the surface, which preferentially heats the fluid in the polar and equatorial regions.  Since the temperature is fixed at the outer surface, this heating increases the normal temperature and entropy gradients.  In other words, the enhanced enthalpy flux $F_e$ in the polar and equatorial regions is converted into an emergent diffusive entropy flux $F_u$.

Heat transport by global convective motions under the influence of rotation and stratification is not taken into account in the analysis of \citet{Zeipel1924a} and \citet{elara11}.  Only global, 3D convection simulations can explicitly capture these effects.  Ours are the first such simulations and they are calling into question common assumptions.  For example, since the imposed heat flux on the bottom boundary is uniform, a non-uniform heat flux across any equipotential surface $\Phi$ necessarily implies a nonzero heat flux along $\Phi$ surfaces somewhere deeper in the convection zone, provided that the simulation is equilibrated.  This violates the assumptions made by both \citet{Zeipel1924a} and \citet{elara11}.

Our results imply that rapidly-rotating, oblate, lower-mass stars with convective envelopes might not have darker equators after all.   However, there is a caveat to this conclusion that has to do with the parameter regimes that we have considered in our simulations.  It is well known that banana cells are the linearly preferred instability modes for convection in rapidly-rotating spherical shells \citep{Miesch2005}.  Thus, as the Rayleigh number Ra is increased beyond the critical value for convection to set in, the equatorial modes are excited first, causing the convective enthalpy flux to peak at the equator.  As Ra is increased further, the polar modes are excited.  Thus, the energy flux profiles for cases $s120$ and $o120$ shown in Figures \ref{fig:Fe_SO} and \ref{fig:Fu_SO}, featuring prominent minima at mid latitudes, are commonly seen in global spherical convection simulatons at moderate Rayleigh numbers \citep[e.g.][]{Gilman77,Miesch2000,Yadav16}.  At the much higher Rayleigh numbers characteristic of real stars, these mid-latitude minima may fill in at least partially, making the emergent heat flux more uniform.   Thus, we may be overestimating the magnitude of the latitudinal variation.  Still, the enhanced brightness at the poles and equator are real physical effects, grounded in the physics of rotating convection.  These are the regions where convective structures can align with the rotation axis, enhancing the efficiency of the convective energy transport.

The simulations of \citet{Yadav16} in particular demonstrate another interesting effect; the suppression of the equatorial heat transport by strong differential rotation.  This occurs at high Rayleigh numbers and moderate rotational influence (Ro $\lesssim 1$), where the relative differential rotation $\Delta \Omega/\Omega_0$ is largest.  The resulting differential rotation peaks strongly at the poles.  If this were to occur in oblate stars, it could enhance the gravity darkening, making the equator relatively dim, as in the classical picture of Von Zeipel, though for different reasons.  However, stars that are rotating rapidly enough to be oblate are likely to possess low Rossby numbers (Ro $<< 1$), where the Yadav et al simulations exhibit an enhanced heat flux at the equator.  Furthermore, oblateness and magnetism may prevent the rotational shear from becoming strong enough to suppress the equatorial heat transport; dynamo simulations by Yadav et al.\ with the same parameters exhibit a mid-latitude minimum in the heat flux as described in the previous paragraph.

More generally, rapidly-rotating stars are known to exhibit vigorous dynamo action which could influence the results reported here.  In addition to altering the convective heat transport, magnetism can influence the latitudinal dependence of the heat flux through polar starspots and faculae.  Another relevant phenomenon that we have neglected is small-scale surface convection.  The steep superadiabatic entropy and density stratification in the surface layers of late-type stars drives small-scale convective motions that cannot be resolved in global models.  Familiar examples include solar granulation and supergranulation.  These will tend to homogenize the outgoing heat flux, possibly mitigating any latitudinal variations that may be established deeper in the convection zone.  Further research is needed to establish how all of these effects contribute to the emergent flux in oblate, late-type stars.  As mentioned previously, the results presented here should be regarded as a baseline for further study and certainly not the final word.

\section{Mean Flows}\label{sec:mean}

Some aspects of the differential rotation were discussed in Section \ref{sec:DR1}, within the context of stellar observations, previous global convection simulations and previous models of oblate stars.  Here we look at the mean flows (differential rotation and meridional circulation) in more detail, considering the influence of oblateness on their spatial structure.

\begin{figure}[!htp]
\subfigure{
\includegraphics[width=0.9 \textwidth, angle=0]{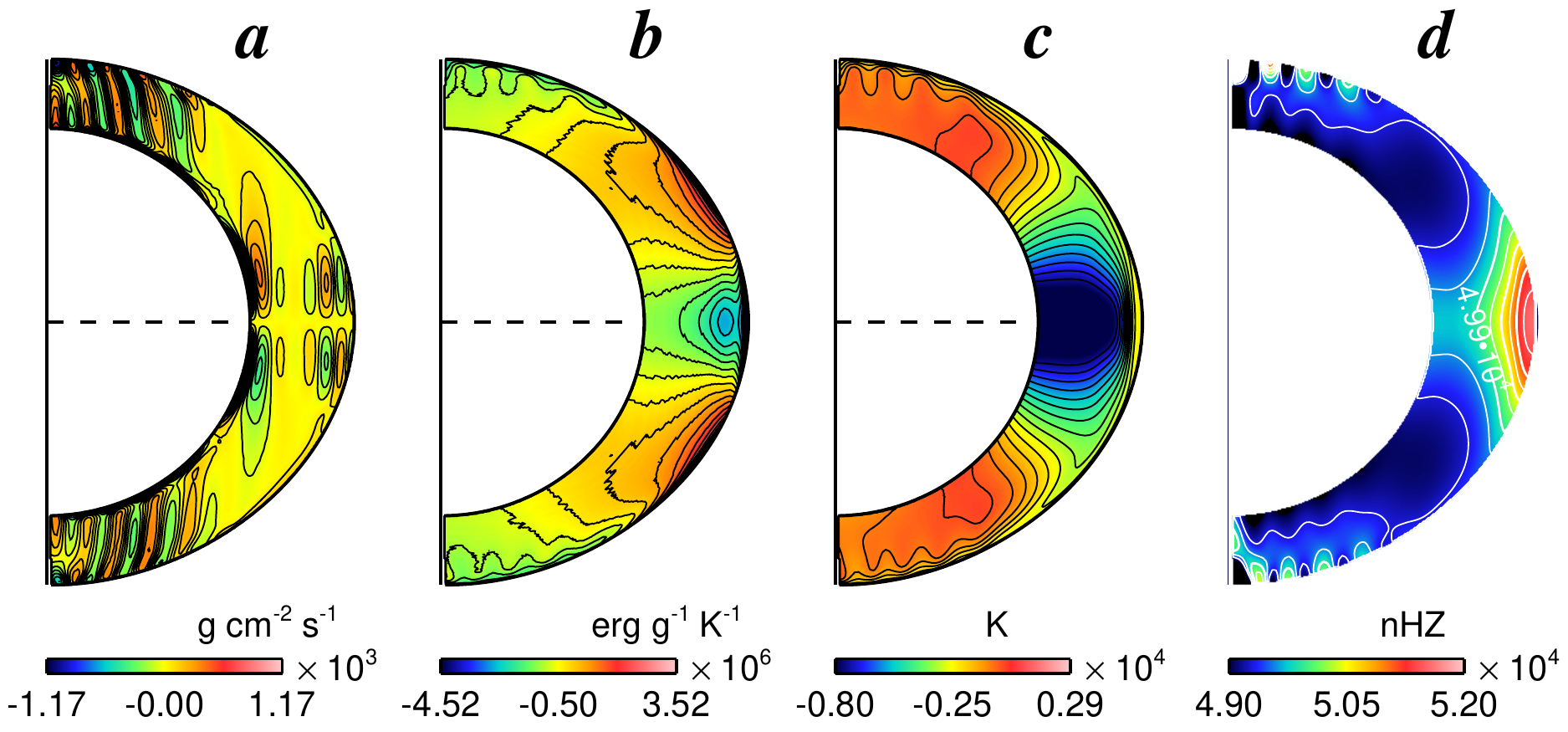}}
\subfigure{
\includegraphics[width=0.9 \textwidth, angle=0]{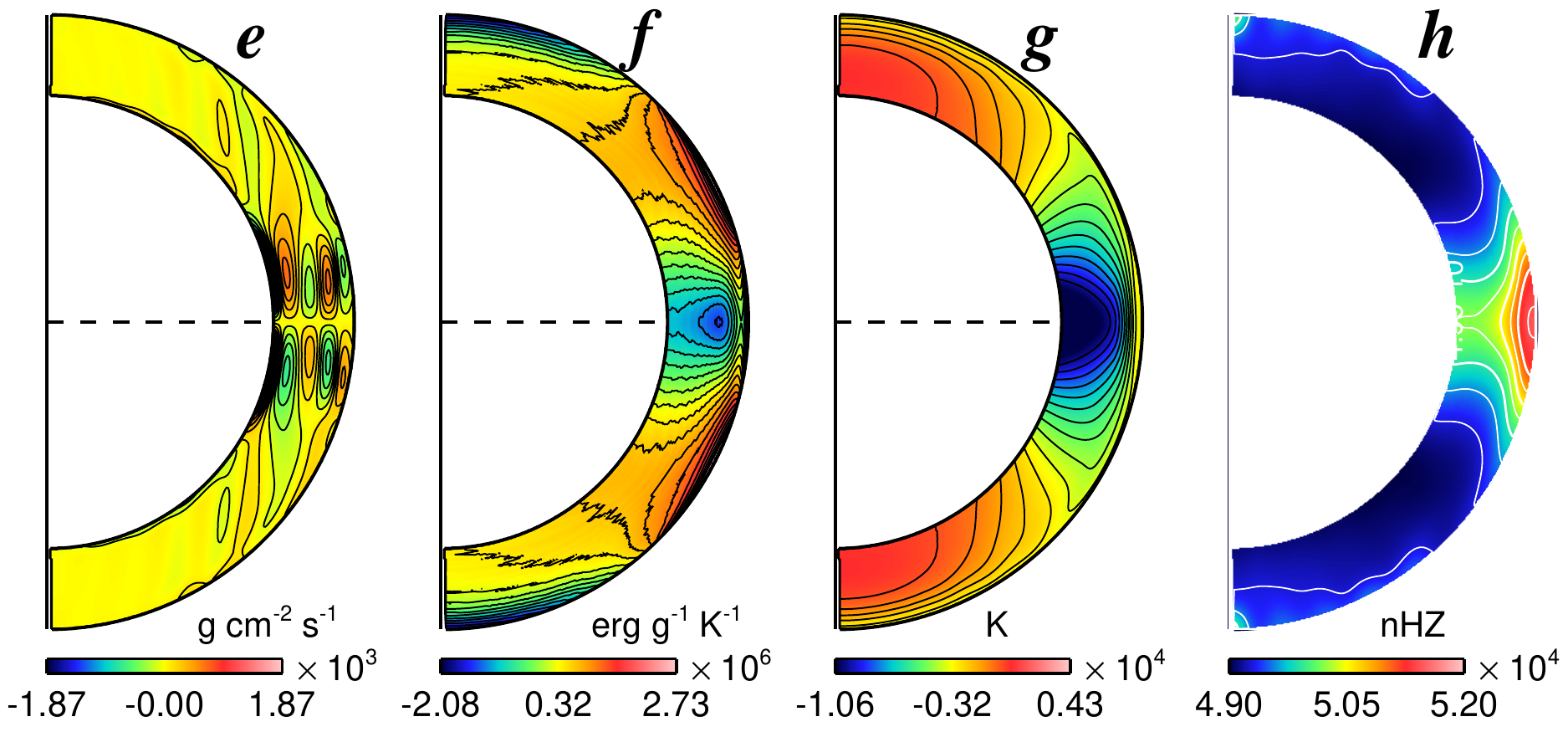}}
\caption{Mean flows averaged over longitude and over a time interval of $10$ days for cases $o120$ (top row) and $s120$ (bottom row). Shown are \textit{(a, e)} latitudinal mass flux $\left<\rho u_{\theta}\right>$, \textit{(b, f)} specific entropy perturbation $S^{'}$, \textit{(c, g)} temperature perturbation $T^{'}$ and \textit{(d, h)} differential rotation, expressed in terms of the angular velocity $\Omega$. Note that the top figures have the same polar radius as those at bottom, but are zoomed out to keep figures in each panel aligned. The color bar in each panel also differs as indicated in order to highlight contours.}
\label{fig:MC_120}
\end{figure}

We define mean flows as averages of the velocity field $\mathbf{u}$ or the mass flux $\rho \mathbf{u}$ over longitude.  Longitudinal averages are denoted by angular brackets, e.g.\ $\langle u_\phi \rangle$.  The differential rotation (in nHz) is expressed in terms of the angular velocity as follows:
\begin{equation}\label{eq:omega}
\Omega = \frac{1}{2\pi} \left ( \frac{\langle u_{\phi} \rangle}{r\sin \theta} + \Omega_{0} \right ).
\end{equation}
We define thermal variations $S^{'}$ and $T^{'}$ by averaging over longitude and time and then subtracting the spherically-symmetric component in order to highlight variations relative to the mean stratification.  The mean flows for cases $o120$ and cases $s120$ are shown in Figure \ref{fig:MC_120}. 

As mentioned in Section \ref{sec:DR1}, the simulations considered here all have a solar-like differential rotation profile, with relatively fast rotation at the equator that decreases toward higher latitudes (Fig.\ \ref{fig:MC_120}$d$,$h$).  This is a common feature of convection simulations at low Rossby number and is established by the convective angular momentum transport associated with banana cells \citep[e.g.][]{Miesch2005,Brown2008}.  These rotation profiles are generally in thermal wind balance, maintained in part by baroclinic torques that offset the tendency for $\Omega$ contours to align with the rotation axis in accordance with the Taylor-Proudman theorem.   It is this thermal wind effect that is responsible for the poleward gradients that dominate the temperature and entropy profiles in Figure \ref{fig:MC_120}, frames $b$, $c$, $f$, and $g$.

However, superposed on this monotonic decrease of $\Omega$ from equator to pole, case $o120$ exhibits a striking series of alternating zonal jets in both polar regions (Fig.\ \ref{fig:MC_120}$d$).  These are associated with a similar series of strong meridional circulation cells that can be seen prominently in the latitudinal mass flux plot of Figure \ref{fig:MC_120}$a$.  Though the zonal jets are confined to the upper convection zone, the meridional cells extend throughout the convection zone, with a significant poleward tilt from bottom to top.   Similar features are also present in cases $o100$ and $o110$ (not shown).

The faint thermal signature in Fig.\ \ref{fig:MC_120}$b$,$c$ suggests that these are axisymmetric convection cells, with the zonal jets arising from the Coriolis-induced deflection of meridional flows toward and away from the rotation axis near the outer surface.  This thermal signature is also seen in the enthalpy flux of Figure \ref{fig:Fe_SO}$b$ and in the emergent flux of Figure \ref{fig:Fu_SO} (as the high-latitude wiggles).   The presence of these convective cells near the polar regions in case $o120$ is likely due in part to the larger entropy gradient since the polar convection zone depth is relatively small compared to the equatorial depth (see Section \ref{sec:flux}).   In any case, they appear to be induced by the oblateness.  

At lower latitudes, outside the tangent cylinder, the meridional flow in both cases $s120$ and $o120$ exhibits a multiple-cell structure characterized by alternating regions of northward and southward mass flux (Fig.\ \ref{fig:MC_120}$a$,$e$).  This too is a common feature in global convection simulations and is attributed to the nature of the convective Reynolds stress induced by banana cells \citep{feath15}.

\begin{figure}[!htp]
\includegraphics[width=0.98 \textwidth, angle=0]{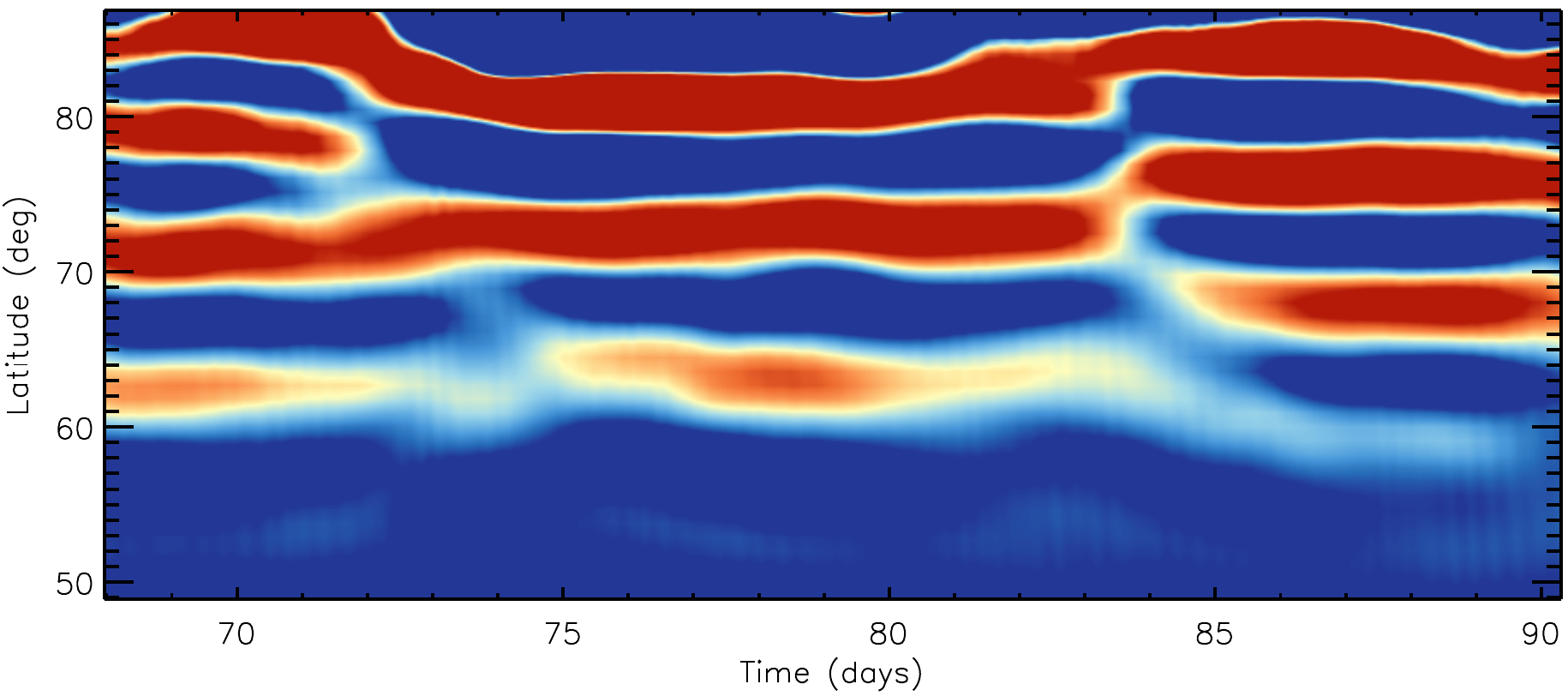}
\caption{Angular velocity $\Omega$ on the outer surface for case $o120$ as a function of latitude and time. The view focuses on the north polar region, latitudes about 50$^\circ$.  The color saturates at $(4.9,5.2) \times 10^{5}$ nHZ, with red denoting larger values and blue smaller values.}
\label{fig:DF_O120}
\end{figure}

To study the temporal evolution of the high-latitude zonal jets in case $o120$, we have plotted in Figure \ref{fig:DF_O120} a latitude-time map of $\Omega$ in the vicinity of the north pole (latitudes $\geq 50^{\circ}$).  Recall that the rotation period in this case is about 0.23 days (a factor of 120 less than the Sun), so the time period covered in the plot spans over 100 rotation periods (and over 20 convective turnover times).  Some of the zonal jets persist for this entire interval, though they meander in latitude.  Each jet is about 4-5$^\circ$ wide.

\section{Summary}\label{Sec:summary}

We have preformed the first global 3D hydrodynamic simulations of convection in the oblate envelopes of rapidly-rotating solar-type stars using the recently introduced compressible high-order unstructured spectral difference (CHORUS) code (WLM15).   The CHORUS code employs unstructured grids, giving it the capability to deal with oblate spheroidal shell geometries.   This unique tool allows us to investigate the influence of the oblateness on the convection and mean flows, including heat transport, differential rotation, and the thermal structure of the convective envelope.

We have found that the influence of rotation and oblateness leads to an enhancement of the emerging heat flux in the polar and equatorial regions.  This is in stark contrast to the conventional concept of gravity darkening and has important observational implications; oblate, low-mass stars with convective envelopes may not exhibit gravity darkening in the same way that higher-mass stars do.  In particular, the minimum suface brightness might occur at mid-latitudes rather than at the equator.

The enhancement of the polar heat flux due to oblateness arises for essentially the same reason as in classical gravity darkening arguments; the stronger effective gravity relative to lower latitudes enhances thermal gradients (see Section \ref{sec:flux}).  This makes the convection more vigorous.  However, the secondary peak in the emergent heat flux at the equator arises from physical effects that are not taken into account in the classical models of \citet{Zeipel1924a}, \citet{lucy67}, and \citet{elara11}.   Oblate stars are rapidly-rotating by nature, implying strong rotational constraints on the convective motions (low Rossby numbers).  In these parameter regimes, the convection in the equatorial region (outside the tangent cylinder) is dominated by banana cells, columnar convective modes aligned with the rotation axis.  These banana cells can transport heat more efficiently than at mid-latitudes.  

In short, the convective heat transport peaks at the poles and at the equator because this is where the convective structures can most easily organize themselves to mitigate the strong rotational constraints: vertical plumes at high latitudes and banana cells at low latitudes.  Furthermore, though banana cells are essentially laminar flow features, this fundamental argument also extends to more turbulent parameter regimes (higher Rayleigh and Reynolds numbers).  For example, in the high-resolution Boussinesq simulations of \citet{kagey08}, the convective columns break up into vortex sheets as the diffusion is decreased.  However, these vortex sheets are still aligned with the rotation axis outside the tangent cylinder and would still be expected to yield an enhancement of the equatorial heat flux.  

Thus, we expect that the general latitudinal dependence of the emergent heat flux in oblate solar-type stars to be similar to that found here, with maxima at the poles and equator.  However, as discussed in Section \ref{sec:darkening}, we may be overestimating the magnitude of the variation, which here is on the order of 40\% (see Figure \ref{fig:Fu_SO}).

The latitudinal dependence of the emergent heat flux is a symptom of the more general tendency for convection in rapidly-rotating spherical or spheroidal shells to separate into distinct polar and equatorial modes, inside and outside the tangent cylinder.  We find that the oblateness enhances this dichotomy, causing the polar regions to largely decouple from the equatorial regions (Figs.\ \ref{fig:con_pattern}, \ref{fig:Vn_SO}).  However, other aspects of the convective structure such as the power spectrum and the width of banana cells, are insensitive to the oblateness.

Another notable result from our simulations is the presence of banded zonal jets at high latitudes (Section \ref{sec:mean}).  Remarkably, these only occur in the oblate simulations, so they are clearly induced by the oblateness.  They appear to be due to narrow, axisymmetric convective modes that are approximately aligned with the rotation axis but significantly tilted (Fig.\ \ref{fig:MC_120}).  Each jet is about 4-5$^\circ$ wide and many persist for over 100 rotation periods, though they meander in latitude (Figure \ref{fig:DF_O120}).

It is unclear whether or not these jets would persist in the presence of magnetic fields.  All stars are ionized and, given the extreme parameter regimes, all stellar convection zones are expected to undergo vigorous dynamo action.   Dynamo-generated magnetic fields might suppress the polar jets directly by means of magnetic tension or indirectly by inhibiting the intrinsic self-organization processes of rapidly-rotating hydrodynamic turbulence (e.g.\ the inverse cascade of kinetic energy and the anisotropic suppression of this cascade that gives rise to the Rhines scale).  However, these jets do not appear to arise from the inverse cascade of smaller-scale vorticies.  Whether or not they occur in the presence of magnetism is a topic for future work.  Though CHORUS does not currently have the capability to solve the full set of MHD equations, we intend to implement this capability in the future.

The amplitude of the differential rotation is not very sensitive to either the rotation rate or the oblateness.  If we decrease $\nu$ and $\kappa$ as we increase the rotation rate in order to mainin a roughly constant supercriticality, we find that the differential rotation scales as $\Delta \Omega \propto \Omega_{0}^{0.89}$.   This result is for spherical (non-oblate) simulations and is qualitatively consistent with previous global convection simulations and stellar observations (Section \ref{sec:DR1}).  Alternatively, if $\nu$ and $\kappa$ are held constant, we find no clear scaling of $\Delta \Omega$ versus $\Omega_0$ (Fig.\ \ref{fig:delta_omega}).  However, the fraction of the total kinetic energy contained in the differential rotation, DRKE/KE, does increase monotonically in these same spherical simulations as $\Omega_0$ is increased from 0.526 in case $s60$ to 0.681 to case $s120$ (Table \ref{Tab:diagnostics}).

It is interesting that the oblateness tends to suppress this increase of DRKE/KE with $\Omega_0$.  For the oblate series of simulations, $o60$ through $o120$, the DRKE/KE ratio saturates at a maximum value of 0.61 for $\Omega_0 \gtrsim 80 \Omega_\odot$ (Table \ref{Tab:diagnostics}).  This trend is also visible in the surface differential rotation $\Delta \Omega$, which decreases slightly beyond about $90 \Omega_\odot$ (Fig.\ \ref{fig:delta_omega}).   The fractional differential rotation $\Delta \Omega / \Omega_0$ decreases with increasing rotation for both the spherical and the oblate stars.  Furthermore, we argue that the differential rotation is likely to be too weak to have a significant effect on the oblateness (Appendix \ref{Appen:A}).

We acknowledge that our simulations certainly have their limitations, the most important being the relatively laminar parameter regimes and the absence of magnetic fields.  However, as the first simulations of their kind, we regard these as a baseline for future work.  We have identified interesting features that warrant further study, such as a departure from the classical gravity darkening paradigm for late-type stars and the presence of polar jets induced by oblateness.  And, we have demonstrated some of the potential of the new CHORUS code, which promises to be a unique and powerful tool for the ongoing investigation of stellar and planetary convection.

\acknowledgments
We thank Matthias Rempel and the anonymous referee for helpful comments on the manuscript. Junfeng Wang is funded by a Newkirk graduate fellowship from the National Center for Atmospheric Center (NCAR). Chunlei Liang is grateful for the faculty start-up fund provided by The George Washington University and a recent CAREER Award (award number 1554005) from the National Science Foundation.  NCAR is operated by the University of Corporation for Atmospheric Research under sponsorship of the National Science Foundation. 
\newpage

\appendix

\section{EFFECT OF DIFFERENTIAL ROTATION ON OBLATENESS}\label{Appen:A}

A fully self-consistent simulation would properly capture the nonlinear interaction between the convection, the differential rotation, and the oblateness, allowing the star to deform in response to the differential rotation that the convection establishes.  However, this poses significant computational challenges and we show in this Appendix that even if we were to overcome these challenges, it would be unlikely to change our results.

In this appendix we investigate the effect of differential rotation on the oblateness of rotating stars in a simple model. In this model, the desired oblate spheroidal shell is obtained by deforming a spherical shell with the uniformly rotating rate $\Omega_{0}$ while the mass and angular momentum in the $z$ direction is kept unchanged.  We neglect convection and meridional circulation in our analysis; the only motion is the differential rotation.  Furthermore, to make the problem more analytically tractable, we assume that the differential rotation follows a conservative rotation law, $\Omega = \Omega(\lambda)$ where $\lambda = r \sin\theta$ is the cylindrical radius, so the centrifugal force can be represented as the gradient of a potential.

The rotation profile in the oblate spheroidal shell is assumed to be a linear function of $\lambda$;
\begin{equation}
\Omega = \Omega_{a} + \delta (\frac{\lambda}{R_{o}})\Omega_{0}
\label{eqn:ro_ln}
\end{equation}
where $R_{o}$ is the outer radius of the original spherical shell, $\delta$ is the fractional differential rotation
and $\Omega_{a}$ is a coefficient that is adjusted to conserve angular momentum as described below.

Thus, the effective gravitational potential is
\begin{equation}
\Phi = -\frac{GM}{r} -(\frac{1}{2}\Omega_{a}^{2}\lambda^{2} + \frac{2}{3}\delta\frac{\Omega_{a}\Omega_{0}}{R_{o}}\lambda^{3}
+ \frac{1}{4}\delta^{2}\frac{\Omega_{0}^{2}}{R_{o}^{2}}\lambda^{4}).
\end{equation}
We then compute the structure of the oblate spheroidal shell as described in Section {\ref{sec:MP}}, using this potential instead of equation (\ref{eqn:potential}).

During the deformation from the spherical shell to the oblate spheroidal shell with the desired linear rotation profile, the mass within the domain is kept unchanged,
\begin{equation}
\int_{o}^{2\pi}\int_{0}^{\pi}\int_{R_{i}}^{R_{o}}\rho \sin \theta r^{2}d\varphi d\theta dr =
\int_{o}^{2\pi}\int_{0}^{\pi}\int_{r_{1}(\theta)}^{r_{2}(\theta)} \rho \sin \theta r^{2}d\varphi d\theta dr,
\label{eqn:mass_def_con}
\end{equation}
where $R_{i}$ is inner radius for the spherical shell and $r_{1}(\theta)$ and $r_{2}(\theta)$ are the inner and outer radius for the oblate spheroidal shell.

Meanwhile, the angular momentum in the $z$ direction is also conserved during the transformation,
\begin{equation}
\int_{o}^{2\pi}\int_{0}^{\pi}\int_{R_{i}}^{R_{o}}\rho \sin \theta r^{2} \Omega_{0} \lambda^{2}d\varphi d\theta dr = \int_{o}^{2\pi}\int_{0}^{\pi}\int_{r_{1}(\theta)}^{r_{2}(\theta)}\rho \sin \theta r^{2} \Omega \lambda^{2}d\varphi d\theta dr.
\label{eqn:ang_def_con}
\end{equation}
Note that the density profile is as described in Section \ref{sec:IC} and the integrals (\ref{eqn:mass_def_con}) and (\ref{eqn:ang_def_con}) are computed numerically, on a grid as described in Section \ref{sec:der_grid}.  By solving Equations (\ref{eqn:mass_def_con}) and (\ref{eqn:ang_def_con}), $\Omega_{a}$ is obtained if $\delta$ is specified.

We vary the amplitude of the differential rotation by varying $\delta$ and, for each value of $\delta$, we compute the oblateness.  The results of this analysis are listed in Table \ref{Tab:oblateness} for $\Omega_0 = 120 \Omega_\odot = 3.12 \times 10^{-4}$ rad s$^{-1}$.  In all cases $\Omega_a < \Omega_0$ because the shell is required to have the same mass and angular momentum as a spherical shell with the same rotation rate.  For example, for no differential rotation ($\delta = 0$), the larger moment of inertia associated with the oblate shell requires $\Omega_a = 2.736\times 10^{-4}$ rad s$^{-1}$ = 105 $\Omega_\odot$ to achieve the same angular momentum as a spherical shell rotating at a rate of $\Omega_0 = 120 \Omega_\odot$.  For $\delta > 0$, $\Omega$ varies from $\Omega_a$ at the rotation axis ($\lambda = 0$) to $\Omega_a + \delta (O+1) \Omega_0$ at the equator ($\lambda = (O+1) R_o$).  Averaging the minimum and maximum values of $\Omega$ then gives us a fiducial rotation rate of $\Omega_f = \Omega_a + \delta (O+1) \Omega_0 / 2$.

The fractional differential rotation can then be defined as $\Delta \Omega / \Omega_f \sim \delta (O+1) \Omega_0 / \Omega_f$.  For $\delta = 0.36$ this gives a fractional differential rotation of 0.5.  Thus, according to Table \ref{Tab:oblateness}, varying the fractional differential rotation from 0 ($\delta = 0$) to 0.5 ($\delta = 0.36$) only changes the oblateness by about 0.22\%.

This can be compared with the fractional differential rotation $\Delta \Omega / \Omega_0$ achieved in our simulations, as shown in Figure \ref{fig:delta_omega}.  As shown there, all cases have a $\Delta \Omega$ of less than $16 \times 10^{-6}$ rad s$^{-1}$.  This corresponds to fractional differential rotation of about 0.10 for $\Omega_0 = 60 \Omega_\odot$ and a fractional differential rotation of about 0.05 for $\Omega_0 = 120 \Omega_\odot$.  Thus, the amplitude of the differential rotation achieved in our simulations is well within the range spanned in Table \ref{Tab:oblateness}.

Though the differential rotation profiles in our simulations are not precisely linear as expressed in equation (\ref{eqn:ro_ln}), these results suggest that the influence of the differential rotation on the oblateness should be negligible.

\begin{deluxetable}{lcccccc}
\footnotesize
\tablecolumns{11}
\tablewidth{0pc}
\tablecaption{$\delta$ and oblateness $O$}
\tablehead{
\colhead{$\delta$} & \colhead{$0.00$}   & \colhead{$0.04$}    & 
\colhead{$0.12$}    & \colhead{$0.20$}    & 
 \colhead{$0.28$}  & \colhead{$0.36$}
}
\startdata
$\Omega_{a}$  & $2.736$ & $2.640$ & $2.448$ & $2.256$ & 
$2.064$ & $1.871$ \\
$O$ & $11.21$\%  & $11.17$\% & $11.11$\% & $11.05$\% & $11.02$\% & $10.99$\%

\enddata
\tablecomments{$\delta$ indicates the degree of the differential rotation and $\Omega_a$ is the axial rotation in units of $10^{-4}$ rad s$^{-1}$. With $\delta$ increasing, both $\Omega_{a}$ and the oblateness $O$ decrease, but the latter does so only slightly.}
\label{Tab:oblateness}
\end{deluxetable}

\section{OUTWARD ENERGY FLUX IN AN OBLATE STAR}\label{Appen:B}

In this Appendix we derive the equation of flux balance in an oblate star (\ref{eq:Fbalance}) and give explicit expressions for $F_e$, $F_k$, $F_r$, and $F_u$.

We begin by integrating Equation (\ref{eqn:energy}) over a volume $V_{\cal S}$ that is bounded on the bottom by the lower boundary of the computational domain, defined by $\Phi = \Phi_i$, and on the top by an arbirtary geopotential within the computational domain, $\Phi = \Phi_{\cal S}$.  This yields
\begin{equation}\label{eq:EVS}
\frac{d}{dt} \int_{V_s} E dV = - \int_A \left[\left(E + p\right) \mathbf{u} - \mathbf{u} \bdot \mathbf{\tau} + \mathbf{f} \right] \bdot d\mathbf{A} + \int_{V_s} \rho \mathbf{u} \bdot \del \Phi dV  ~~~.
\end{equation}
In deriving equation (\ref{eq:EVS}) we have used Stokes' theorem to express the volume integral of the flux divergences as a surface integral and we have expressed the effective gravity as the gradient of the Roche potential, eq.\ (\ref{eqn:potential}).  The area $A$ in the surface integral includes both the inner boundary at $\Phi = \Phi_i$, which we call surface ${\cal S}_i$, and the upper boundary of $V_s$ at $\Phi = \Phi_{\cal S}$, which we call surface ${\cal S}$.

We proceed by assuming that the flow is in a statistically steady state.  This implies that the time derivative on the left-hand-side of eq.\ (\ref{eq:EVS}) can be neglected.  Together with equation (\ref{eqn:mass_conservation}), it also implies that the mass flux is approximately divergence-free, $\del \bdot (\rho \mathbf{u}) = 0$.  The divergence-free nature of the mass flux also follows from the low Mach number and nearly adiabatic, hydrostatic background stratification, which forms the basis for the anelastic approximation \citep[e.g.][]{Jones2011}.  This allows us to write the last term in eq.\ (\ref{eq:EVS}) as
\begin{equation}
\int_{V_s} \rho \mathbf{u} \bdot \del \Phi dV = 
\int_{V_s} \del \bdot \left(\rho \mathbf{u} \Phi\right) dV = 
\Phi_i \int_{{\cal S}_i} \rho \mathbf{u} \bdot d\surf_i + 
\Phi_{\cal S} \int_{{\cal S}} \rho \mathbf{u} \bdot d\surf = 0 ~~~. 
\end{equation}
The final equality follows from the impenetrable lower boundary condition, $\mathbf{u} = 0$ on surface ${\cal S}_i$, and the divergence-free nature of the mass flux.  Note in particular that
\begin{equation}
\Phi_{\cal S} \int_{{\cal S}} \rho \mathbf{u} \bdot d\surf = 
\Phi_{\cal S} \int_{A} \rho \mathbf{u} \bdot d\mathbf{A} = 
\Phi_{\cal S} \int_{V_s} \del \bdot \left(\rho \mathbf{u}\right) dV = 0  ~~~.
\end{equation}
Thus, the buoyant and centrifugal acceleration do not change the total energy contained within the volume $V_s$.

Note also that, since $\mathbf{u} = 0$ on surface ${\cal S}_i$, the only contribution to the integrated flux on that surface is from the diffusive fluxes $\mathbf{f}$.  Equation (\ref{eq:Fbot}) then implies
\begin{equation}
- \int_{{\cal S}_i} \left[\left(E + p\right) \mathbf{u} - \mathbf{u} \bdot \mathbf{\tau} + \mathbf{f} \right] \bdot d\mathbf{A} = L  ~~~.
\end{equation}
Thus, in what follows we need only concern ourselves with the energy flux across surface ${\cal S}$.

Now consider the integrated enthalpy and kinetic energy flux, which can be written as
\begin{equation}
\int_{\cal S}  (E + p) \mathbf{u} \bdot d\surf = 
\int_{\cal S}  \left[\frac{\rho}{2} u^2 
+ \left(\frac{\gamma}{\gamma-1}\right) p\right] \mathbf{u} \bdot d\surf = 
\int_{\cal S}  \left[\frac{\rho}{2} u^2 + C_P ~ \rho T \right] ~ \mathbf{u} \bdot d\surf = (F_k + F_e) A_{\cal S} ~~~,
\end{equation}
where
\begin{equation}\label{eq:Fe}
F_e = \frac{1}{A_{\cal S}} ~ \int_{\cal S} \overline{\rho}C_{p} u_{n} (T - \overline{T}) ~ d{\cal S} ~~~,
\end{equation}
and
\begin{equation}\label{eq:Fk}
F_k = \frac{1}{A_{\cal S}} ~ \int_{\cal S} \frac{1}{2} \overline{\rho} u^2 ~ u_n ~ d{\cal S} ~~~.
\end{equation}

The overbars on $\overline{\rho}$ and $\overline{T}$ denote averages over the surface ${\cal S}$ and 
$A_{\cal S}$ denotes the total area of the surface ${\cal S}$.  In this derivation we have used the expressions $E = \rho u^2/2 + p/(\gamma-1)$, $\gamma = C_p/C_v$, and the ideal gas law $p = (C_p - C_v) \rho T$, where $C_p$ and $C_v$ are the specific heat at constant pressure and volume respectively.  We have also made use of the expression $\del \bdot (\rho \mathbf{u}) = 0$ which, as noted above, implies that there is no net mass flux through the surface ${\cal S}$.  So, $\int_{\cal S} \rho \overline{T} \mathbf{u} \bdot d\surf = \overline{T} \int_{\cal S} \rho \mathbf{u} \bdot d\surf = 0$.  In equations (\ref{eq:Fe}) and (\ref{eq:Fk}) we have also assumed that $\rho \approx \overline{\rho}$ so that the results can be easily compared with anelastic simulations (see WLM15).

The thermal flux terms arising from radiative and entropy diffusion follow straightforwardly from the expressions for $\mathbf{f}$ given in Section \ref{sec:hydro}:  
\begin{equation}
F_{r} = - \frac{1}{A_{\cal S}} \int_{\cal S} \kappa_{r} \rho C_{p} \del T \bdot d\surf \approx
- \kappa_{r} \overline{\rho} C_{p} \del \overline{T} \bdot \uvn ~~~,
\end{equation}
and
\begin{equation}
F_{u} = - \frac{1}{A_{\cal S}} \int_s \kappa \rho T \del S \bdot d\surf
\approx - \kappa \overline{\rho}\overline{T}\del\overline{S} \bdot \uvn ~~~.
\label{eqn:Fu}
\end{equation}
We neglect the viscous flux $\mathbf{u} \bdot \mathbf{\tau}$ which is generally small in global stellar convection simulations \citep[e.g.][]{Miesch2005,Brown2008}. 

Putting all of these results together then implies that we can rewrite the energy balance equation (\ref{eq:EVS}) as the balance of fluxes in eq.\ (\ref{eq:Fbalance}).


\end{document}